\documentclass[aps,footinbib, notitlepage, longbibliography,eqsecnum]{revtex4-1}
\pdfoutput=1
\usepackage{graphicx}
\usepackage{amssymb}
\usepackage{amsmath}
\usepackage{amsfonts}
\usepackage{hyperref}
\usepackage{color}
\usepackage[normalsize]{subfigure}
\usepackage{xcolor,xspace}
\usepackage{bm}
\usepackage[capitalize]{cleveref}

\DeclareFontFamily{OT1}{pzc}{}
\DeclareFontShape{OT1}{pzc}{m}{it}{<-> s * [1.10] pzcmi7t}{}
\DeclareMathAlphabet{\mathpzc}{OT1}{pzc}{m}{it}

\DeclareMathAccent{\ring}{\mathalpha}{operators}{"17}
\providecommand{\st}[1]{_{\text{#1}}}

\providecommand{\sfrac}[2]{#1/#2}
\providecommand{\ut}[1]{^{\text{#1}}}

\def\onehalf{\frac{1}{2}}

\def\bra{\ensuremath{\langle}}
\def\ket{\ensuremath{\rangle}}
\def\eq{\st{eq}}

\def\ueff{\ut{eff}}

\def\const{\mathrm{const}}

\def\pd{\partial}

\def\im{\mathrm{i}}

\def\dt{\delta t}

\def\b0{\bv{0}}

\def\Ccal{\mathcal{C}}

\def\Hcal{\mathcal{H}}
\def\Mcal{\mathcal{M}}

\def\Ocal{\mathcal{O}}
\def\Pcal{\mathcal{P}}
\def\Scal{\mathcal{S}}
\def\pbc{\ut{(p)}}
\def\Dbc{\ut{(D)}}

\def\DirNoFl{^{(\mathrm{D}')}}
\def\DirCP{^{(\mathrm{D})}}
\def\hyp13{{_1 F_3}}
\def\bcs{boundary conditions\xspace}
\def\frict{\eta}
\def\kbT{\Theta}

\def\dps{\displaystyle}
\def\d{\mathrm{d}}
\def\mass{\mathcal{A}}
\def\lattsp{\Delta x}
\def\Mred{\Mcal}

\def\cro{_\times}

\def\paperI{Ref.\ \cite{gross_first-passage_2017}\xspace}

\newcommand{\bitem}{\begin{itemize}}
\newcommand{\eitem}{\end{itemize}}
\newcommand{\benum}{\begin{enumerate}}
\newcommand{\eenum}{\end{enumerate}}
\newcommand{\bblock}[1]{\begin{block}{#1}}
\newcommand{\eblock}{\end{block}}
\newcommand{\bmini}[1]{\begin{minipage}{#1}}
\newcommand{\emini}{\end{minipage}}
\newcommand{\btab}[1]{\begin{tabular}{#1}}
\newcommand{\etab}{\end{tabular}}
\newcommand{\btabn}[1]{\begin{tabular}{#1}}
\newcommand{\etabn}{\end{tabular}}
\newcommand{\beq}{\begin{equation}}
\newcommand{\eeq}{\end{equation}}
\newcommand{\bv}[1]{\mathbf{#1}}

\begin{document}
\title{First-passage dynamics of linear stochastic interface models:\\ numerical simulations and entropic repulsion effect}
\author{Markus Gross}
\email{gross@is.mpg.de}
\affiliation{Max-Planck-Institut f\"{u}r Intelligente Systeme, Heisenbergstra{\ss}e 3, 70569 Stuttgart, Germany}
\affiliation{IV.\ Institut f\"{u}r Theoretische Physik, Universit\"{a}t Stuttgart, Pfaffenwaldring 57, 70569 Stuttgart, Germany}
\date{\today}

\begin{abstract}
A fluctuating interfacial profile in one dimension is studied via Langevin simulations of the Edwards-Wilkinson equation with non-conserved noise and the Mullins-Herring equation with conserved noise. 
The profile is subject to either periodic or Dirichlet (no-flux) boundary conditions.
We determine the noise-driven time-evolution of the profile between an initially flat configuration and the instant at which the profile reaches a given height $M$ for the first time.
The shape of the averaged profile agrees well with the prediction of weak-noise theory (WNT), which describes the \emph{most-likely} trajectory to a \emph{fixed} first-passage time.
Furthermore, in agreement with WNT, on average the profile approaches the height $M$ algebraically in time, with an exponent that is essentially independent of the boundary conditions.
However, the actual value of the dynamic exponent turns out to be significantly smaller than predicted by WNT. 
This ``renormalization'' of the exponent is explained in terms of the entropic repulsion exerted by the impenetrable boundary on the fluctuations of the profile \emph{around} its most-likely path. 
The entropic repulsion mechanism is analyzed in detail for a single (fractional) Brownian walker, which describes the anomalous diffusion of a tagged monomer of the interface as it approaches the absorbing boundary.
The present study sheds light on the accuracy and the limitations of the weak-noise  approximation for the description of the full first-passage dynamics.
\end{abstract}


\maketitle

\section{Introduction}
\label{sec_intro}
In the present study, first-passage events arising in the Edwards-Wilkinson and the Mullins-Herring equation for various \bcs are investigated based on Langevin simulations. 
The obtained results for the spatio-temporal evolution of the profile are confronted to WNT --- which has been discussed in a preceding paper (\paperI) --- and to reduced models of (fractional) Brownian walkers.
In order to make the present study self-contained, the relevant models are briefly reviewed in the following.

We consider a one-dimensional interfacial profile $h(x,t)$, defined on a domain of size $L$ ($0\leq x\leq L$) governed by either the Edwards-Wilkinson (EW) equation \cite{edwards_surface_1982}
\beq \pd_t h = \frict \pd_x^2 h + \zeta\,,
\label{eq_EW}
\eeq
or the stochastic Mullins-Herring (MH) equation \cite{mullins_theory_1957, herring_effect_1950,krug_origins_1997}
\beq \pd_t h = -\frict \pd_x^4 h + \pd_x \zeta.
\label{eq_MH}
\eeq
The noise $\zeta$ is a Gaussian random variable of zero mean and correlation 
\beq \bra \zeta(x,t)\zeta(x',t')\ket = 2D\delta(x-x')\delta(t-t')\,.
\label{eq_noise}
\eeq
The ratio between the friction coefficient $\frict$ and the noise strength $D$ is related to the temperature via a fluctuation-dissipation relation (see below). 
The initial configuration is generally taken to be flat, 
\beq h(x,t=0)=0,
\label{eq_init_cond}\eeq 
and the profile is assumed to fulfill either periodic \bcs (p)
\beq h\pbc(x,t)=h\pbc(x+L,t),
\label{eq_H_pbc}\eeq 
or Dirichlet \bcs (D)
\beq h\Dbc(0,t)= 0=h\Dbc(L,t).
\label{eq_H_Dbc}\eeq 
When using the latter in conjunction with the MH equation, we additionally impose a no-flux condition at the boundaries, 
\beq \pd_x^3 h\DirNoFl(0,t) = 0 = \pd_x^3 h\DirNoFl(L,t),
\label{eq_H_noflux}\eeq 
which is indicated by a primed superscript (D').
For the MH equation with periodic or Dirichlet no-flux \bcs, the area under $h$, 
\beq \mass([h],t) \equiv \int_0^L \d x \, h(x,t)\,, \label{eq_mass}
\eeq
henceforth called the ``mass'', is conserved in time:
\beq \mass([h],t)=0.
\label{eq_zero_vol}\eeq 
In contrast, due to the presence of $\zeta$ instead of $\pd_x \zeta$ in \cref{eq_EW}, the mass is generally not conserved for the EW equation. 
In particular, for periodic \bcs, $\mass([h],t)$ behaves diffusively at large times \cite{gross_interfacial_2013, krug_origins_1997}, while for Dirichlet \bcs, $\bra \mass([h],t)\ket=0$ holds only as a time-average. 
In order to enforce \cref{eq_zero_vol} also for EW dynamics with periodic \bcs, we consider in this case instead of $h\pbc$ the profile 
\beq \tilde h\pbc(x,t)\equiv h\pbc(x,t) - \mass([h\pbc],t)/L, \qquad \text{(EW)}
\label{eq_height_redef}\eeq
which fulfills $\mass([\tilde h\pbc],t)=0$. 
In the simulations discussed here, the prescription in \cref{eq_height_redef} is applied at each time step.
In order to simplify notation, the tilde will be dropped henceforth.
We emphasize that \cref{eq_height_redef} is rather artificial from a physical point of view and is imposed here mainly in order to compare the different models under the common condition  $\bra \mass([h],t)\ket=0$. 

We focus on the stochastic evolution of $h(x,t)$ until the (random) first-passage time $T$, at which the profile has reached a given maximum height $M>0$ for the first time: 
\beq  \max_x h(x,T)= M.
\label{eq_firstpsg_cond}\eeq
The resulting (random) coordinate $x$ will be denoted in the following by $x_M$. 
\Cref{eq_firstpsg_cond} implies an \emph{absorbing boundary condition} for the profile at the height $M$ \cite{gardiner_stochastic_2009,redner_guide_2001}.
The absorbing boundary condition acts over the whole domain $[0,L]$ and represents an impenetrable repulsive barrier to the profile (see also Refs.\ \cite{delfino_interfaces_2013,belardinelli_thermal_2016}).
For a highly correlated system, such as an profile in the presence of a mass constraint [\cref{eq_zero_vol}], analytical solutions of the first-passage problem are technically difficult and are available only in certain limits (see, e.g., Refs.\ \cite{krug_persistence_1997, likthman_first-passage_2006, guerin_non-markovian_2012, bray_persistence_2013, cao_large_2015, meerson_macroscopic_2016}).
The first-passage dynamics of the profile is thus addressed here via numerical simulation of \cref{eq_EW,eq_MH}, as well as by relying on reduced descriptions of the effective (fractional) Brownian dynamics of a ``tagged monomer'', i.e., of $h(x_M,t)$.
Note that, in the absence of an absorbing boundary, the stochastic process governed by \cref{eq_EW,eq_MH} is fully Gaussian and underlies the well-studied phenomenon of interfacial roughening (see, e.g., Refs.\ \cite{krug_origins_1997, halpin-healy_kinetic_1995} as well as \cref{sec_roughening}).

A tractable approximation to the first-passage problem discussed here is provided by weak-noise theory (WNT), also known as macroscopic fluctuation theory \cite{freidlin_random_1998, bertini_macroscopic_2015, meerson_macroscopic_2016}. WNT of \cref{eq_EW,eq_MH} has been discussed in detail in \paperI.
WNT represents a leading-order saddle-point approximation to the first-passage problem and describes the \emph{most-likely} (``optimal'') trajectory between two states.
Specifically, within WNT, \cref{eq_firstpsg_cond} is replaced by a height constraint, $h(x,T)=M$, and the first-passage time $T$ is taken as a free, but constant, parameter. 
Accordingly, WNT neither takes into account fluctuation-induced interactions with the absorbing boundary nor the fact that the first-passage time $T$ follows a certain probability distribution.
However, it is shown here that, despite these limitations, WNT accurately captures the scaling functions of the averaged profile shape.
A significant difference nevertheless arises in the value of the dynamic exponent characterizing the time-dependence of the first-passage profile. 
Based on insights from models of (fractional) Brownian walkers, this difference is argued to be a genuine consequence of the fluctuations around the most-likely path near an impenetrable boundary. 

The first-passage problem of the MH equation discussed here and in \paperI is, \emph{inter alia}, physically relevant for noise-driven rupture of liquid films on substrates.
So far, typically films have been considered which are either linearly unstable with respect to small fluctuations of the interface or where the rupture proceeds via hole nucleation in the presence of disjoining pressure \cite{bausch_lifetime_1994, bausch_critical_1994,blossey_nucleation_1995,foltin_critical_1997, seemann_dewetting_2001,thiele_dewetting:_2001,thiele_importance_2002,tsui_views_2003,becker_complex_2003,gruen_thin-film_2006, fetzer_thermal_2007, croll_hole_2010, blossey_thin_2012, nguyen_coexistence_2014, duran-olivencia_instability_2017}. 
Here and in \paperI, we focus on linearized models in one dimension and assume absence of any deterministic force beside surface tension. In particular, we neglect the influence of disjoining pressure, which is experimentally justified for colloidal fluids \cite{lekkerkerker_life_2008,aarts_droplet_2008}. 
Accordingly, in this case film rupture is solely driven by noise.
This situation is analogous to the noise-driven breakup of a liquid nanojet, which has been analyzed within WNT in Ref.\ \cite{eggers_dynamics_2002} and studied experimentally and by simulations in Refs.\ \cite{petit_break-up_2012,hennequin_drop_2006,mo_mesoscopic_2015}.
Physical realizations of one-dimensional interfaces occur, e.g., in lipid bilayer membranes below their demixing transition \cite{honerkamp-smith_line_2008, thiam_biophysics_2013}.
The extension of the present study to two-dimensional interfaces as well as the incorporation of an interface potential are reserved for future work.

\section{Model and simulations}
\label{sec_sim}

\subsection{General aspects}

In the following, a number of relevant properties of the considered models are summarized.
It is useful to note that, dimensionally $[\frict]\sim [L]^z/[T]$, $[D]\sim [M]^2 [L]^{z-1} / [T]$, $[D/\eta] = [M]^2/[L]$, where $[M]$, $[L]$, and $[T]$ represent the fundamental dimensions of height, length, and time, respectively.
In order to facilitate the analysis of the first-passage dynamics, we recall the phenomenology of \emph{interfacial roughening} (see \cref{sec_roughening} for details).
To this end, we consider Eqs.\ \eqref{eq_EW} and \eqref{eq_MH} in the absence of an absorbing boundary condition.
In this situation, one can analytically determine the trajectory $h(x,t)$ as well as the \emph{roughness} $\bra |\delta h(x,t)|^2\ket$, where $\delta h(x,t)\equiv h(x,t) - h(x,0)$ is the relative height fluctuation. We consider either a flat initial condition, $h(x,0)=0$, or a thermal one. In the latter case, the roughness is calculated as an average over an ensemble of equilibrium profiles $h(x,0)$.
Since, for Dirichlet \bcs, the variance depends on position, we evaluate in the following $\bra h(x,t)\ket$ at a fixed location $x_M$ far from the boundaries (the precise value of $x_M$, however, is irrelevant for the general scaling behavior).
The roughness resulting from \cref{eq_EW,eq_MH} is characterized by three regimes \cite{majaniemi_kinetic_1996, abraham_dynamics_1989,racz_scaling_1991,antal_dynamic_1996, barabasi_fractal_1995, majaniemi_kinetic_1996, flekkoy_fluctuating_1995,flekkoy_fluctuating_1996,krug_origins_1997,taloni_generalized_2010, taloni_generalized_2012, taloni_correlations_2010,panja_probabilistic_2011,gross_interfacial_2013, halpin-healy_kinetic_1995}:
\beq \bra |\delta h(x_M,t)|^2\ket  \sim  \begin{cases} t, \quad &  t\lesssim \tau\cro, \\
t^{1/z}, & \tau\cro \lesssim t \lesssim \tau,\\
\const, &  t\gtrsim \tau,
\end{cases}\label{eq_roughness}\eeq
where
\beq z\equiv
\begin{cases}
2,\qquad \text{EW equation},\\ 
4,\qquad \text{MH equation},
\end{cases}
\label{eq_dynindex}\eeq 
is the dynamic index and $\tau$ denotes the roughening time. The latter coincides with the relaxation time of the (eigen-)mode with the largest ``wavelength'' that can be accommodated in the system: 
\beq 
\tau = \left(\frac{L}{\omega_1}\right)^z,\qquad
\omega_1 \equiv 
\begin{cases}
2\pi,\qquad &\text{periodic,}\\ 
\pi,\qquad &\text{standard Dirichlet,}\\
4.73\ldots \qquad &\text{Dirichlet no flux \bcs.}
\end{cases}
\label{eq_relaxtime}\eeq 
Dirichlet no-flux \bcs are considered only for the MH equation ($z=4$), in which case the value $\omega_1\simeq 4.73$ represents the smallest positive solution of the associated eigenvalue equation, $\cos\omega \cosh\omega=1$ [see \cref{app_prev_results}].
Within WNT, $\tau$ is in fact the characteristic time scale for the development of the first-passage profile in an equilibrium system (see \paperI). This property is confirmed by the present simulations.
Furthermore, $\tau\cro$ in \cref{eq_roughness} represents a \emph{cross-over time} related to the presence of a microscopic cutoff. While $\tau\cro=0$ in the continuum limit, for a one-dimensional lattice one has (see \cref{app_implement})
\beq \tau\cro=\tau \left(\frac{\omega_1}{\omega_{k\cro}}\right)^z,
\label{eq_crossover_time}\eeq 
with $\omega_{k\cro}\pbc = 2L/\lattsp$ and $\omega_{k\cro}\Dbc = L/\lattsp$ for periodic and (standard) Dirichlet \bcs, respectively, where $\lattsp$ is the lattice spacing. For Dirichlet no-flux \bcs, a numerical analysis yields $k\cro \gtrsim 0.5 L/\lattsp$, with the actual value depending on the particular problem under study [see \cref{app_bench}; the corresponding value of $\omega\DirNoFl_{k\cro}$ follows from the eigenvalue equation in \cref{eq6_DirNoFl_eigenvaleq}]. 

According to \cref{eq_roughness}, a tagged monomer of the profile exhibits standard Brownian diffusion at early times, followed by a subdiffusive regime characterized by a Hurst exponent \cite{family_dynamics_1991, meakin_growth_1993}
\beq H= \frac{1}{2z}.
\label{eq_Hurst}\eeq  
For a sufficiently large system, the latter regime dominates the roughening behavior.
A tagged monomer thus diffuses the distance $M$ approximately within the time [see \cref{eqR_tdiff}]
\beq \tau_D = \frac{M^{1/H}}{2[(2/\pi)\kbT\Gamma(1-z^{-1})]^z \frict},
\label{eq_tdiff}\eeq 
where $\kbT$ is the temperature [see \cref{eq_FDT} below].
The numerical prefactors in \cref{eq_tdiff} arise from a detailed analysis (see \cref{sec_roughening}) along with the two-time correlation function of the relative height fluctuations $\delta h(x,t)$ [see \cref{eqR_hr_correl}],
\begin{subequations}\begin{align} 
\bra \delta h(x,t) \delta h(x,s)^* \ket\st{flat} &\simeq (2/\pi)\frict^{1/z}\Gamma(1-z^{-1}) \kbT \left[ (t+s)^{1/z} - |t-s|^{1/z} \right], \label{eq_hr_rough}\\
\bra \delta h(x,t) \delta h(x,s)^* \ket\st{th} &\simeq (2/\pi)\frict^{1/z}\Gamma(1-z^{-1})  \kbT \left[ t^{1/z} + s^{1/z} - |t-s|^{1/z} \right], \label{eq_hr_fbm}
\end{align}\label{eq_hr_correl}\end{subequations}
corresponding to flat and thermal initial conditions, respectively.
The Gaussian stochastic process described by \cref{eq_hr_fbm} is a fractional Brownian motion (fBM) \cite{kolmogoroff_wienersche_1940, mandelbrot_fractional_1968, McCauley_hurst_2007, jeon_fractional_2010} \footnote{Note that, occasionally, different definitions of fractional Brownian motion are used in the literature, see, e.g., Refs.\ \cite{lim_fractional_2001,lim_self-similar_2002}.}.

For times $t\gtrsim \Ocal(\tau)$, all memory of the initial condition has been lost and the interface has reached its equilibrium roughness.
In this regime, the profile $h(x,t)$ follows a time-independent joint Gaussian distribution with a temperature (see \cref{app_equil_pdf})
\beq \kbT=\frac{D}{2 \frict} .
\label{eq_FDT}\eeq
This equation represents a fluctuation-dissipation relation for \cref{eq_EW,eq_MH}. For periodic \bcs, the one-point variance $\bra h(x,t)^2\ket$ is independent of position $x$ and is in equilibrium given by [see \cref{eq_Pss_var}]
\beq \bra |h\pbc|^2\ket = \frac{1}{6}\kbT L.
\label{eq_var_pbc}\eeq 
For Dirichlet \bcs, the equilibrium variance at the mid-point $x=L/2$ is given by [see \cref{eqBM_var_bridge,eqBM_var_areaconstr}]
\begin{subequations}\begin{align}
 \bra [ h\Dbc(L/2,t) ])^2\ket &= \frac{1}{2}\kbT L 
 \intertext{and}
 \bra [ h\DirNoFl(L/2,t) ]^2\ket &= \frac{7}{8}\kbT L,
\end{align}\label{eq_var_Dir}\end{subequations}
in the cases without and with an additional mass constraint [\cref{eq_zero_vol}], respectively.

The first-passage dynamics is generally distinct in the transient and the equilibrium regime, which, within WNT, correspond to $T/\tau \ll 1$ and $T/\tau \gg 1$, respectively.
However, for the actual stochastic equations \eqref{eq_EW} and \eqref{eq_MH}, the first-passage time $T$ is a random quantity and $T/\tau$ is therefore not an appropriate parameter \footnote{In fact, the mean first-passage time $\bra T\ket$ is a complicated function of the system parameters and is not exactly known for the models considered here.}.
We thus define instead the \emph{reduced height} 
\beq \Mred \equiv \frac{M}{\sqrt{\kbT L}} \sim \frac{M}{\bra h^2\ket^{1/2}} ,
\label{eq_hm_red}\eeq 
which is essentially the ratio between the maximum height $M$ and the equilibrium variance of the profile [\cref{eq_var_pbc,eq_var_Dir}].
For $\Mred\ll 1$, the profile is likely to reach the height $M$ within its roughening phase, whereas for $\Mred\gg 1$, the profile is fully equilibrated before the first-passage event occurs.
The definition in \cref{eq_hm_red} is consistent with the fact that the transient regime corresponds to diffusion times $\tau_D\ll\tau$ [see \cref{eq_tdiff,eq_relaxtime}].
We henceforth take $\tau_D$ and $\tau$ as the fundamental time scales for the first-passage dynamics in the transient and equilibrium regimes, respectively.

\begin{figure}[t]\centering
	\includegraphics[width=0.45\linewidth]{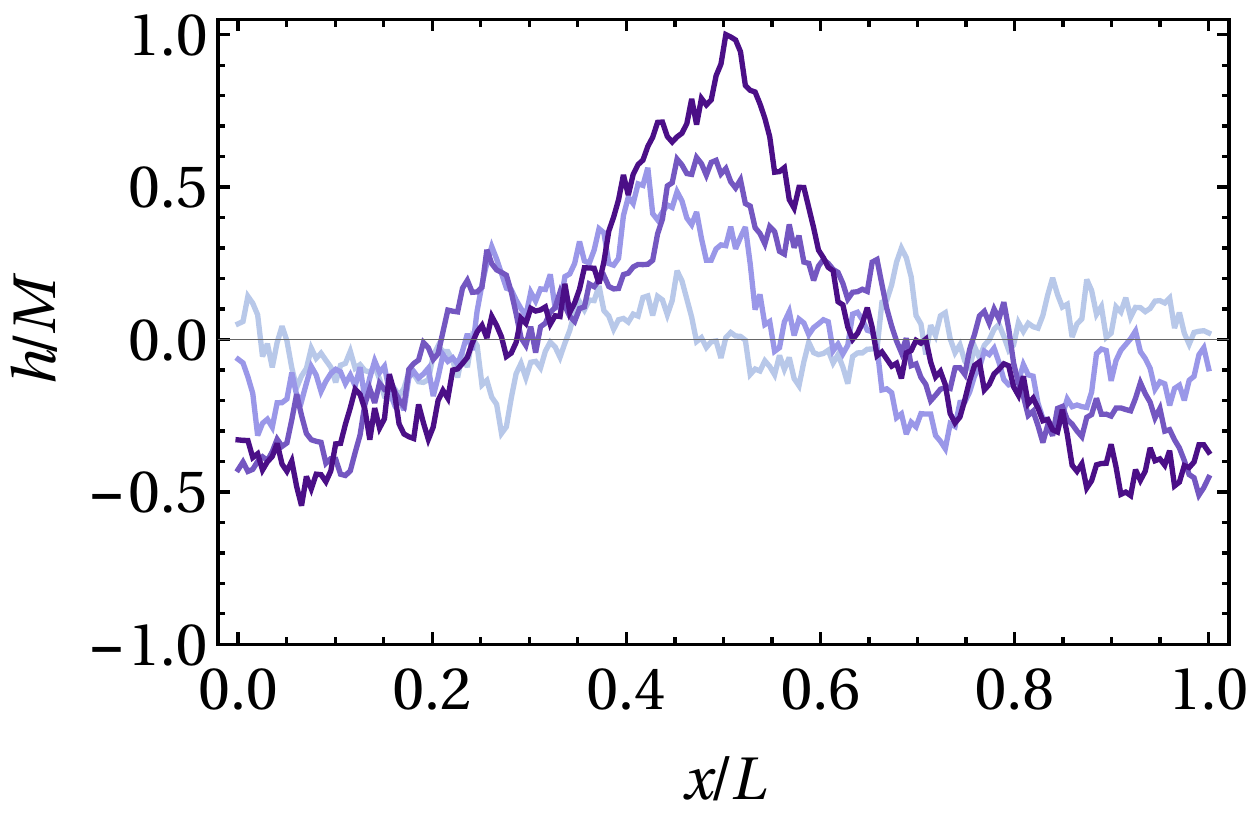}
	\caption{Typical time evolution of a profile $h(x,t)$ until the first-passage event (time $T$), at which the height $M$ is reached for the first time [see \cref{eq_firstpsg_cond}]. The plotted profiles represent snapshots at times $(T-t)/\tau\pbc\simeq 0.5, 0.05, 0.02, 5\times 10^{-5}$ (from center bottom to top) obtained from a simulation of \cref{eq_EW} for periodic \bcs [and with \cref{eq_height_redef} imposed at each time step] on a lattice size of $L=200\Delta x$. $\tau\pbc$ denotes the fundamental relaxation time reported in \cref{eq_relaxtime}. The initial profile is flat [\cref{eq_init_cond}, not shown]. Owing to translational invariance in the case of periodic \bcs, the individual profiles are shifted such that the maximum occurs at the center of the box, i.e., $h(x_M=L/2,T)=M$. }
    \label{fig_prof_evol}
\end{figure}

\subsection{Implementation}
\label{sec_impl}
The stochastic equations in \cref{eq_EW,eq_MH} are discretized on a one-dimensional lattice comprising $N=L/\Delta x$ nodes with spacing $\Delta x$ and are solved using a standard forward Euler scheme with time step $\Delta t$ (see, e.g., Refs.\ \cite{krug_persistence_1997, karma_phase-field_1999}):
\beq h(x_i,t+\Delta t) = h(x_i,t) -  \eta\,\Delta t\,(-\nabla^2)^{z/2} h(x_i,t) + \sqrt{2D \Delta t} \nabla^{z/2-1}\tilde \zeta(x_i,t),
\label{eq_discretized}\eeq 
with $i=0,\ldots,N-1$.
The noise variables $\tilde \zeta(x_i)$ are uncorrelated Gaussian variables of zero mean and unit variance, $\bra \tilde \zeta(x_i,t) \tilde \zeta(x_j,t')\ket =\delta_{i,j}\delta_{t,t'}$. 
The discretized forms of  the derivative operators $\nabla^2$ and $\nabla^4$ as well as further technical details on the numerical simulations are provided in \cref{app_implement}.
In the simulations, a profile is generally initialized in a flat configuration [\cref{eq_init_cond}]. If an equilibrated system is required at the first-passage event, the height $M$ is chosen sufficiently large such that $T\gg \tau$ (see also \cref{sec_firstpsg}). 
\Cref{fig_prof_evol} exemplifies a typical time evolution of a profile governed by \cref{eq_EW} close to the first-passage event. 

The main object of the present study is the \emph{averaged} profile $\bra h(x,\dt)\ket$, which is obtained in the following way: let $\{ h^{(s)}(x,t) \}$, $s=1,\ldots ,S$ be an ensemble of profiles obtained from a total number of $S$ simulations. Let $T^{(s)}$ be the corresponding first-passage time, such that $h^{(s)}(x_M^{(s)}, T^{(s)})\geq M$ for the first time for any $x_M^{(s)}$. The averaged profile is defined as 
\beq \bra h(x,\dt)\ket \equiv \frac{1}{N(T\geq \dt)} \sum_{s=1}^{N(T\geq \dt)} h^{(s)}(x-X^{(s)},T^{(s)}-\dt),
\label{eq_def_avgprof}\eeq 
where $N(T\geq \dt)\leq S$ denotes the number of profiles for which $T^{(s)}\geq \dt$. 
Note that the averaged profile is a function of the time variable $\dt$, which is defined such that the first-passage event corresponds to $\dt=0$, i.e., $ \bra h(x_M,0)\ket =M$.
Depending on the model and the regime considered, we set either $X^{(s)}\equiv 0$ or $X^{(s)}\equiv x^{(s)}_M-L/2$, where the latter choice induces a shift of the location of the maximum $x_M^{(s)}$ to the center $L/2$ \footnote{As will be justified in the corresponding sections, we set $X^{(s)}=x^{(s)}_M-L/2$ generally in the transient regime and for periodic \bcs also in the equilibrium regime. For Dirichlet \bcs, we set $X^{(s)}=0$ in the equilibrium regime.}.

The finite value of the time step in \cref{eq_discretized} gives rise to two potentials errors: first, a profile can ``overshoot'' the boundary, i.e., instead of \cref{eq_firstpsg_cond} one finds $h^{(s)}(x_M,T)=M+\delta M^{(s)}$ with $\delta M^{(s)}>0$. This effect is taken into account by subtracting the individual $\delta M^{(s)}$ on the r.h.s.\ of \cref{eq_def_avgprof}. 
While the overshoot leads to slight changes of the observed scaling of the peak $\bra h(x,\dt)\ket$ for small $\dt$, 
it turns out to not significantly affect the intermediate asymptotics. 
Second, there is a certain finite probability that between two discrete time steps the profile has crossed the boundary \cite{clifford_simulation_1986,peters_efficient_2002}. Performing simulations with a decreased time step in a few cases indicate that the results here are essentially insensitive to this effect. 

\section{First-passage time}
\label{sec_firstpsg}

\begin{figure}[t]\centering
      \subfigure[]{\includegraphics[width=0.45\linewidth]{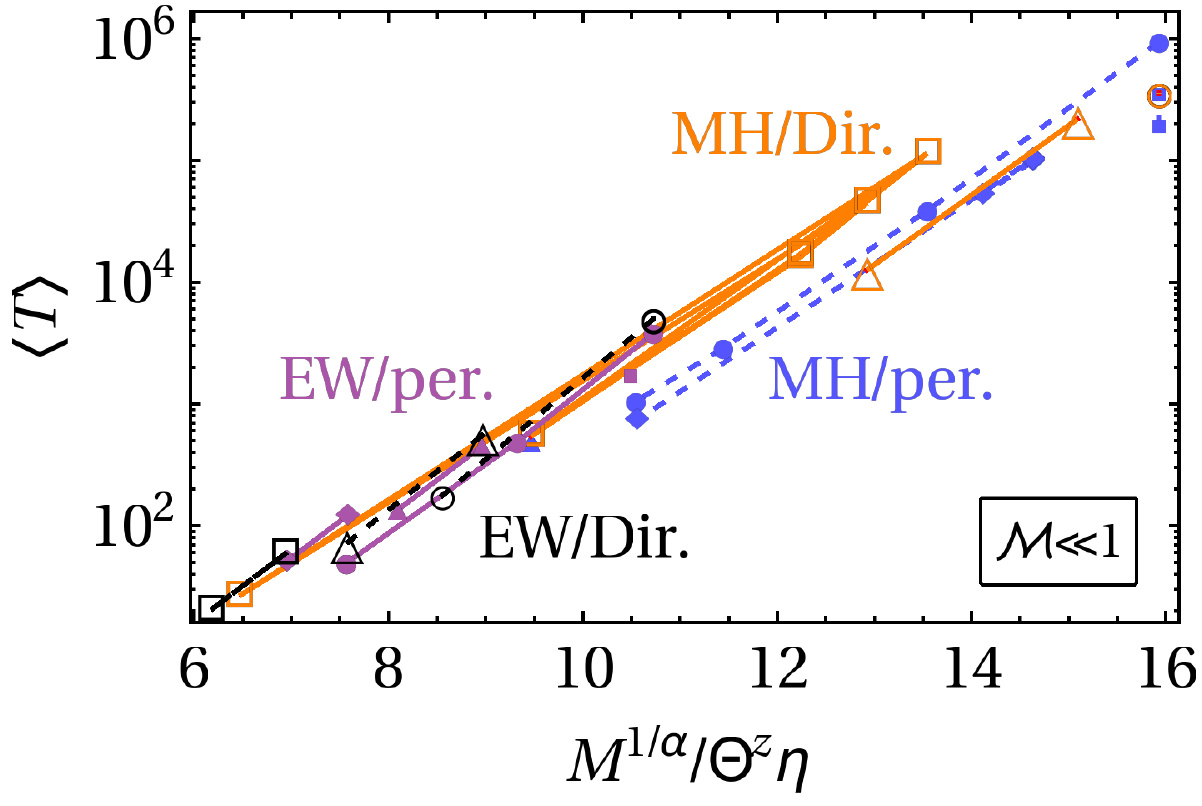} \label{fig_meanfpt_tr}} \qquad 
      \subfigure[]{\includegraphics[width=0.45\linewidth]{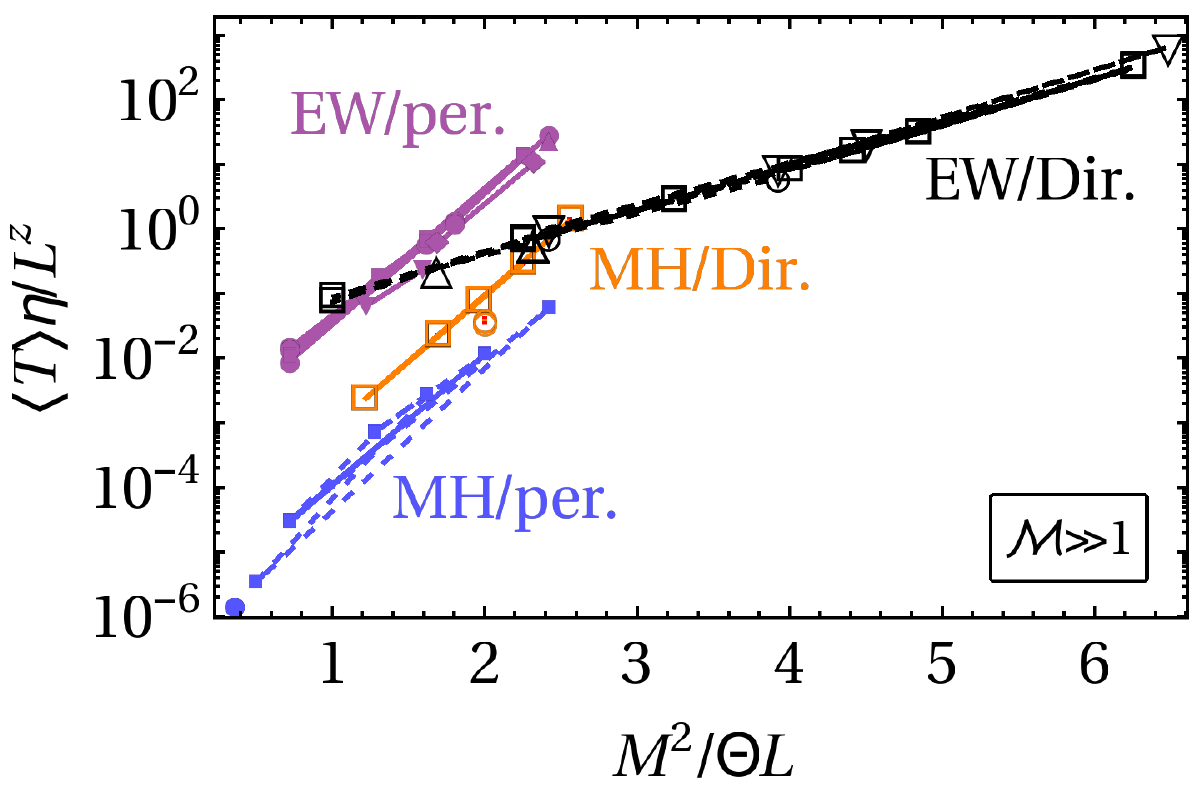} \label{fig_meanfpt_eq}} 
      \caption{Mean first-passage time $\bra T\ket$ for a profile with flat initial configuration [\cref{eq_init_cond}] in (a) the transient and (b) the equilibrium regime, corresponding to $\Mred\ll 1$ and $\Mred\gg 1$, respectively [see \cref{eq_hm_red}]. In (a), effective values $\alpha\st{EW}=0.29$ and $\alpha\st{MH}=0.17$ are used for the exponent $\alpha$ in the scaling relation \eqref{eq_meanfpt_tr_scaling}. Time is expressed in units of the simulation time step $\Delta t$. In (b), the axes are scaled according to \cref{eq_meanfpt_eq_approx}. The bulk dynamic equation and the \bcs are indicated by the labels near the data (purple solid symbols connected by solid lines: EW equation with periodic bc.s; black open symbols connected by dashed lines: EW equation with Dirichlet bc.s; blue solid symbols connected by dashed lines: MH equation with periodic bc.s; orange open symbols connected by solid lines: MH equation with Dirichlet no-flux bc.s).}
    \label{fig_meanfpt}
\end{figure}

Before addressing the profile dynamics, we briefly turn to the first-passage time $T$, i.e., the time at which the profile, starting from the initial configuration in \cref{eq_init_cond}, reaches the given height $M$ for the first time.
We remark that related first-passage problems of linear interface and polymer models have been studied previously in, e.g., Refs.\ \cite{krug_persistence_1997, kantor_anomalous_2007,chatelain_probability_2008,amitai_first-passage_2010, bray_persistence_2013}. Closed analytical expressions are, however, available only within certain approximations \cite{guerin_non-markovian_2012, cao_large_2015, likthman_first-passage_2006}.

The first-passage distribution $\Pcal_1(T)$ is discussed separately in \cref{app_fp_distr}.
For the models considered here, we find that $\Pcal_1(T)$ decays either exponentially or algebraically for large $T$, with an exponent smaller than $-2$.
Consequently, the mean first-passage time 
\beq \bra T\ket = \int_0^\infty \d T \, T \Pcal_1(T)
\label{eq_meanfp_time}\eeq 
is finite.
In order to obtain an estimate for $\bra T\ket$ in the transient regime, we recall that a tagged monomer traverses the distance between $h=0$ and $M$ within a time of order of $t^\alpha$, with $\alpha=1/(2z)$.
Specifically, based on \cref{eq_tdiff} one expects 
\beq \bra T\ket \sim \frac{M^{1/\alpha}}{\kbT^z \eta}.
\label{eq_meanfpt_tr_scaling}\eeq 
However, instead of the naive value $\alpha=1/(2z)$, we use in \cref{eq_meanfpt_tr_scaling} the effective values $\alpha\st{EW}\simeq 0.27-0.3$ and $\alpha\st{MH}\simeq 0.16-0.18$ in the case of EW and MH dynamics, respectively, which coincide with the values of the exponent characterizing the averaged path (see \cref{sec_EW,sec_MH}).
As demonstrated in \cref{fig_meanfpt_tr}, the scaling behavior of the mean first-passage time in the transient regime is well captured by the scaling relation \eqref{eq_meanfpt_tr_scaling} \footnote{We remark that, formally, using a value of $\alpha \neq 1/(2z)$ requires a factor with dimension $[M]^{2z-1/\alpha}$ to be present on the r.h.s.\ of \cref{eq_meanfpt_tr_scaling}.}.

In the equilibrium regime, \cref{eq_meanfpt_tr_scaling} does not provide a satisfactory description of the mean first-passage time.
Instead we recall that the steady-state probability distribution of the profile is Gaussian with a single-site variance given in \cref{eq_var_pbc,eq_var_Dir}.
We can thus consider a tagged monomer $h(x_M,t)$ as a fractional Brownian walker [$H=1/(2z) < 1/2$, see \cref{eq_hr_fbm}] in an effective harmonic potential $U(h) \simeq \sfrac{h^2}{\kbT L}$.
To leading order, the monomer dynamics can be approximated by a Markovian Brownian process ($H=1/2$), such that the present first-passage problem reduces to the well-known Kramers escape problem \cite{gardiner_stochastic_2009, malakhov_exact_1996}.
Accordingly, the mean-first-passage time of a tagged monomer in the equilibrium regime is expected to behave as
\beq \bra T\ket \simeq c_1 L^z \eta^{-1} \exp\left(c_2 \frac{M^2}{\kbT L}\right),
\label{eq_meanfpt_eq_approx} \eeq 
where $c_1$ and $c_2$ are fit parameters (independent of $M$, $L$ and $\kbT$) \footnote{The prefactor of the exponential in \cref{eq_meanfpt_eq_approx} can be motivated based on dimensional considerations: noting that  $[\eta]\sim L^z/T$ and $[D]\sim M^2 L^{z-1}/T$, a dimensionally consistent  ansatz for the prefactor is given by $L^a M^b \eta^{-1+b/2} D^{-b/2}$, with $a=z-b/2$. It turns out that a satisfactory scaling collapse of the data is possible with the simplest choice, $b=0$, which implies \cref{eq_meanfpt_eq_approx}.}. 
Essentially the same form as in \cref{eq_meanfpt_eq_approx} has been obtained in Ref.\ \cite{sliusarenko_Kramers-like_2010} for a fBM in a parabolic potential as well as in Ref.\ \cite{cao_large_2015} in the case of a Rouse polymer chain.
As demonstrated in \cref{fig_meanfpt_eq}, the simulation data pertaining to each model falls onto distinct master curves described by \cref{eq_meanfpt_eq_approx}. 

\section{Edwards-Wilkinson equation}
\label{sec_EW}

\begin{figure}[t]\centering
	\includegraphics[width=0.6\linewidth]{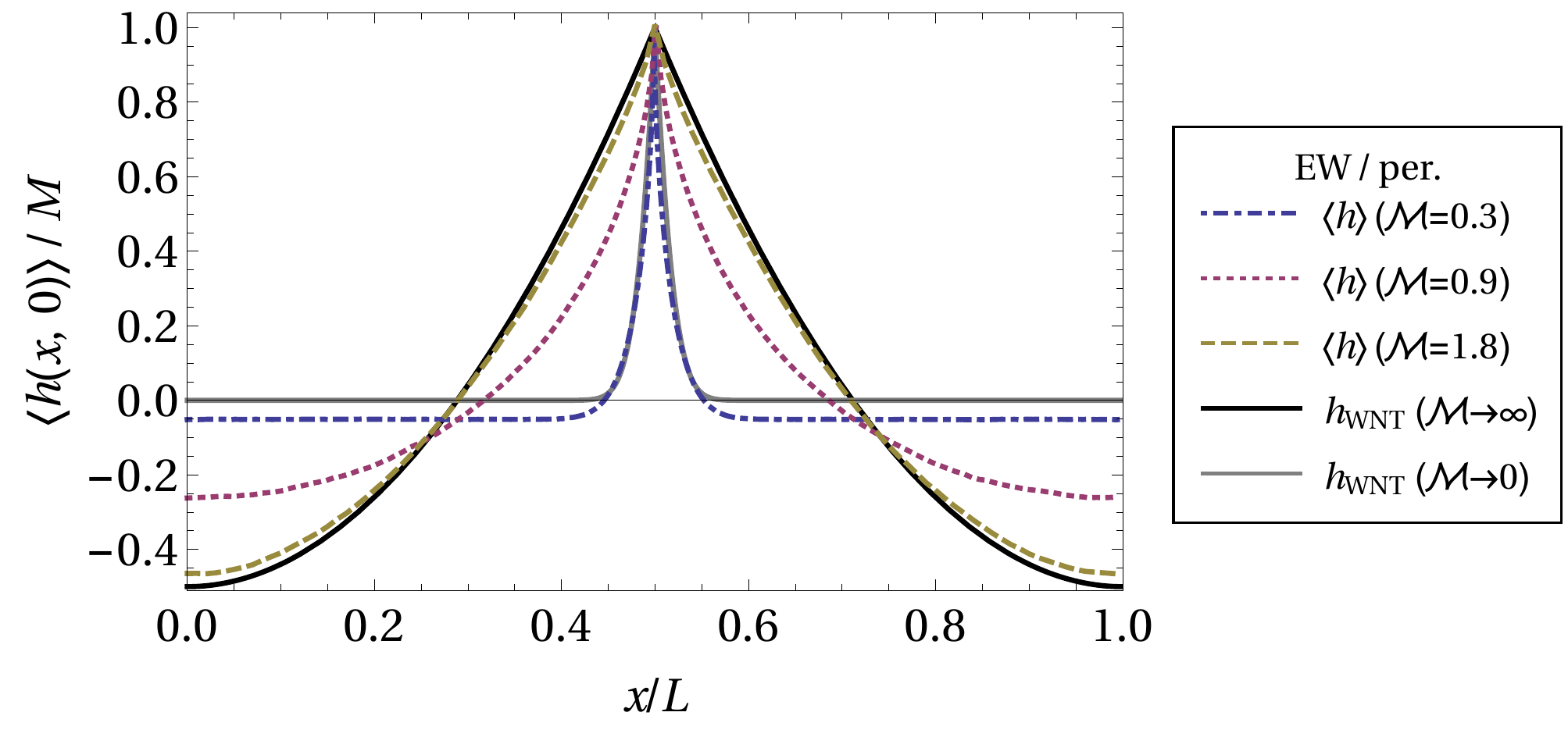}
	\caption{Averaged profile $\bra h(x,\dt=0)\ket$ (broken lines) \emph{at} the first-passage event, as obtained from simulations of the EW equation [\cref{eq_EW}] with periodic \bcs and for various reduced heights $\Mred$ [\cref{eq_hm_red}]. A constraint of zero mass [\cref{eq_zero_vol}] is imposed via \cref{eq_height_redef}. The solid lines represent the asymptotic scaling profiles predicted by WNT in the transient regime [$\Mred\to 0$, \cref{eq_h_shortT_EW}, sharply peaked curve] and in the equilibrium regime [$\Mred\to\infty$, \cref{eq_opt2_finalprof_eq_pbc}, broadly peaked curve]. In the former case, the parameter $T$ results from a fit as $T\simeq 1.5\times 10^{-9} \tau\pbc$.}
    \label{fig_avgprof_nc_pbc_eq_finalT}
\end{figure}

\begin{figure}[t!]\centering
	\subfigure[]{\includegraphics[width=0.4\linewidth]{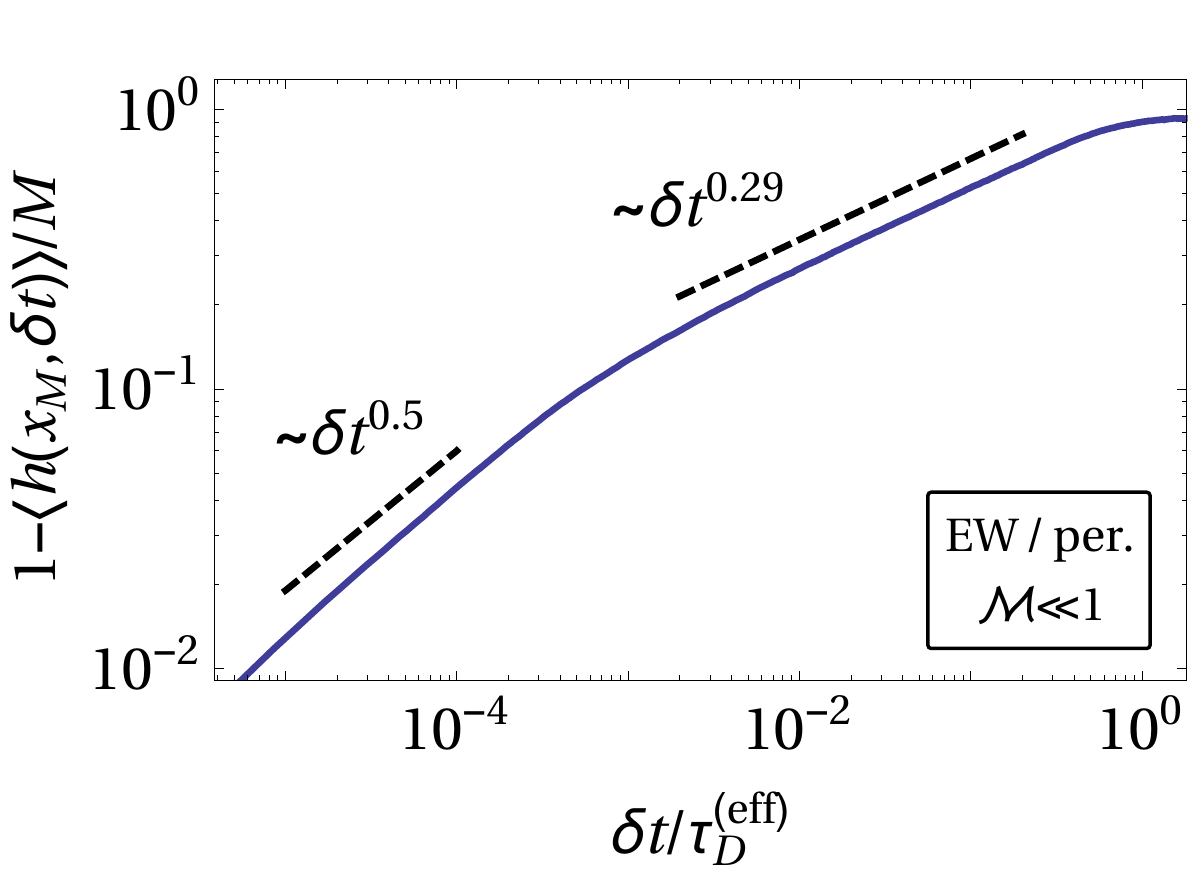} \label{fig_peakevol_nc_pbc_neq}}\quad
	\subfigure[]{\includegraphics[width=0.4\linewidth]{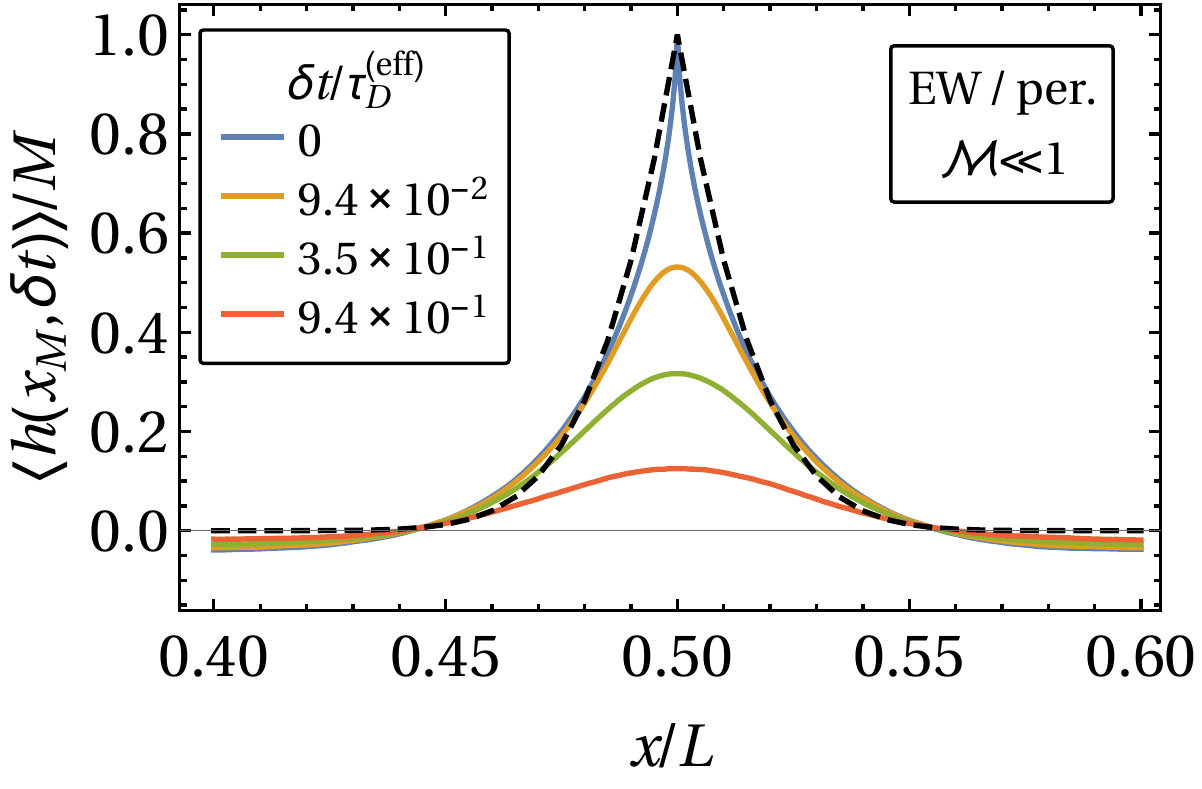}}\quad
	\subfigure[]{\includegraphics[width=0.4\linewidth]{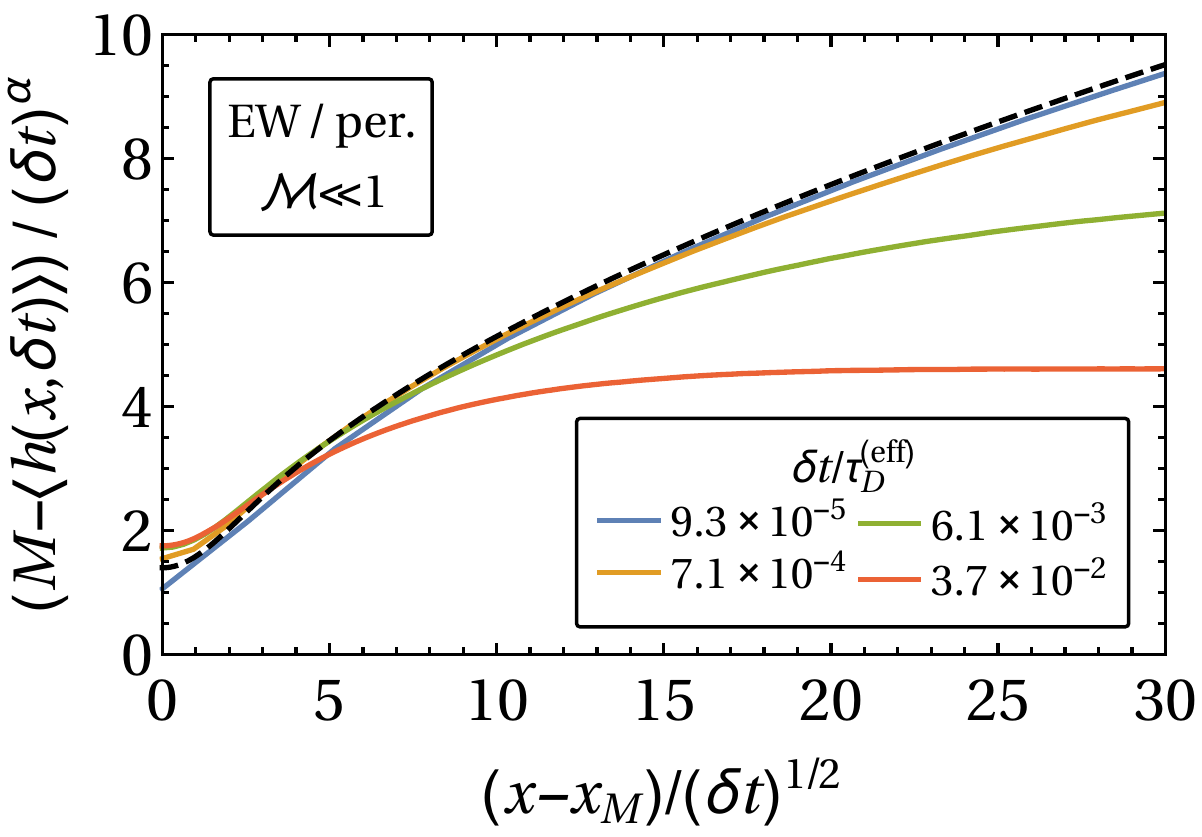}}
	\caption{Averaged profile $\bra h(x,\dt)\ket$ obtained for the EW equation [\cref{eq_EW}] with periodic \bcs in the transient regime ($\Mred\ll 1$). The first passage of the height $M$ occurs at the time $\dt=0$. Utilizing translational invariance, the individual profiles obtained from simulation are shifted such that the height $M$ is reached at location $x_M=L/2$. Time is normalized to the diffusion time scale $\tau_D$ [\cref{eq_tdiff}] using for the exponent $z$ an \emph{effective} value of $1/(2\alpha)\simeq 1.7$ ($\alpha\simeq 0.29)$ instead of $2$, as implied by panel (a).
	(a) Time-evolution of the peak of the profile, $\bra h(x_M,\dt)\ket$, which exhibits an intermediate asymptotic regime $M-\bra h(x_M,\dt)\ket\propto \dt^\alpha$ with $\alpha\simeq 0.29$. 
	(b) Spatio-temporal evolution of the averaged profile. The solid curves represent the profiles obtained from numerical simulations, while the dashed curve indicates the asymptotic profile predicted by WNT in \cref{eq_h_shortT_EW}, taking $T$ as a fit parameter. (c) Test of the dynamic scaling behavior of $\bra h(x,\dt)\ket$ as predicted by WNT according to \cref{eq_h2_shortTdyn_asympt}, using a value of $1/z\simeq 0.28$. The dashed curve represents the scaling function $c \tilde \Hcal$ in \cref{eq_hscalf_asympt_EW}, with a prefactor $c\simeq 1.4$ determined from a fit. }
	\label{fig_avgprof_nc_pbc_neq}
\end{figure}

We now turn to the first-passage dynamics of a profile governed by \cref{eq_EW} with periodic and Dirichlet \bcs. 

\subsection{Summary of WNT}
Before discussing the simulation results, we summarize a few relevant predictions of WNT of the EW equation (see \paperI for details).
The following expressions for $h(x,\dt)$ are to be understood as the leading-order contribution to the averaged profile $\bra h(x,\dt)\ket$.
Note that, differently from \paperI, we use $\dt=T-t$ as the time variable.
Within WNT, the first-passage time $T$ is a fixed parameter and the transient and the equilibrium regime are distinguished by the value of $T/\tau$.
In the \emph{transient} regime ($T\ll \tau$), a scaling profile at time $\dt=0$ results from WNT as
\beq h(x,\dt=0)\big|_{T\ll \tau} = M \Hcal\left(\frac{x-L/2}{(2T)^{1/z}}\right),\qquad z=2,
\label{eq_h_shortT_EW}\eeq 
with the scaling function
\beq \Hcal(\xi)=
\exp\left(-\frac{\xi^2}{4}\right) + \onehalf \sqrt{\pi} |\xi| \left[\mathrm{erf}\left(\frac{|\xi|}{2}\right)-1 \right].
\label{eq_h_shortT_scalF_EW}\eeq 
For $0<\delta t\ll T$, one obtains the dynamic scaling profile 
\beq h(x,\delta t)\Big|_{\substack{T\ll \tau\\\delta t\ll T}} = M - M \left(\frac{\delta t}{2T}\right)^{1/z} \tilde \Hcal\left(\frac{x-L/2}{\delta t^{1/z}}\right),\qquad z=2,
\label{eq_h2_shortTdyn_asympt}\eeq 
with the scaling function 
\beq \tilde \Hcal(\xi) = 
\exp\left(-\frac{\xi^2}{4}\right) + \onehalf \sqrt{\pi}\, \xi\, \mathrm{erf}\left(\frac{\xi}{2}\right). \label{eq_hscalf_asympt_EW}\eeq 
When applying \cref{eq_h_shortT_EW,eq_h2_shortTdyn_asympt} to simulation results, we consider the quantity $T$ as a fit parameter.
In the \emph{equilibrium} regime ($T\gg \tau$) for $\dt=0$, one finds the following asymptotic first-passage profiles for periodic and Dirichlet \bcs, respectively:
\begin{subequations}\begin{align}
h\pbc(x,\dt=0)\big|_{T\to\infty}/M &=   1-6\Bigg|\frac{x}{L}-\onehalf\Bigg| + 6\left(\frac{x}{L}-\onehalf\right)^2, \label{eq_opt2_finalprof_eq_pbc} \\
h\Dbc(x,\dt=0)\big|_{T\to\infty}/M  &= 1-\left|1-\frac{2x}{L}\right|. \label{eq_opt2_finalprof_eq_Dir}
\end{align}\label{eq_opt2_finalprof_eq}\end{subequations}
These profiles attain their maximum at $x_M=L/2$.
They follow readily from the constrained minimization of the corresponding equilibrium free energy. 
For times $0<\delta t\ll T$, one finds a dynamic scaling form,
\beq h(x,\delta t)\big|_{T\gg \tau} \simeq M -  M (\delta t)^{1/z} \Gamma(1-1/z) \tilde\Hcal\left(\frac{x-L/2}{\delta t^{1/z}}\right),\qquad z=2,
\label{eq_h_lateT_dynscal_EW}\eeq 
with the same scaling function as in \cref{eq_hscalf_asympt_EW}. 
Note that, unless otherwise indicated, the above scaling forms apply to all \bcs considered here.
Exact analytical expressions for the profile $h(x,t)$ obtained within WNT can be found in \paperI and are not repeated here.

We emphasize that the above expressions pertain to a continuum system.
As shown in \paperI, the presence of a microscopic cutoff (e.g., a lattice constant) modifies the dynamics for times $\dt\lesssim \tau\cro$, where $\tau\cro$ is the crossover time in \cref{eq_crossover_time}. 
Upon taking this effect into account, the time-evolution of the profile $h(x,\dt)$ at $x=x_M$ is given within WNT by
\beq 1- h(x_M,\delta t)/M \propto 
\begin{cases} 
	\delta t,\qquad & \dt\lesssim \tau\cro,\\
	\delta t^{1/z}, \qquad & \dt\gtrsim \tau\cro.
\end{cases}
\label{eq_peakevol_EW}\eeq 
This result is independent of the \bcs and applies to both the transient and equilibrium regime [see \cref{eq_h2_shortTdyn_asympt,eq_h_lateT_dynscal_EW}].

\begin{figure}[t]\centering
	\subfigure[]{\includegraphics[width=0.4\linewidth]{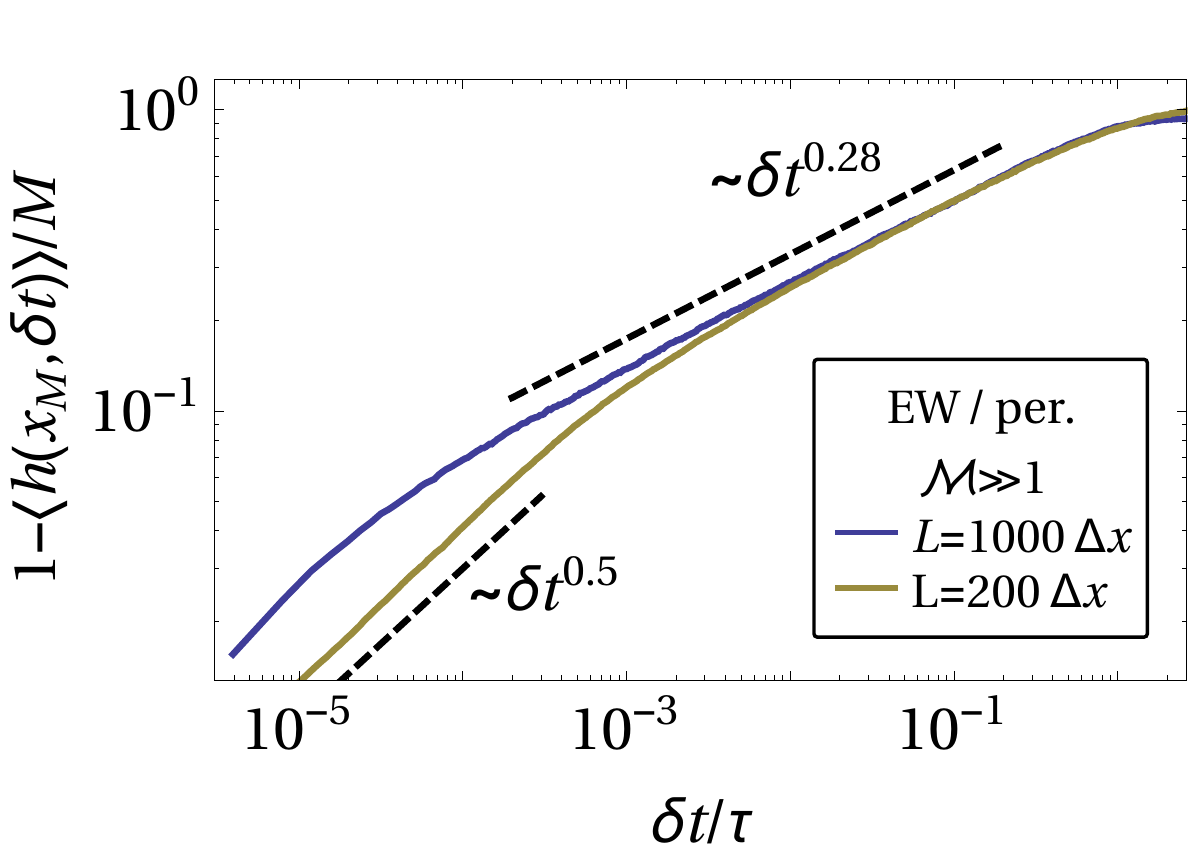} \label{fig_peakevol_nc_pbc_eq}}\quad
	\subfigure[]{\includegraphics[width=0.4\linewidth]{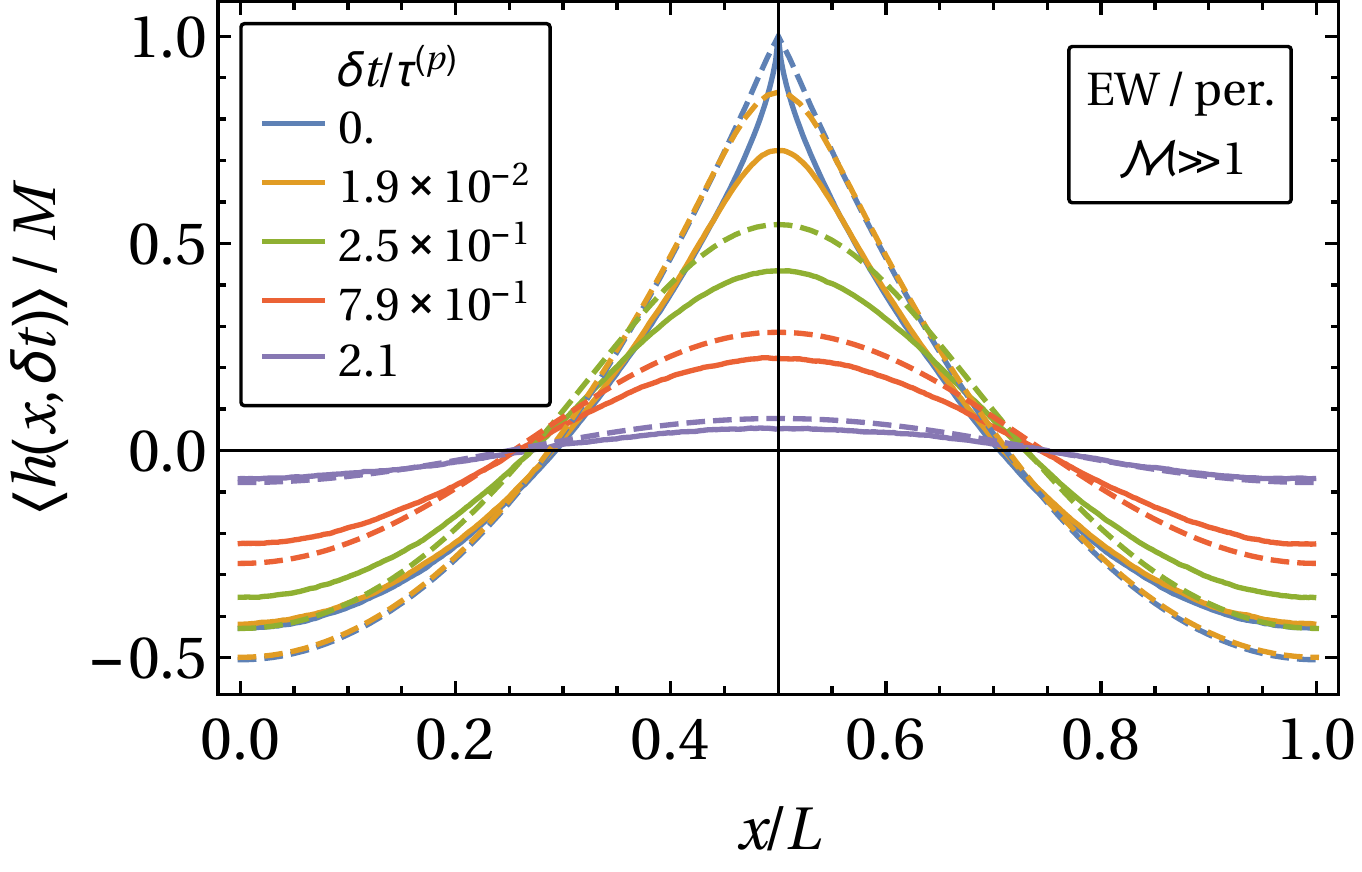}}\quad
	\subfigure[]{\includegraphics[width=0.4\linewidth]{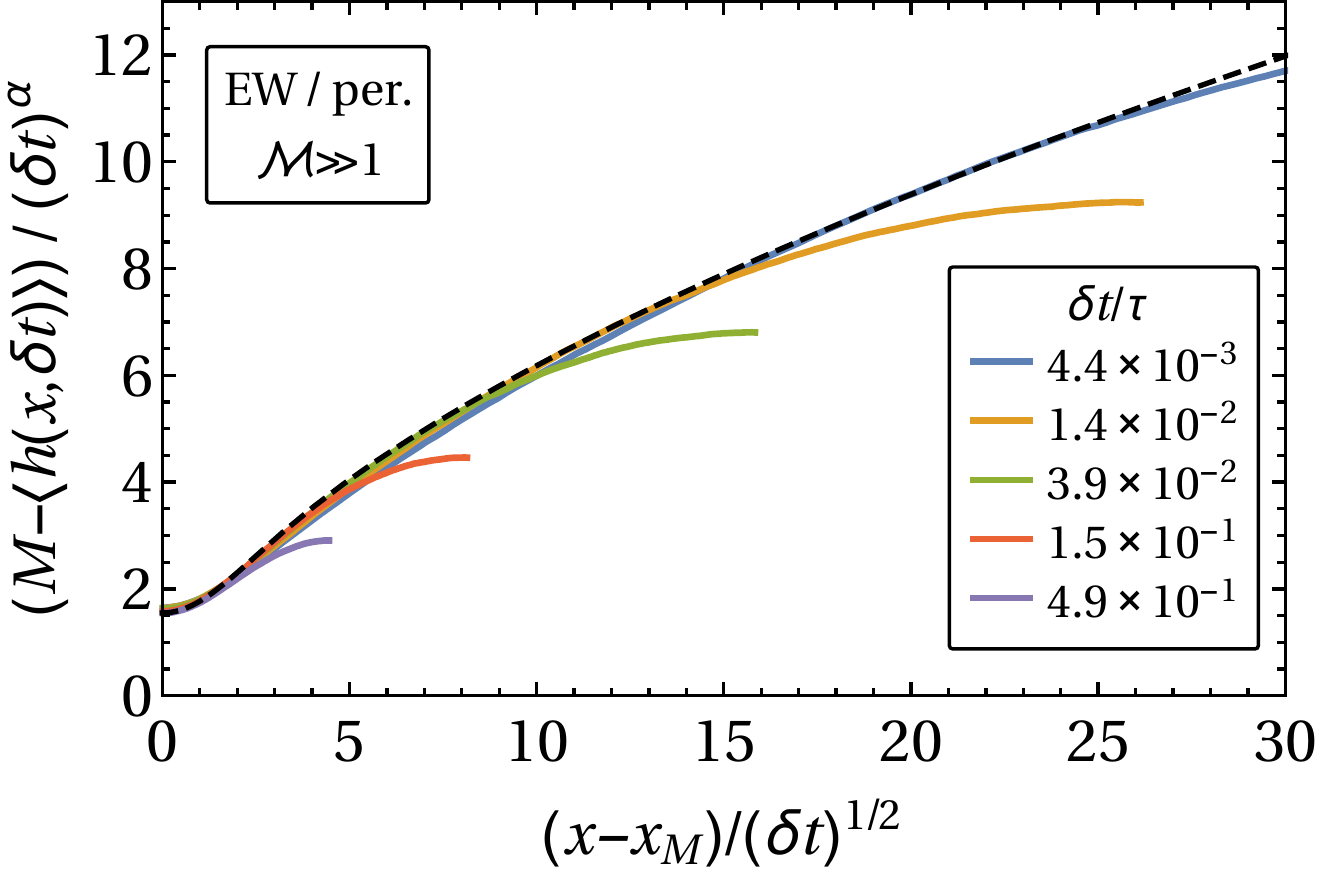}}
	\caption{Averaged profile $\bra h(x,\dt)\ket$ obtained for the EW equation [\cref{eq_EW}] with periodic \bcs in the equilibrium regime ($\Mred\gg 1$). The first passage of the height $M$ occurs at the time $\dt=0$. Utilizing translational invariance, the individual profiles obtained from simulation are shifted such that the height $M$ is reached at location $x_M=L/2$. Time is normalized to the relaxation time scale $\tau$ [see \cref{eq_relaxtime}]. 
	(a) Time-evolution of the peak of the profile, $\bra h(x_M,\dt)\ket$, for two different system sizes (in units of the lattice spacing $\lattsp$). In the intermediate asymptotic regime one has $M-\bra h(x_M,\dt)\ket\propto \dt^\alpha$ with $\alpha\simeq 0.28$. 
	(b) Spatio-temporal evolution of the averaged profile. The solid curves represent the profiles obtained from numerical simulations with $L=400\lattsp$, $\Mred\simeq 1.6$, while the dashed curves indicate the prediction of WNT [see Eq.~(2.19) in \paperI]. (c) Test of the dynamic scaling behavior of $\bra h(x,\dt)\ket$ as predicted by WNT according to \cref{eq_h_lateT_dynscal_EW}, using a value of $1/z \simeq 0.3$. The dashed curve represents the scaling function $c \tilde \Hcal$ in \cref{eq_hscalf_asympt_EW}, with a prefactor $c\simeq 1.6$ determined from a fit.}
	\label{fig_avgprof_nc_pbc_eq}
\end{figure}

\subsection{Periodic \bcs}
\label{sec_EW_pbc}

We now turn to the discussion of the first-passage properties of a profile governed by the EW equation [\cref{eq_EW}] with periodic \bcs. 
We recall that, in this case, the constraint of zero mass [\cref{eq_zero_vol}] is imposed via \cref{eq_height_redef} at each time step in the simulation. 
(Within WNT, this constraint is reflected by the absence of the zero mode in the series solution for the profile, see \paperI.)
\Cref{fig_avgprof_nc_pbc_eq_finalT} illustrates the spatial shape of the averaged profile \emph{at} the first-passage event, $\bra h(x,\dt=0)\ket$, for various reduced heights $\Mred$ \footnote{Simulations in the equilibrium regime are found to be computationally feasible only for reduced maximum heights of $\Mred\sim \Ocal(1)$, since the probability [\cref{eq_Pss}] to observe significantly larger height fluctuations becomes exponentially small, see \cref{app_equil_pdf}.}.
The asymptotic scaling profiles predicted by WNT in the transient and the equilibrium regime [\cref{eq_opt2_finalprof_eq_pbc,eq_h_shortT_EW}, solid lines] agree well with the numerical results in the limits $\Mred\ll 1$ and $\Mred\gg 1$. 
According to \cref{eq_h_shortT_EW}, the analytical profile in the transient regime still depends on $T$, which is considered here as a fit parameter and effectively controls the width of the profile. Furthermore, since \cref{eq_h_shortT_EW} is obtained by neglecting the mass constraint [\cref{eq_zero_vol}], it applies only to an inner region of the profile.
In contrast, the full solution of WNT provides an accurate description for $\Mred\lesssim 1$ also in the outer regions, as is illustrated below. 
Part of the remaining discrepancies between the analytically and numerically obtained profiles in \cref{fig_avgprof_nc_pbc_eq_finalT} can be attributed to the fact that WNT neglects fluctuations \emph{around} the saddle point solution. Such fluctuations can give rise to an effective repulsion from the boundary. We will return to this aspect in \cref{sec_discussion}.

In \cref{fig_avgprof_nc_pbc_neq,fig_avgprof_nc_pbc_eq}, the spatio-temporal evolution of the averaged profile approaching the first-passage event $\bra h(x=x_M,0)\ket=M$ is illustrated in the transient and equilibrium regimes, respectively.
As observed in Fig.\ \ref{fig_avgprof_nc_pbc_neq}(a) and \ref{fig_avgprof_nc_pbc_eq}(a) \footnote{The data underlying the time-evolution of the peak shown here and in the other figures occasionally stem from two separate simulations, which have been performed with identical parameters but different time resolutions.}, both in the transient and the equilibrium regime, the peak of the profile, $\bra h(x_M,\dt)\ket$ (with $x_M=L/2$), approaches the maximum height $M$ algebraically, 
\beq M-\bra h(x_M,\dt)\ket\propto \dt^\alpha.
\label{eq_peakevol_EW_pbc}\eeq 
For times $\delta t$ larger than a cross-over time $\tau\cro$ (see below), one obtains an exponent  
\beq \alpha\simeq 0.28-0.3,
\label{eq_peakevol_alpha_EW_pbc}\eeq 
while $\alpha= \alpha_0\simeq 0.5$ for $\dt\lesssim \tau\cro$.
The extent of the intermediate asymptotic regime described by \cref{eq_peakevol_alpha_EW_pbc} grows upon increasing the system size $L$, as illustrated in \cref{fig_peakevol_nc_pbc_eq}.
Notably, the above values of the exponent $\alpha$ differ significantly from the values $\alpha\st{WNT}=1/z=1/2$ and $\alpha\st{0,WNT}=1$ predicted by WNT in \cref{eq_peakevol_EW}.
An explanation of these findings, which are analogously obtained also for the other models considered in this study, is provided in \cref{sec_discussion}.
As seen in \cref{fig_avgprof_nc_pbc_eq}, in the equilibrium regime, the first-passage evolution of the profile happens essentially within a timescale of the order of $\tau\pbc$ [see \cref{eq_relaxtime}], as predicted by WNT.
In the transient regime, the characteristic time scale is taken here to be the effective diffusion time $\tau_D\ueff$. The latter is defined by \cref{eq_tdiff}, using for the dynamic exponent $z$ the effective value $1/(2\alpha)\simeq 1.7$ with $\alpha$ given in \cref{eq_peakevol_alpha_EW_pbc}.
Using instead the value $z=2$ predicted by WNT leads to a significant underestimation of the first-passage time scale.
The non-vanishing cross-over time $\tau\cro$ arises due to the finite lattice spacing $\lattsp$ in the simulations.
In agreement with the numerical data, \cref{eq_crossover_time} predicts $\tau\cro/\tau\pbc \sim 10^{-5}$ ($L=1000\lattsp$) and $\tau\cro/\tau\pbc \sim 2\times 10^{-4}$ ($L=200\lattsp$) for the two system sizes considered in \cref{fig_peakevol_nc_pbc_eq}.

In Figs.\ \ref{fig_avgprof_nc_pbc_neq}(b) and \ref{fig_avgprof_nc_pbc_eq}(b), the shape of the averaged profile is illustrated for various times $\dt$ (solid lines).
In \cref{fig_avgprof_nc_pbc_eq}(b) the dashed lines represent the time-dependent profiles obtained within WNT [Eq.~I-(2.19)].
Since the actual time-dependence of $\bra h(x,t)\ket$ differs from the prediction of WNT due to a different value of the dynamic exponent $\alpha$, analytical profiles do in general not match the numerical solutions well for $\dt>0$. 
These discrepancies are found to be more severe in the transient regime [\cref{fig_avgprof_nc_pbc_neq}(b)], where we show only the scaling profile given in \cref{eq_h_shortT_EW} (dashed line).

In  Figs.\ \ref{fig_avgprof_nc_pbc_neq}(c) and \ref{fig_avgprof_nc_pbc_eq}(c), the dynamic scaling behavior asymptotically predicted by WNT [see \cref{eq_h_lateT_dynscal_EW,eq_h2_shortTdyn_asympt}] is tested. 
To this end, the profile height $\bra h\ket$ and the coordinate $x$ are rescaled accordingly and the scaling function $c\tilde H$ in \cref{eq_hscalf_asympt_EW} is fitted via the parameter $c$.
In order to account for the renormalization of the dynamic exponent $\alpha$, we use for $1/z$ in \cref{eq_h_lateT_dynscal_EW,eq_h2_shortTdyn_asympt} an effective value which is close to the value for $\alpha$ reported in \cref{eq_peakevol_alpha_EW_pbc} \footnote{While within WNT, the exponent $z$ is identified with $1/\alpha$ [see, e.g., \cref{eq_peakevol_EW}], beyond WNT, it turns out that $z$ has to be identified with a value close to $1/(2\alpha)$. This fact is also used in the definition of $\tau_D\ueff$. See \cref{sec_discussion} for further discussion.}.
As shown in Figs.\ \ref{fig_avgprof_nc_pbc_neq}(c) and \ref{fig_avgprof_nc_pbc_eq}(c), this results in a satisfactory matching (in an inner region) of the numerical profiles with the scaling function $\tilde H$ in \cref{eq_hscalf_asympt_EW} (dashed line). 
The outer parts of the profiles deviate from the scaling function due to the influence of the boundary conditions.

\subsection{Dirichlet \bcs}

\begin{figure}[t]\centering
	\includegraphics[width=0.45\linewidth]{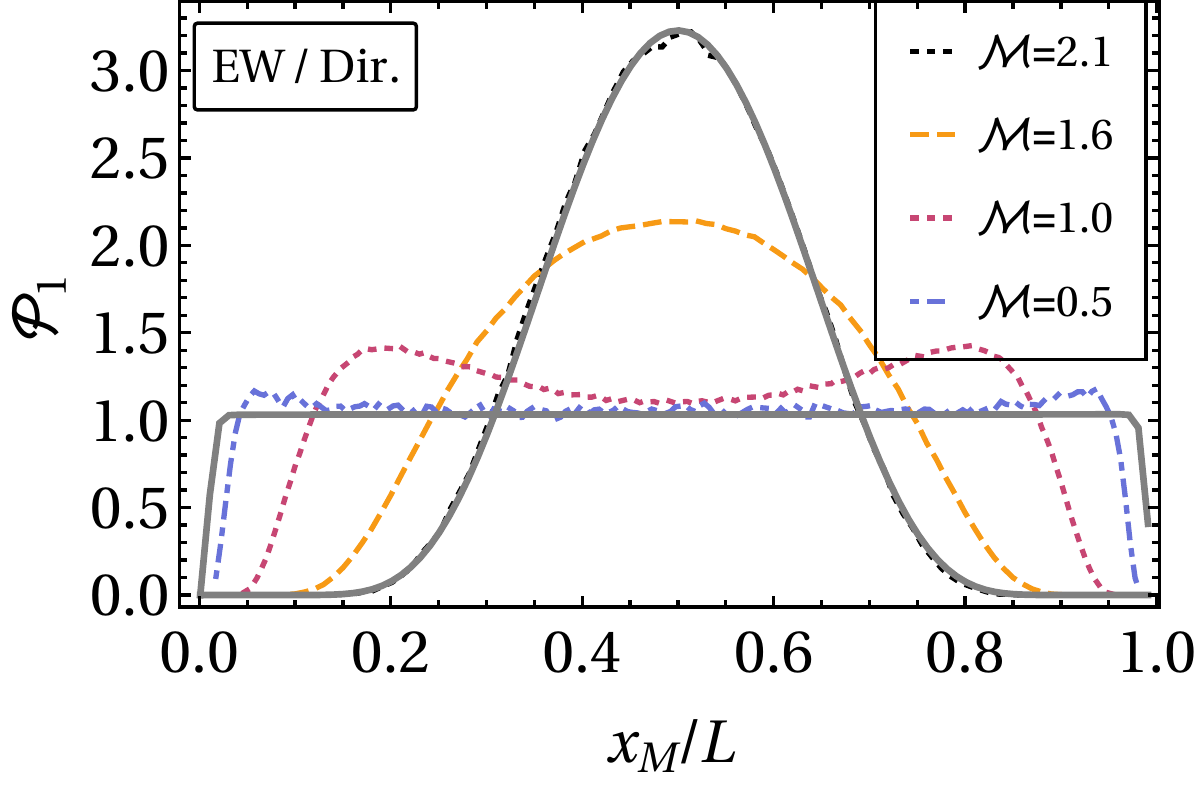}
	\caption{Distribution $\Pcal_1(x_M)$ of the spatial location $x_M$ at which the profile passes the height $M$ for the first time in the case of the EW equation [\cref{eq_EW}] with Dirichlet \bcs. The broken lines represent simulation data for various reduced heights $\Mred$. The solid curve with the flat center represents the prediction of WNT asymptotically in the transient regime ($\Mred\to 0$, see \paperI). The bell-shaped solid curve pertains to the equilibrium regime ($\Mred\gtrsim 1$) and is given by Eq.~I-(2.16), evaluated using $\frict M^2/DL\simeq 1.8$ and $T/\tau\Dbc\gg 1$, as determined from a fit.}
    \label{fig_hittingloc_nc_Dir}
\end{figure}

\begin{figure}[t]\centering
	\subfigure[]{\includegraphics[width=0.4\linewidth]{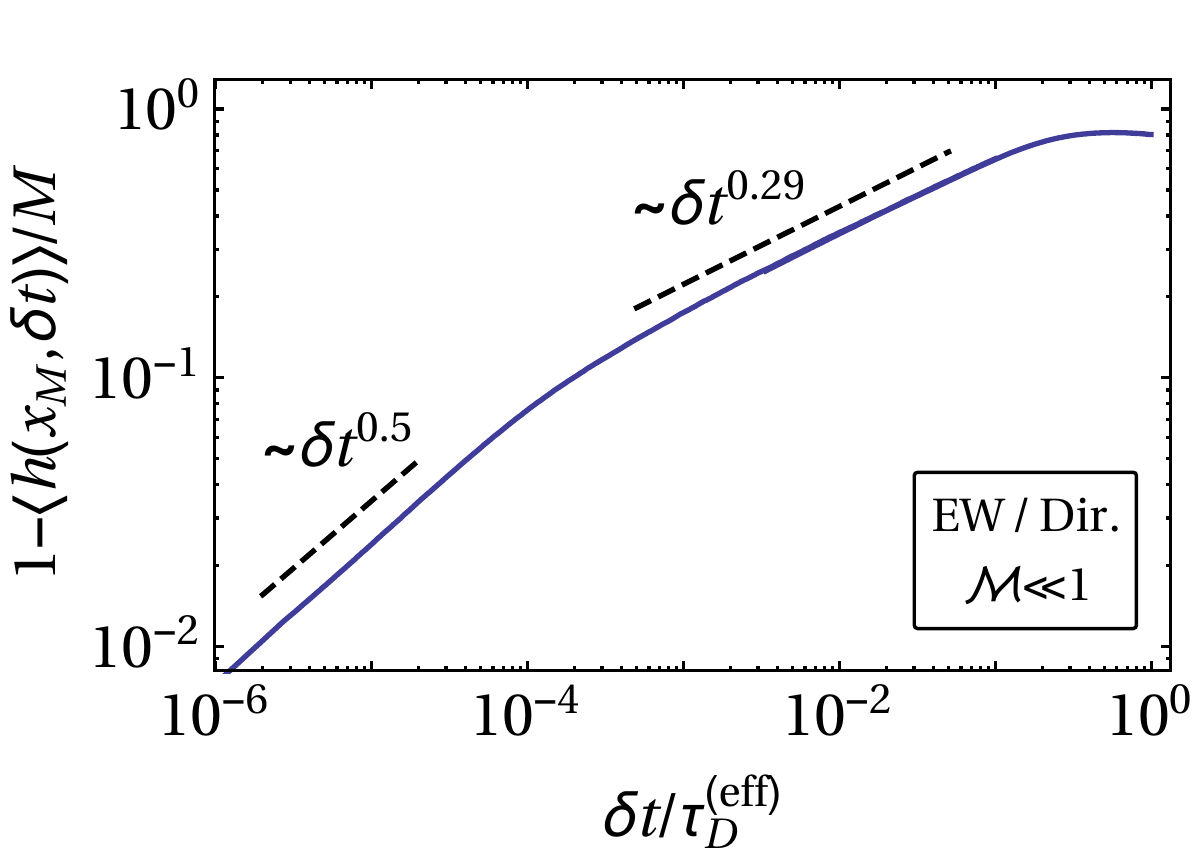} \label{fig_peakevol_nc_Dir_neq}}\qquad
	\subfigure[]{\includegraphics[width=0.4\linewidth]{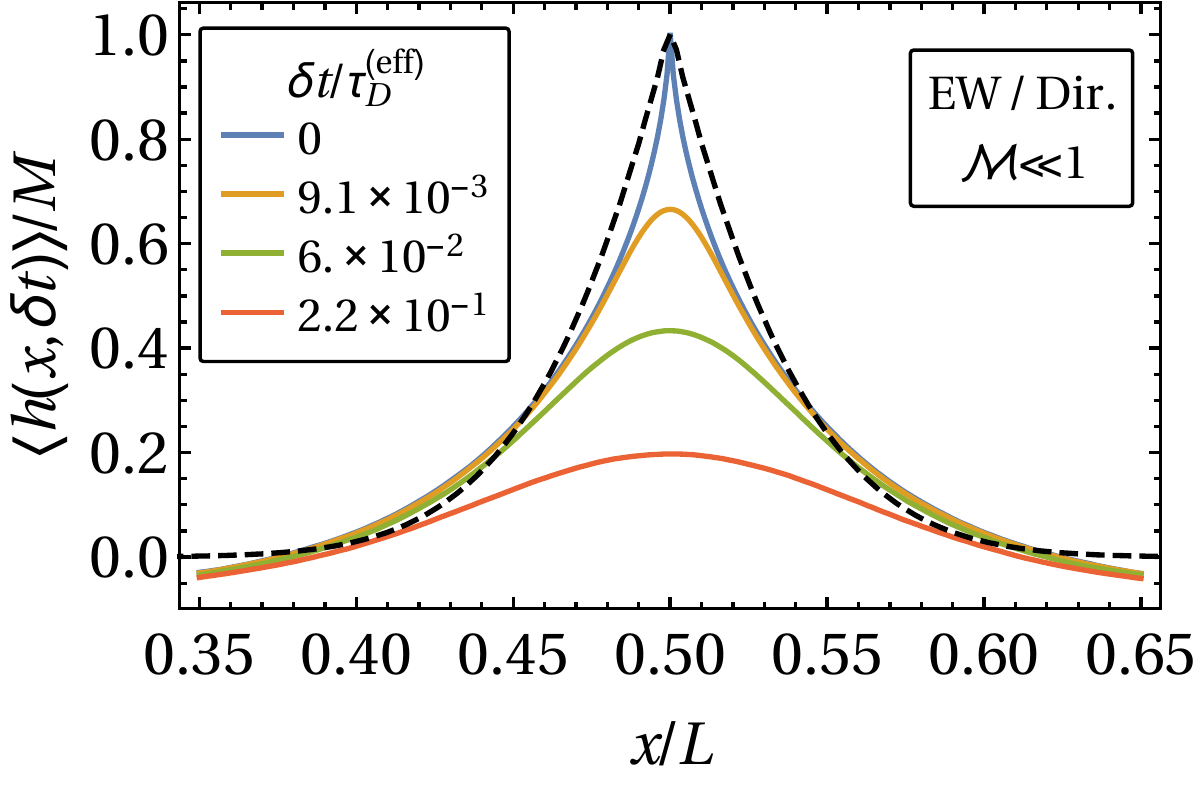}}
	\subfigure[]{\includegraphics[width=0.4\linewidth]{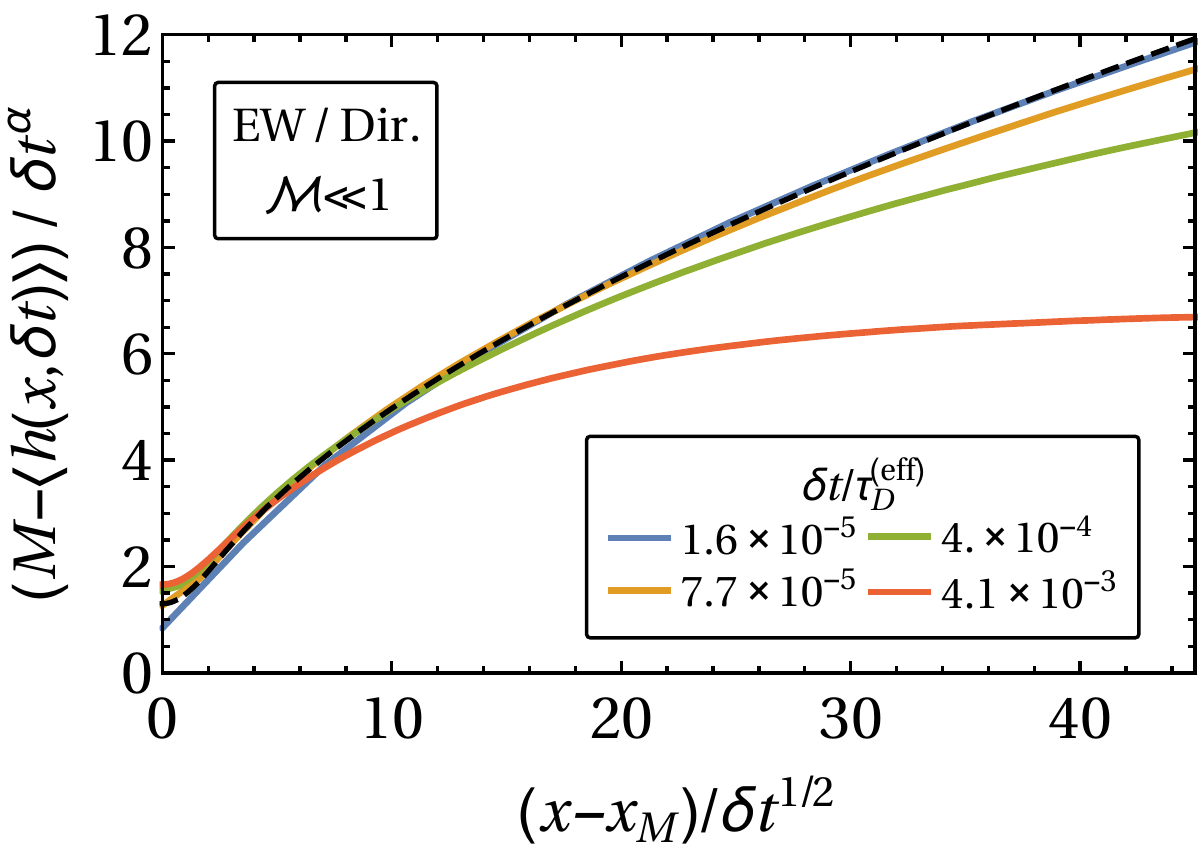}}
	\caption{Averaged profile $\bra h(x,\dt)\ket$ obtained for the EW equation [\cref{eq_EW}] with Dirichlet \bcs in the transient regime ($\Mred\ll 1$). The first passage of the height $M$ occurs at the time $\dt=0$. Time is normalized to the diffusion time scale $\tau_D$ [\cref{eq_tdiff}] using for the exponent $z$ an \emph{effective} value of $1/(2\alpha)\simeq 1.7$ ($\alpha\simeq 0.29)$ instead of $2$, as implied by panel (a).
	(a) Time-evolution of the peak of the profile, $\bra h(x_M,\dt)\ket$, which exhibits an intermediate asymptotic regime $M-\bra h\ket\propto \dt^\alpha$ with $\alpha\simeq 0.29$. 
	(b) Spatio-temporal evolution of the averaged profile. The solid curves represent the profiles obtained from numerical simulations, while the dashed curve indicates the asymptotic profile predicted by WNT in \cref{eq_h_shortT_EW}, taking $T$ as a fit parameter. (c) Test of the dynamic scaling behavior of $\bra h(x,t)\ket$ as predicted by WNT in \cref{eq_h2_shortTdyn_asympt}, using a value of $1/z \simeq 0.29$. The dashed curve represents the scaling function $c \tilde \Hcal$ in \cref{eq_hscalf_asympt_EW}, with a prefactor $c\simeq 1.3$ determined from a fit. In order to account for the localized nature of the profiles in the transient regime, in all panels the individual profiles are shifted before averaging such that $h^{(s)}(L/2,T^{(s)})=M$ [see \cref{eq_def_avgprof}].}
	\label{fig_avgprof_nc_Dir_neq}
\end{figure}

\begin{figure}[t]\centering
	\subfigure[]{\includegraphics[width=0.4\linewidth]{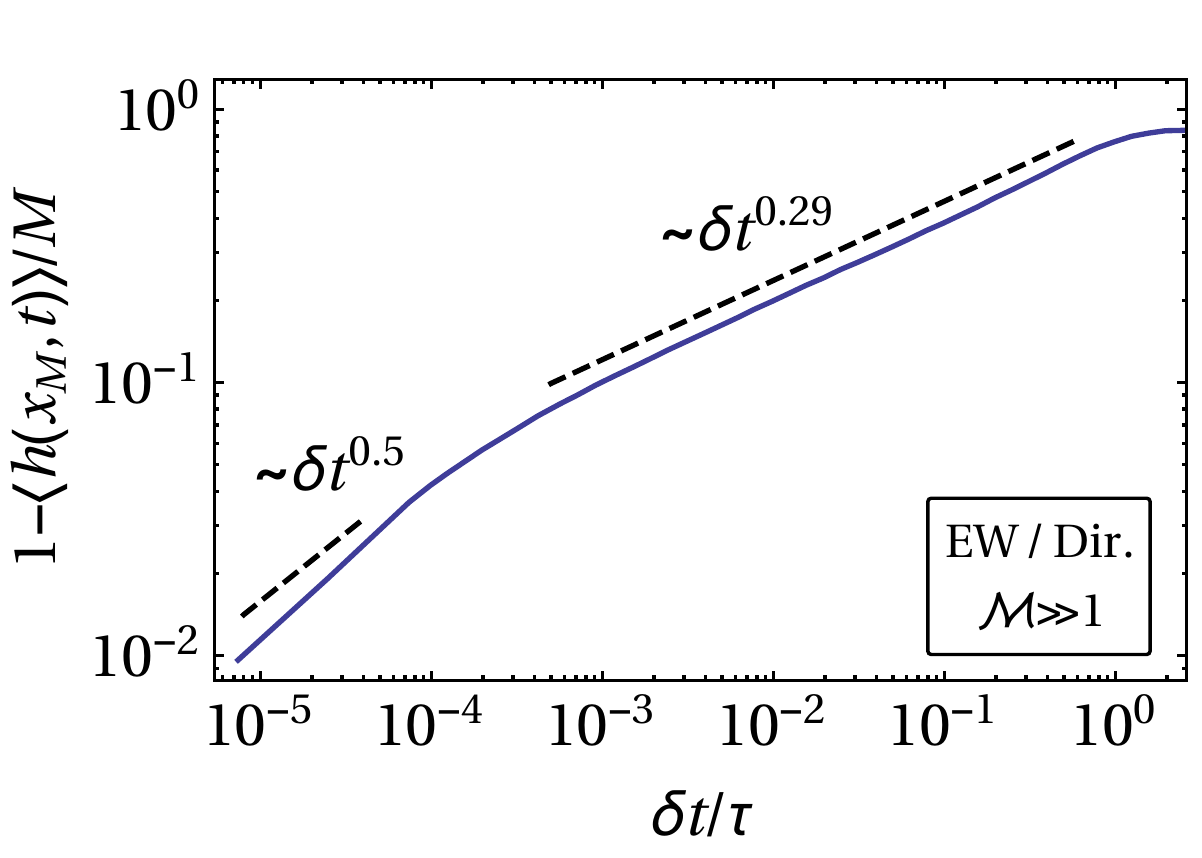} \label{fig_peakevol_nc_Dir_eq}}\qquad
	\subfigure[]{\includegraphics[width=0.4\linewidth]{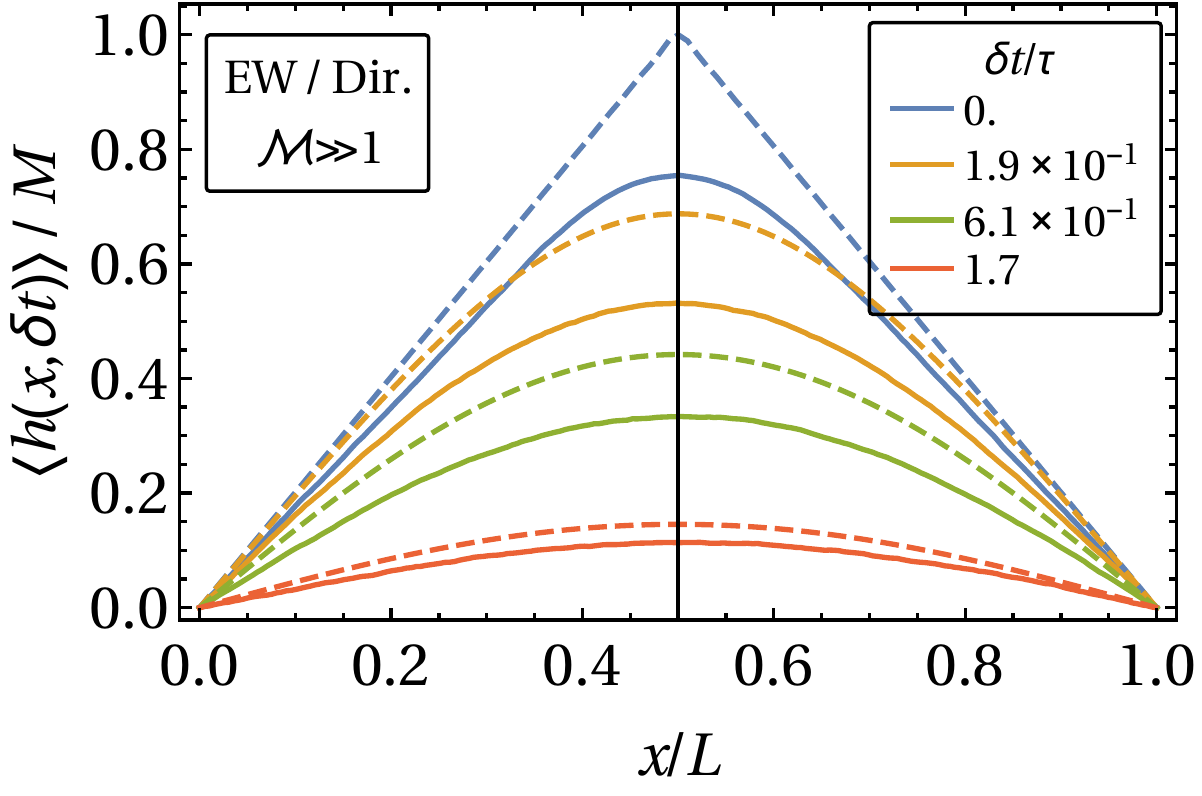}}
	\subfigure[]{\includegraphics[width=0.4\linewidth]{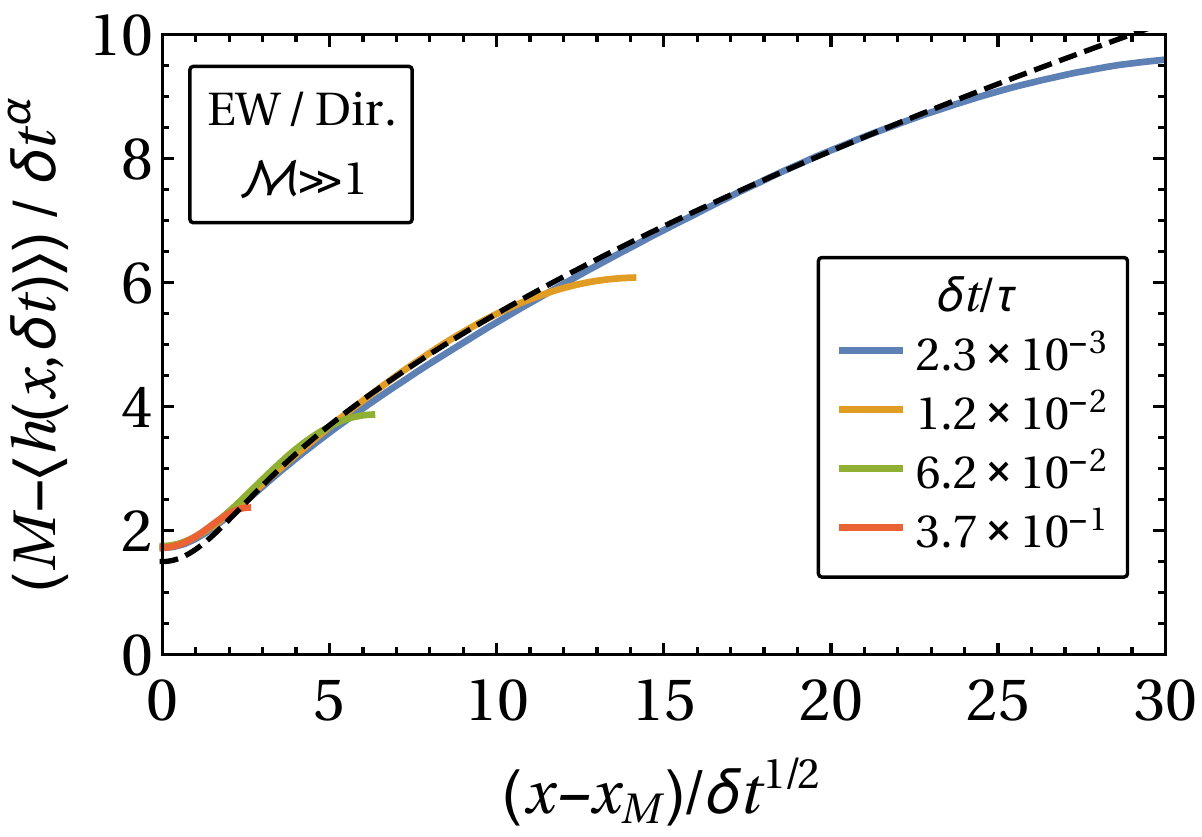}}
	\caption{Averaged profile $\bra h(x,\dt)\ket$ for the EW equation [\cref{eq_EW}] with Dirichlet \bcs in the equilibrium regime ($\Mred\gg 1$). The first passage of the height $M$ occurs at the time $\dt=0$. Time is normalized to the relaxation time $\tau\Dbc$ [see \cref{eq_relaxtime}].
	(a) Time-evolution of the peak of the profile, $\bra h(x_M,\dt)\ket$, which exhibits an intermediate asymptotic regime $M-\bra h\ket\propto \dt^\alpha$ with $\alpha\simeq 0.29$. 
	(b) Spatio-temporal evolution of the averaged profile. The solid curves represent the profiles obtained from numerical simulations, while the dashed curves indicate the prediction of WNT [see Eq.~I-(2.21)]. (c) Test of the dynamic scaling behavior of $\bra h(x,t)\ket$ as predicted by WNT in \cref{eq_h_lateT_dynscal_EW}, using a value of $1/z \simeq 0.29$. The dashed curve represents the scaling function $c \tilde \Hcal$ in \cref{eq_hscalf_asympt_EW}, with a prefactor $c\simeq 1.5$ determined from a fit. In order to properly exhibit the scaling behavior, in panels (a) and (c) the individual profiles are shifted before averaging such that $h^{(s)}(L/2,T^{(s)})=M$  [see \cref{eq_def_avgprof}]. }
	\label{fig_avgprof_nc_Dir_eq}
\end{figure}

We now turn to the rare event dynamics of a profile governed by the EW equation with standard Dirichlet \bcs [\cref{eq_H_Dbc}]. We recall that, in this case, the mass constraint in \cref{eq_zero_vol} is not fulfilled by the individual realizations of the profile.
The probability distribution $\Pcal_1(x_M)$ of the location $x_M$ of the first-passage event [see \cref{eq_firstpsg_cond}] is shown in \cref{fig_hittingloc_nc_Dir} for various reduced heights $\Mred$.
For $\Mred\ll 1$, $\Pcal_1$ is essentially flat, in agreement with the prediction of WNT in the transient regime (see \paperI).
For $\Mred\gg 1$, instead, the first-passage event is most likely to occur at the center of the system. In this regime, $\Pcal_1$ can be well fitted by the analytical expression reported in Eq.~I-(2.16), using a value of $\frict M^2/DL\simeq 1.8$ and $T/\tau\Dbc\gg 1$ (the precise value of the latter parameter is immaterial since $\Pcal_1$ becomes independent of it provided it is sufficiently large).
In the crossover region between the transient and the equilibrium regime, $\Pcal_1$ depends within WNT on both $T/\tau\Dbc$ and $\frict M^2/DL$ and, therefore, a fit is less meaningful.
Differently from WNT, $\Pcal_1$ develops two maxima near the boundaries for $\Mred\sim \Ocal(1)$. 

In \cref{fig_avgprof_nc_Dir_neq,fig_avgprof_nc_Dir_eq}, the spatio-temporal evolution of the averaged profile in the transient and equilibrium regimes, respectively, is illustrated.
Since the distribution $\Pcal_1(x_M)$ of the first-passage location is flat in the transient regime, the averaged profiles shown in \cref{fig_avgprof_nc_Dir_neq} are obtained by shifting each realization such that the first-passage event occurs at $x_M=L/2$ [see \cref{eq_def_avgprof}]. Since the profile is strongly localized in the transient regime, such a shift does not significantly affect its averaged shape. 
As shown in \cref{fig_avgprof_nc_Dir_neq}(b), a fit via the parameter $T$ of the asymptotic profile of WNT reported in \cref{eq_h_shortT_EW} yields satisfactory agreement with the data.
In the equilibrium regime, the averaged profile is computed according to \cref{eq_def_avgprof} without a shift ($X^{(s)}=0$). 
In this case, the finite width of $\Pcal_1(x_M)$ [see \cref{fig_hittingloc_nc_Dir}] is reflected by the rather strong deviation of $\bra h(x,\dt)\ket$ from the prediction of WNT [\cref{eq_opt2_finalprof_eq_Dir}, dashed lines in \cref{fig_avgprof_nc_Dir_eq}(b)] as well as  by the fact that $\bra h(x_M,\dt=0)\ket <M$.
These deviations diminish upon increasing $\Mred$. 

As shown in \cref{fig_peakevol_nc_Dir_neq,fig_peakevol_nc_Dir_eq}, both in the transient and equilibrium regime, the peak $\bra h(x_M,\dt)\ket$ follows the same algebraic time-evolution as in \cref{eq_peakevol_EW_pbc} and is characterized by two distinct dynamic exponents.
Similarly to periodic \bcs [see \cref{eq_peakevol_alpha_EW_pbc}], we obtain $\alpha\simeq 0.29$ and $\alpha=\alpha_0\simeq 0.5$ for the values of the dynamic exponent at late and early times $\dt$, respectively, which are different from the prediction of WNT in \cref{eq_peakevol_EW}.
Despite this discrepancy, the time-dependent averaged profiles of WNT qualitatively match the simulation results in the equilibrium regime [see \cref{fig_avgprof_nc_Dir_eq}(b)].
Deviations are more significant in the transient regime (not shown), although the qualitative behavior agrees with WNT. 

In Fig.\ \ref{fig_avgprof_nc_Dir_neq}(c) and \ref{fig_avgprof_nc_Dir_eq}(c), the dynamic scaling behavior predicted in \cref{eq_h_lateT_dynscal_EW,eq_h2_shortTdyn_asympt}, respectively, is tested.
Using an effective value of $1/z= \alpha\simeq 0.29$ for the dynamic exponent, a satisfactory fit of the numerical profiles with the scaling function in \cref{eq_hscalf_asympt_EW} is obtained.
The agreement between WNT and simulations generally improves as $\dt\to 0$.

\section{Mullins-Herring equation}
\label{sec_MH}

\begin{figure}[t]\centering
	\subfigure[]{\includegraphics[width=0.4\linewidth]{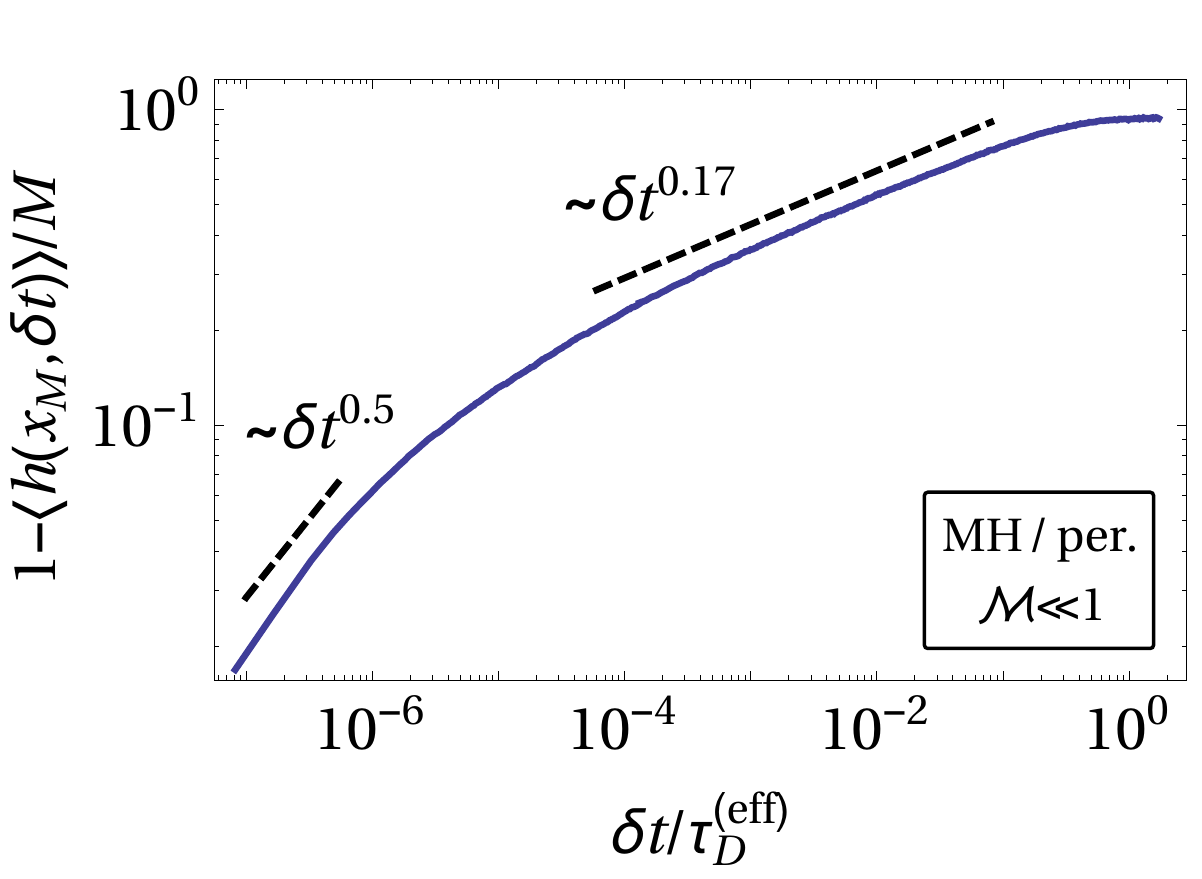} \label{fig_peakevol_c_pbc_neq}}\qquad 
	\subfigure[]{\includegraphics[width=0.4\linewidth]{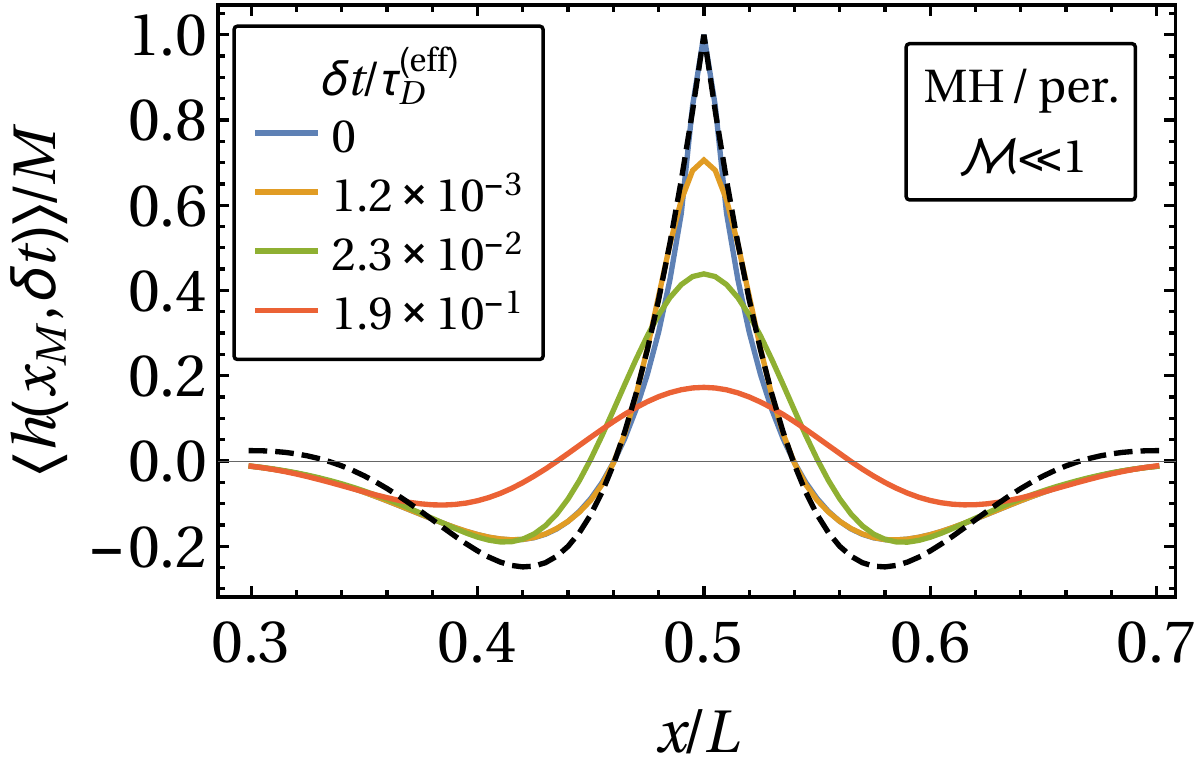}}
	\subfigure[]{\includegraphics[width=0.4\linewidth]{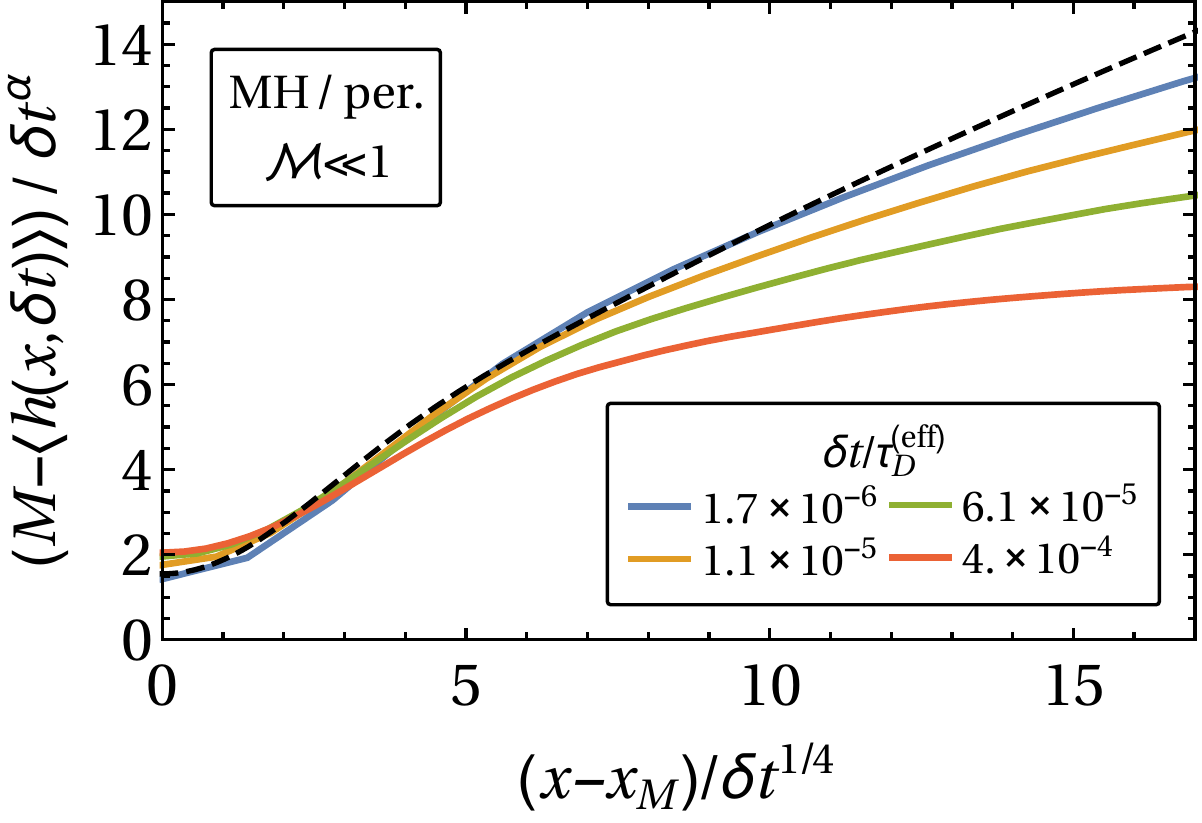}}
	\caption{Averaged profile $\bra h(x,\dt)\ket$ for the MH equation [\cref{eq_MH}] with periodic \bcs in the transient regime ($\Mred\ll 1$). The first passage of the height $M$ occurs at the time $\dt=0$. Utilizing translational invariance, the individual profiles obtained from simulation are shifted such that the height $M$ is reached at location $x_M=L/2$. Time is normalized to the diffusion time scale $\tau_D$ [\cref{eq_tdiff}] using for the exponent $z$ an \emph{effective} value of $1/(2\alpha)\simeq 2.9$ ($\alpha\simeq 0.17)$ instead of $4$, as implied by panel (a).
	(a) Time-evolution of the peak of the profile, $\bra h(x_M,\dt)\ket$, which exhibits an intermediate asymptotic regime, $M-\bra h(x_M,\dt)\ket\propto \dt^\alpha$ with $\alpha\simeq 0.17$. 
	(b) Spatio-temporal evolution of the averaged profile. The solid curves represent numerical simulations, while the dashed curve indicates the asymptotic profile predicted by WNT in \cref{eq_h_shortT_MH}, taking $T$ as a fit parameter. 
	(c) Test of the dynamic scaling behavior of $\bra h(x,\dt)\ket$ as predicted by WNT in \cref{eq_h4_shortTdyn_asympt}, using a value of $1/z\simeq 0.19$. The dashed curve represents the scaling function $c \tilde \Hcal$ in \cref{eq_hscalf_asympt_MH}, with a prefactor $c\simeq 1.4$ determined from a fit. In order to account for the localized nature of the profile in the transient regime, in all panels the individual profiles are shifted before averaging such that $h^{(s)}(L/2,T^{(s)})=M$ [see \cref{eq_def_avgprof}].}
	\label{fig_avgprof_c_pbc_neq}
\end{figure}

\begin{figure}[t]\centering
	\subfigure[]{\includegraphics[width=0.4\linewidth]{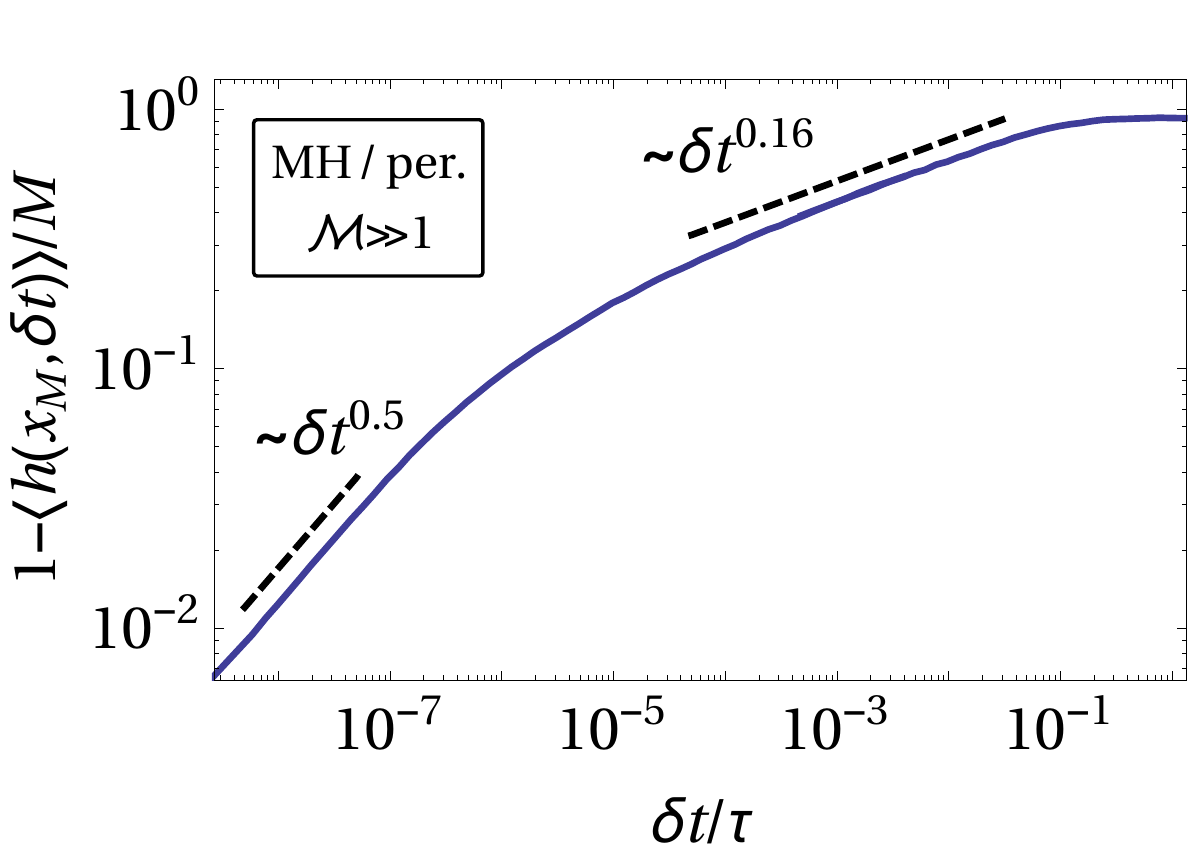} \label{fig_peakevol_c_pbc_eq}}\qquad 
	\subfigure[]{\includegraphics[width=0.4\linewidth]{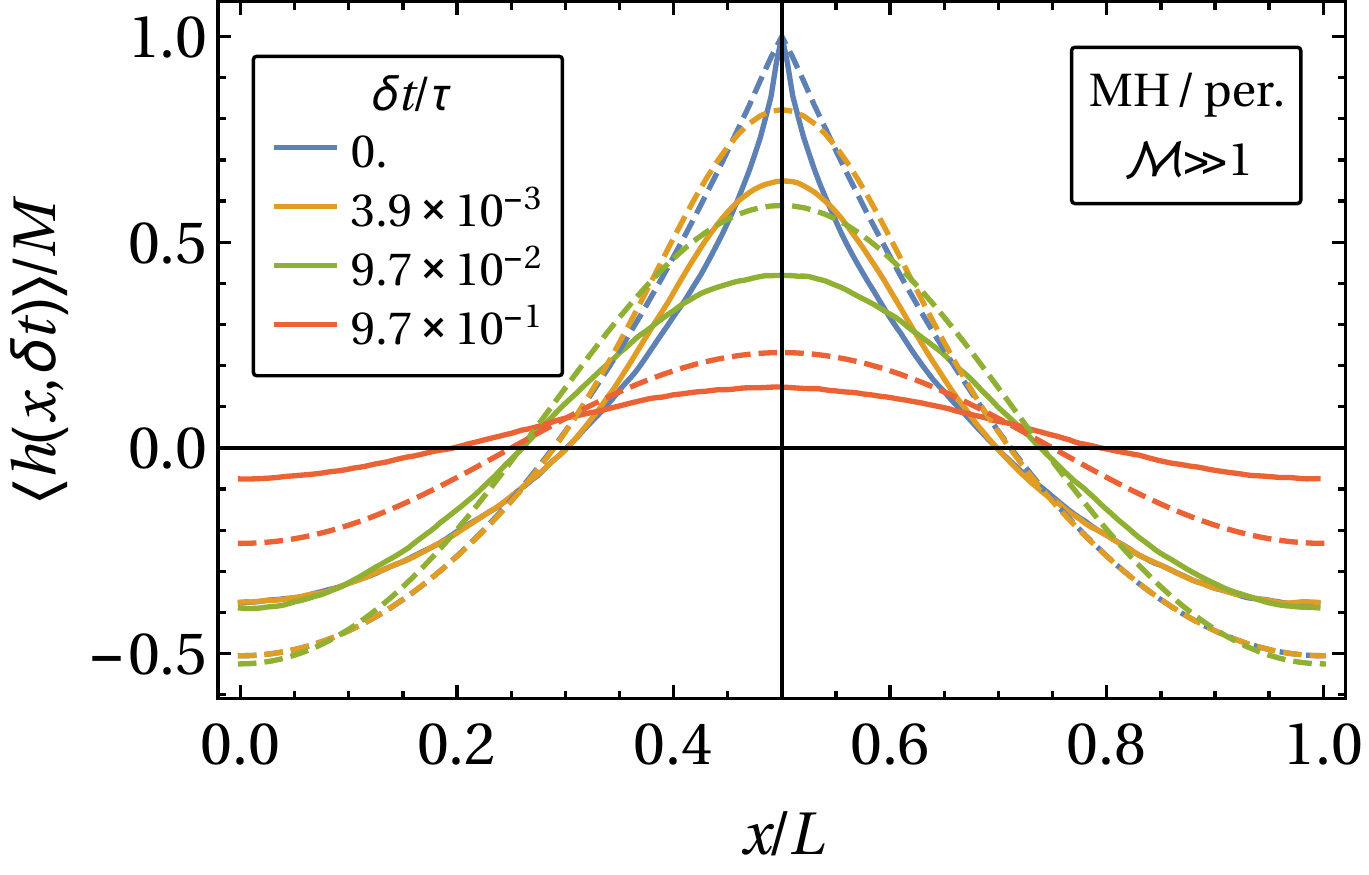}}
	\subfigure[]{\includegraphics[width=0.4\linewidth]{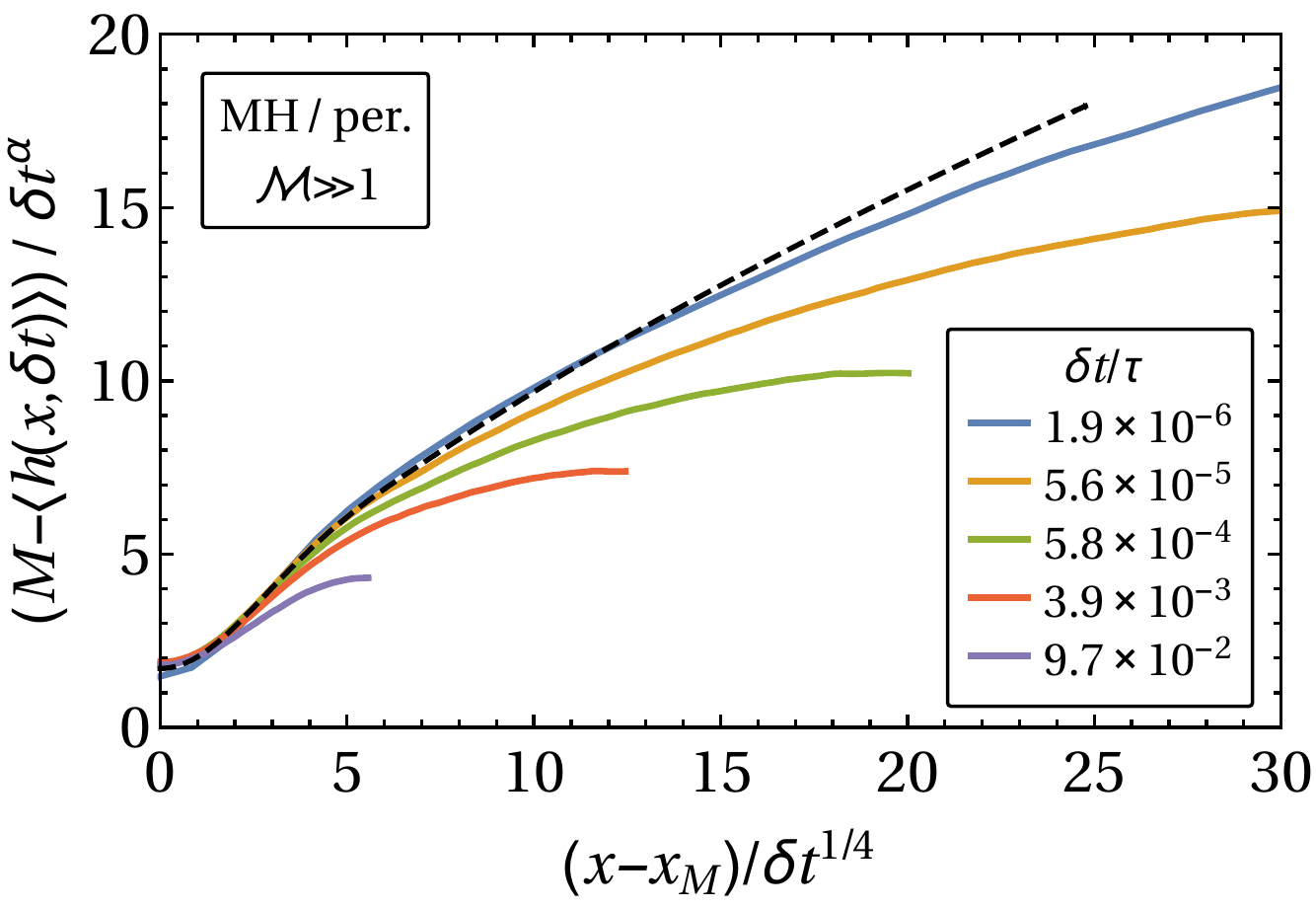}}
	\caption{Averaged profile $\bra h(x,\dt)\ket$ for the MH equation [\cref{eq_MH}] with periodic \bcs in the equilibrium regime ($\Mred\gg 1$). The first passage of the height $M$ occurs at the time $\dt=0$. Utilizing translational invariance, the individual profiles obtained from simulation are shifted such that the height $M$ is reached at location $x_M=L/2$. Time is normalized to the relaxation time $\tau\pbc$ [\cref{eq_relaxtime}].
	(a) Time-evolution of the peak of the profile, $\bra h(x_M,\dt)\ket$, which exhibits an intermediate asymptotic regime $M-\bra h(x_M,\dt)\ket\propto \dt^\alpha$ with $\alpha\simeq 0.16$. 
	(b) Spatio-temporal evolution of the averaged profile. The solid curves represent the  numerical simulations, while the dashed curves indicate the prediction of WNT [see Eq.~I-(3.17) as well as Ref.\ \cite{meerson_macroscopic_2016}]. (c) Test of the dynamic scaling behavior of $\bra h(x,t)\ket$ as predicted by WNT according to  \cref{eq_h_lateT_dynscal_MH}, using a value of $1/z \simeq 0.17$. The dashed curve represents the scaling function $c \tilde \Hcal$ in \cref{eq_hscalf_asympt_MH}, with a prefactor $c\simeq 1.7$ determined from a fit.}
	\label{fig_avgprof_c_pbc_eq}
\end{figure}

We proceed with the discussion of the first-passage dynamics for the MH equation [\cref{eq_MH}].
For the considered \bcs, the mass [\cref{eq_mass}] is conserved in time and, in fact, $\mass([h],t)=0$ owing to the initial condition in \cref{eq_init_cond}.
Due to the larger value $z=4$ of the dynamic index [see \cref{eq_relaxtime}], simulations are more time-demanding than for the EW equation. Moreover, it turns out that the cross-over regions between the different asymptotic regimes are broader, making it more difficult to identify clear power-laws.

\subsection{Summary of WNT}

Before proceeding to the simulation results, we summarize the essential predictions of WNT (see \paperI, as well as Ref.\ \cite{meerson_macroscopic_2016} in the case of periodic \bcs). 
As before, we use $\dt=T-t$ as the time variable and the following expressions for $h$ are to be understood as the leading-order contributions to the averaged profile $\bra h\ket$.
Asymptotically for $T\to 0$ in the \emph{transient} regime, one obtains the following static scaling profile at the first-passage event:
\beq h(x,\dt=0)\big|_{T\ll \tau} = M \Hcal\left(\frac{x-L/2}{(2T)^{1/z}}\right),\qquad z=4,
\label{eq_h_shortT_MH}\eeq 
with the scaling function
\beq \Hcal(\xi)={}_1 F_3\left(-\frac{1}{4}; \frac{1}{4},\frac{1}{2}, \frac{3}{4}; \frac{\xi^4}{256}\right) + \xi^2 \frac{\Gamma\left(\frac{1}{4}\right)}{8\Gamma\left(\frac{3}{4}\right)} {}_1 F_3\left(\frac{1}{4}; \frac{3}{4}, \frac{5}{4}, \frac{3}{2}; \frac{\xi^4}{256}\right) - \frac{\pi}{2\Gamma\left(\frac{3}{4}\right)}|\xi| ,
\label{eq_h_shortT_scalF_MH}\eeq 
which applies to periodic as well as Dirichlet no-flux \bcs. ${}_1 F_3$ is a hypergeometric function \cite{olver_nist_2010}.
A dynamic scaling profile for times $\dt>0$ with $\delta t\ll T$ is given, to leading order in $\dt/T$, by
\beq h(x,\dt)\big|_{\substack{T\ll \tau\\\dt\ll T}} = M - M \left(\frac{\delta t}{2T}\right)^{1/z} \tilde \Hcal\left(\frac{x-L/2}{(\delta t)^{1/z}}\right),\qquad z=4,
\label{eq_h4_shortTdyn_asympt}\eeq 
with the scaling function
\beq \tilde \Hcal(\xi) = 
 {}_1 F_3\left(-\frac{1}{4}; \frac{1}{4},\frac{1}{2}, \frac{3}{4}; \frac{\xi^4}{256}\right) + \xi^2 \frac{\Gamma\left(\frac{1}{4}\right)}{8\Gamma\left(\frac{3}{4}\right)} {}_1 F_3\left(\frac{1}{4}; \frac{3}{4}, \frac{5}{4}, \frac{3}{2}; \frac{\xi^4}{256}\right) .
 \label{eq_hscalf_asympt_MH}\eeq 
In the \emph{equilibrium} regime, the static profile $h\pbc(x,\dt=0)|_{T\to\infty}$ minimizing the corresponding free energy for periodic \bcs (see \paperI) coincides with the one in \cref{eq_opt2_finalprof_eq_pbc}.
For Dirichlet no-flux \bcs, instead, one finds
\beq h\DirNoFl(x,\dt=0)\big|_{T\to\infty} = h\pbc(x + L/2-x_M,\dt=0)\big|_{T\to\infty}
\label{eq_opt4_finalprof_eq_DirNoFl}
\eeq
with 
\beq 
x_M\DirNoFl\big|_{T\to \infty} = \frac{L}{2}\left(1 \pm \frac{1}{\sqrt{3}}\right).
\label{eq_xM_DirNoFl}
\eeq 
For definiteness, we choose henceforth the smaller value for $x_M\DirNoFl$, such that \cref{eq_opt4_finalprof_eq_DirNoFl} can be explicitly written as
\beq h\DirNoFl(x,\dt=0)\big|	_{T\to\infty}/M = \begin{cases}\displaystyle 6\frac{x}{L} \left(\frac{x}{L} + \frac{1}{\sqrt{3}}\right) ,\qquad &x\leq x_M\DirNoFl,\\
\displaystyle 6\left(\frac{x}{L}-1\right)\left(\frac{x}{L}-1+\frac{1}{\sqrt{3}}\right) ,&x > x_M\DirNoFl.
\end{cases}\label{eq_opt4_finalprof_eq_DirNoFl_spec}
\eeq
In the equilibrium regime for times $\dt>0$ with $\delta t\ll T$, a dynamic scaling profile for periodic and Dirichlet \bcs is given by
\beq h(x,\dt)\big|_{T\gg\tau} \simeq M -  M (\delta t)^{1/z} \Gamma(1-1/z) \tilde\Hcal\left(\frac{x-x_M}{\delta t^{1/z}}\right),\qquad z=4,
\label{eq_h_lateT_dynscal_MH}\eeq 
with the same scaling function as in \cref{eq_hscalf_asympt_MH}.
Note that the above expressions pertain to a continuum system.
In the presence of an upper bound to the eigenmode spectrum, the time evolution of the peak $h(x_M,\dt)$ of the profile exhibits two regimes:
\beq 1- h(x_M,\delta t)/M \propto 
\begin{cases} 
	\dt,\qquad & \dt\lesssim \tau\cro,\\
	\dt^{1/z}, \qquad & \dt\gtrsim \tau\cro,
\end{cases}
\label{eq_peakevol_MH}\eeq
where $\tau\cro$ is the crossover time [see \cref{eq_crossover_time}].  
As was the case for the EW equation [see \cref{eq_peakevol_EW}], \cref{eq_peakevol_MH} is independent of the \bcs and applies to both the transient and the equilibrium regime.
Explicit expressions for the first-passage profiles obtained within WNT for all times are reported in \paperI.

\subsection{Periodic \bcs}
\label{sec_MH_pbc}
Here, we discuss simulation results obtained for the MH equation with periodic \bcs.
\Cref{fig_avgprof_c_pbc_neq,fig_avgprof_c_pbc_eq} illustrate the time evolution of the averaged profile $\bra h(x,\dt)\ket$ towards the first-passage event in the transient and equilibrium regimes, respectively.
As shown in panels (a), in both regimes, the peak $\bra h(x_M=L/2,\dt)\ket$ approaches the height $M$ via a power-law, $M-\bra h(x_M=L/2,\dt)\ket \propto \dt^\alpha$, with $\alpha \simeq 0.16-0.17$ at intermediate times ($\dt\gtrsim \tau\cro$) and  $\alpha= \alpha_0 \simeq 0.5$ at early times ($\dt\lesssim \tau\cro$).
Analogously to the finding for EW dynamics (see \cref{sec_EW}), these values of the dynamical exponent are significantly smaller than the prediction $\alpha=1/4$ and $\alpha_0=1$ obtained from WNT [\cref{eq_peakevol_MH}]. This finding is  rationalized in \cref{sec_discussion} below.
In order to account for this quantitative change in the dynamics, in \cref{fig_peakevol_c_pbc_neq} we rescale time by an effective diffusion time scale $\tau_D\ueff$, which results from \cref{eq_tdiff} by replacing $z$ by the value $1/(2\alpha)\simeq 2.9-3.1$ [cf.\ \cref{sec_EW_pbc}]. 
For the systems considered in \cref{fig_peakevol_c_pbc_neq,fig_peakevol_c_pbc_eq}, the crossover time defined in \cref{eq_crossover_time} follows as $\tau\cro/\tau\pbc\simeq 2\times 10^{-7}$ and  $6\times 10^{-8}$, respectively, which is in good agreement with the simulation data.

Figs.\ \ref{fig_avgprof_c_pbc_neq}(b) and \ref{fig_avgprof_c_pbc_eq}(b) illustrate the spatio-temporal evolution of the averaged profile. 
The deviations from the prediction of WNT (dashed curves) can be mainly attributed to the fact that simulations operate in the finite-noise regime. 
As shown in \ref{fig_avgprof_c_pbc_eq}(b), in the equilibrium regime, the time-dependent profile shapes obtained from simulations are qualitatively similar to WNT, although the difference in the value of the dynamic exponent $\alpha$ leads to a faster time evolution in the latter case. 

Figs.\ \ref{fig_avgprof_c_pbc_neq}(c) and \ref{fig_avgprof_c_pbc_eq}(c) demonstrate that, in an inner region, the profiles follow the scaling behavior implied by \cref{eq_h4_shortTdyn_asympt,eq_h_lateT_dynscal_MH}. The agreement improves upon decreasing $\dt$. Scaling collapse is obtained here by using in \cref{eq_h4_shortTdyn_asympt,eq_h_lateT_dynscal_MH} for $1/z$ an effective value of $0.17-0.19$, consistent with the value of the exponent ($\alpha$) that governs the time-evolution of the peak of the profile [see \cref{fig_peakevol_c_pbc_eq,fig_peakevol_c_pbc_neq}].

\subsection{Dirichlet no-flux \bcs}

\begin{figure}[t]\centering
	\includegraphics[width=0.53\linewidth]{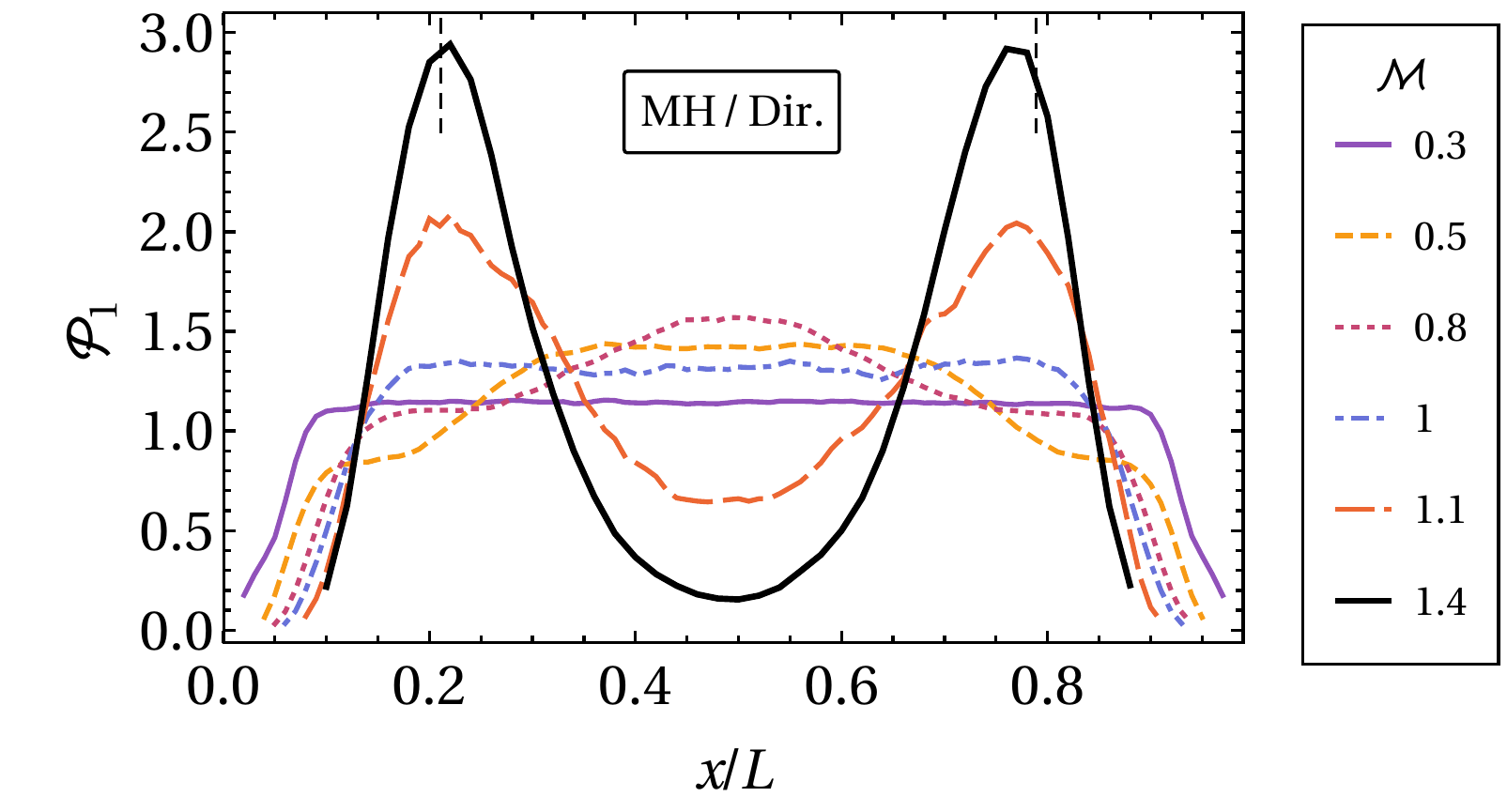}\qquad 
	\caption{Probability distribution $\Pcal_1(x_M)$ obtained for the MH equation [\cref{eq_MH}] with Dirichlet no-flux \bcs and for various reduced heights $\Mred$ [\cref{eq_hm_red}]. $x_M$ denotes the spatial location at which the profile passes the height $M$ for the first time. Asymptotically for $\Mred\to 0$ in the transient regime, $\Pcal_1$ is generally flat (except at the boundaries). For Dirichlet no-flux \bcs in the equilibrium regime ($\Mred\gtrsim 1$), two peaks emerge at the locations given in \cref{eq_xM_DirNoFl}.}
	\label{fig_hittingpdf_c_Dir}
\end{figure}

\begin{figure}[t]\centering
	\subfigure[]{\includegraphics[width=0.4\linewidth]{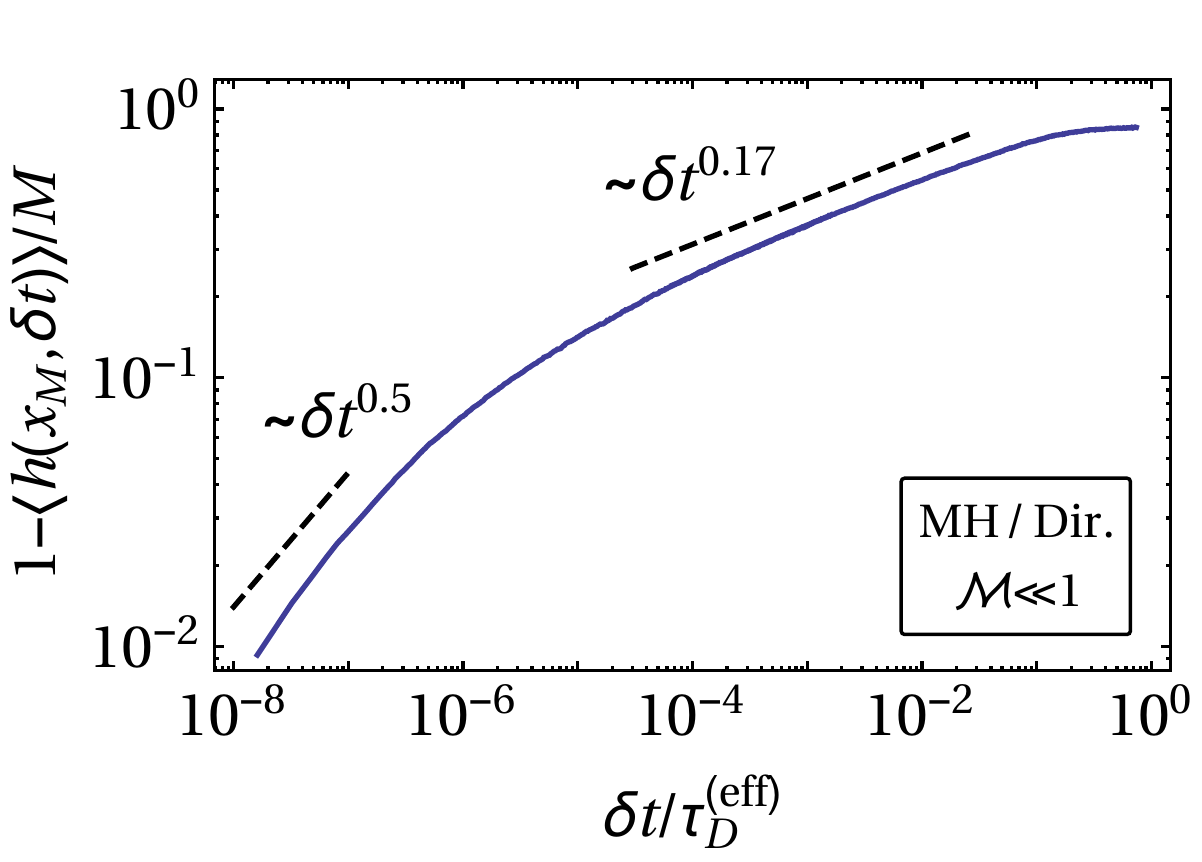} \label{fig_peakevol_c_Dir_neq}}\qquad
	\subfigure[]{\includegraphics[width=0.4\linewidth]{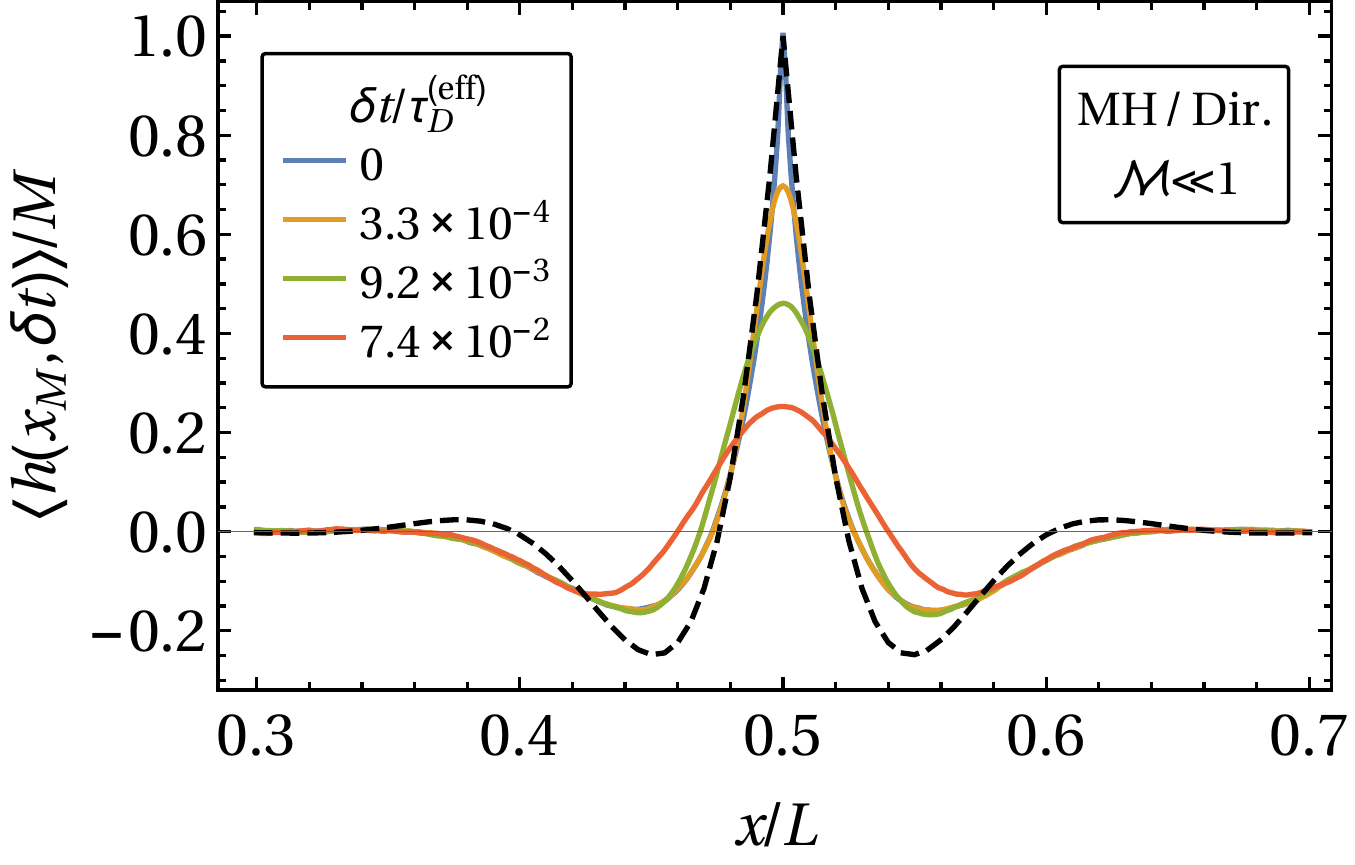}}
	\subfigure[]{\includegraphics[width=0.4\linewidth]{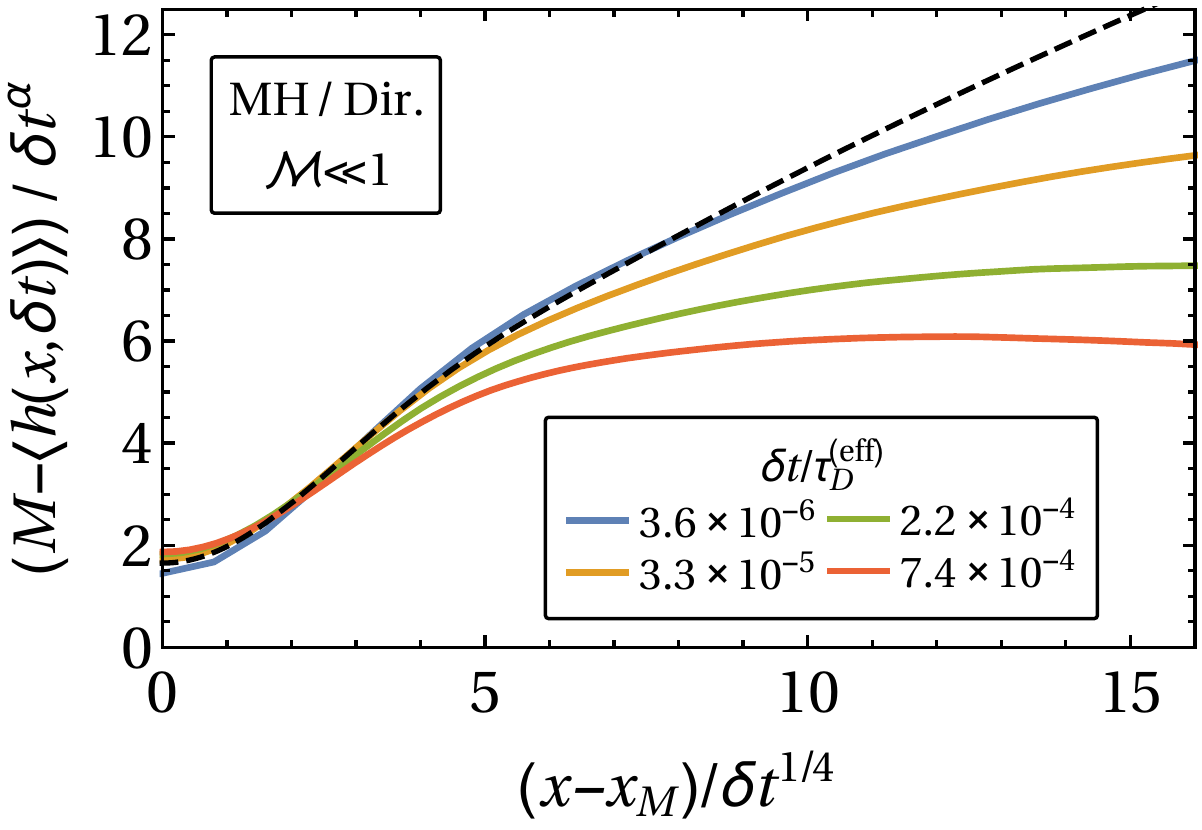}}
	\caption{Averaged profile $\bra h(x,\dt)\ket$ for the MH equation [\cref{eq_MH}] with Dirichlet no-flux \bcs in the transient regime ($\Mred\ll 1$). The first passage of the height $M$ occurs at the time $\dt=0$. Time is normalized to the diffusion time scale $\tau_D$ [\cref{eq_tdiff}] using for the exponent $z$ an \emph{effective value} of $1/(2\alpha)\simeq 2.9$ ($\alpha\simeq 0.17)$ instead of $4$, as suggested by panel (a).
	(a) Time-evolution of the peak of the profile, $\bra h(x_M,\dt)\ket$, which exhibits an intermediate asymptotic regime $M-\bra h(x_M,\dt)\ket\propto \dt^\alpha$ with $\alpha\simeq 0.17$. 
	(b) Spatio-temporal evolution of the averaged profile. The solid curves represent the  numerical simulations, while the dashed curve indicates the asymptotic profile predicted by WNT in \cref{eq_h_shortT_MH}, taking $T$ as a fit parameter.  
	(c) Test of the dynamic scaling behavior of $\bra h(x,\dt)\ket$ as predicted by WNT in \cref{eq_h4_shortTdyn_asympt}, using an effective value of $1/z \simeq 0.17$. The dashed curve represents the scaling function $c \tilde \Hcal$ in \cref{eq_hscalf_asympt_MH}, with a prefactor $c\simeq 1.3$ determined from a fit. In order to account for the localized nature of the individual profiles, data in panels (a) and (c) are obtained by shifting the individual profiles before averaging such that $h^{(s)}(L/2,T^{(s)})=M$ [see \cref{eq_def_avgprof}].}
	\label{fig_avgprof_c_Dir_neq}
\end{figure}

\begin{figure}[t]\centering
	\subfigure[]{\includegraphics[width=0.4\linewidth]{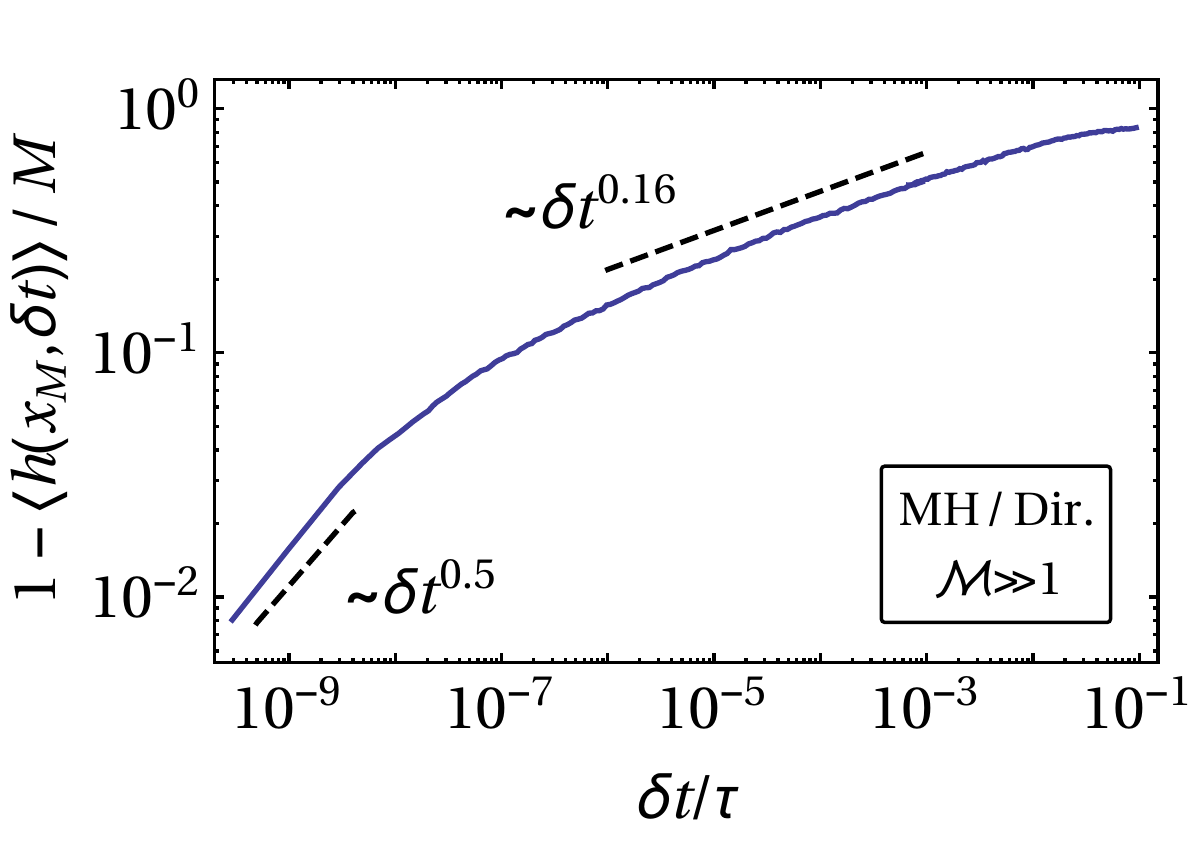} \label{fig_peakevol_c_Dir_eq}}\qquad
	\subfigure[]{\includegraphics[width=0.4\linewidth]{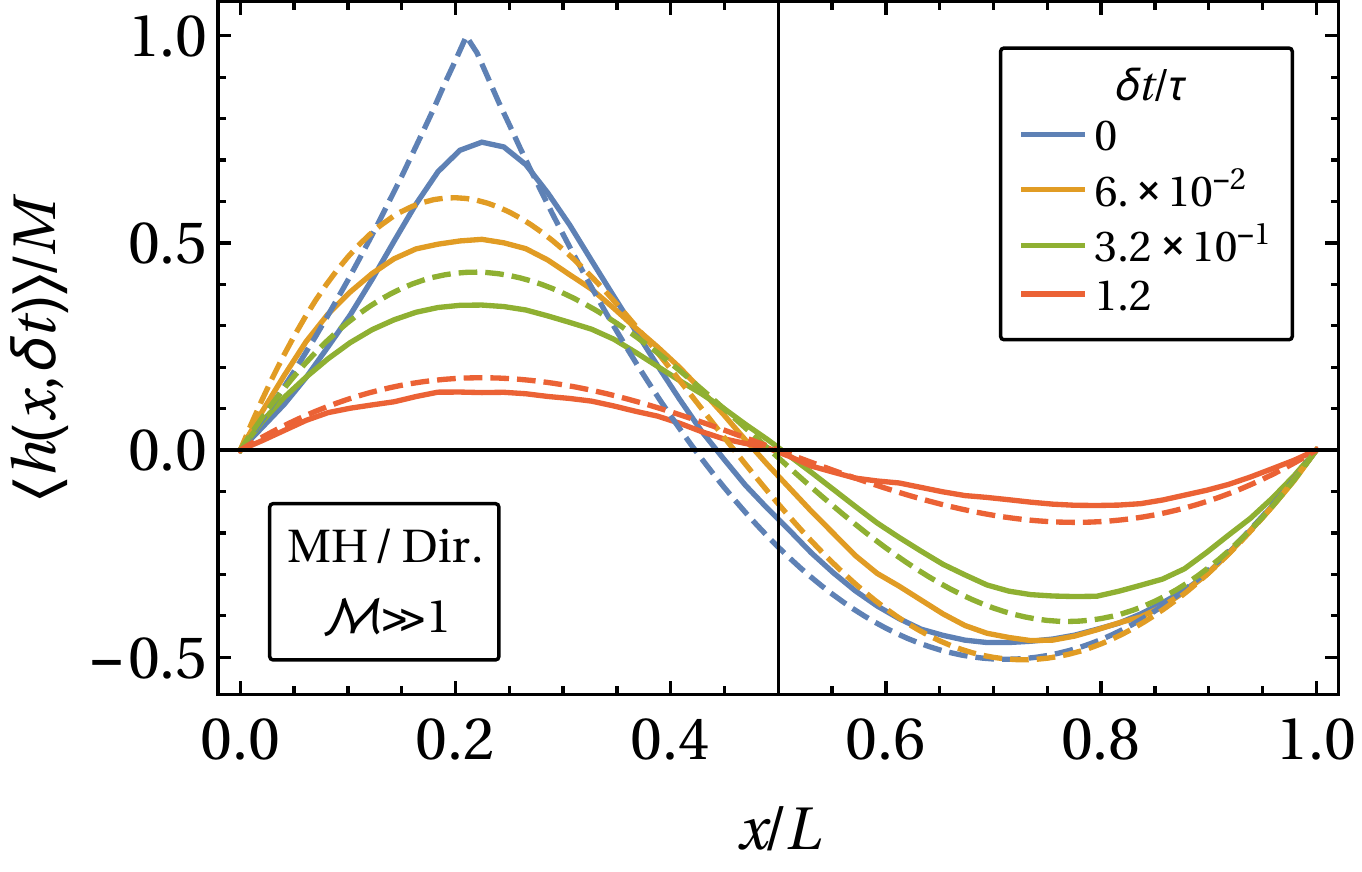}} 
	\subfigure[]{\includegraphics[width=0.4\linewidth]{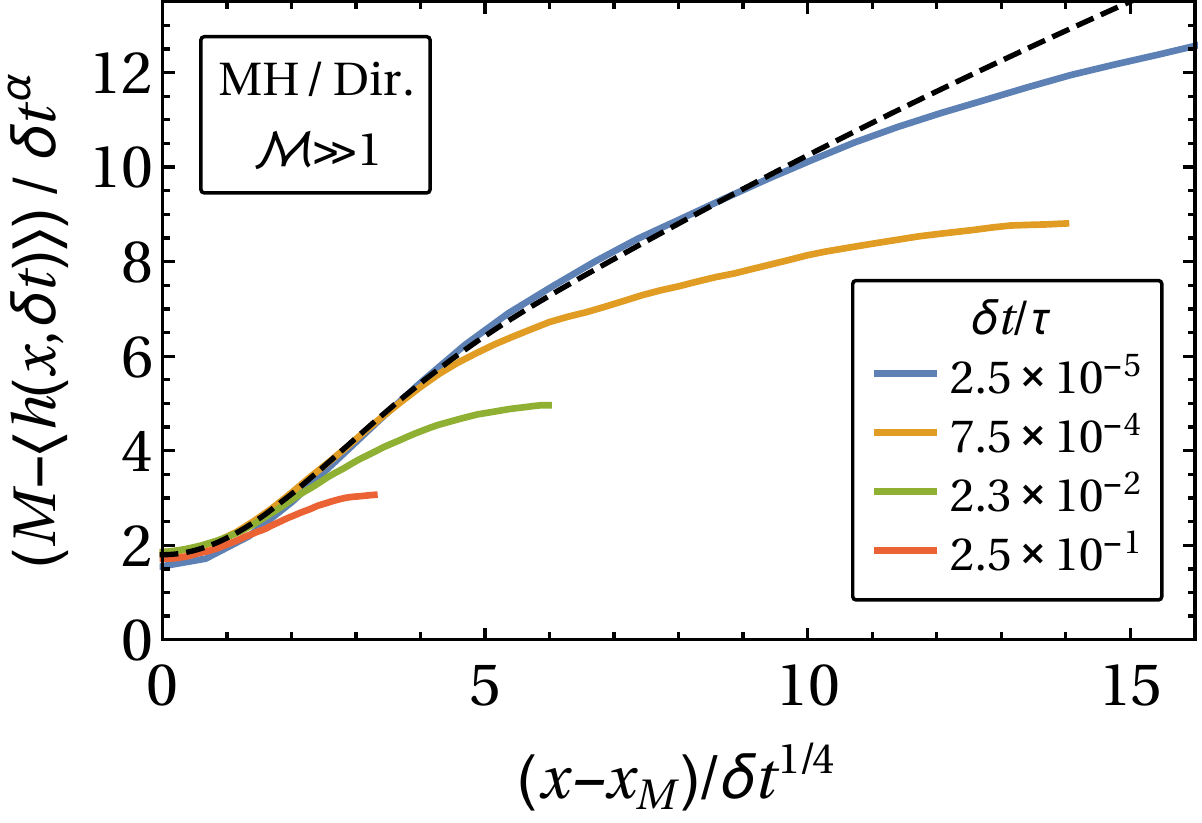}}
	\caption{Averaged profile $\bra h(x,\dt)\ket$ for the MH equation [\cref{eq_MH}] with Dirichlet no-flux \bcs in the equilibrium regime ($\Mred\gg 1$). The first passage of the height $M$ occurs at the time $\dt=0$. Time is normalized to the relaxation time $\tau\DirNoFl$ [\cref{eq_relaxtime}].
	(a) Time-evolution of the peak of the profile, $\bra h(x_M,\dt)\ket$, which exhibits an intermediate asymptotic regime $M-\bra h(x_M,\dt)\ket\propto \dt^\alpha$ with $\alpha\simeq 0.16$. 
	(b) Spatio-temporal evolution of the averaged profile. The solid curves represent the numerical simulations, while the dashed curves indicate the prediction of WNT [see Eq.~I-(3.19)]. (c) Test of the dynamic scaling behavior of $\bra h(x,\dt)\ket$ as predicted by WNT in \cref{eq_h_lateT_dynscal_MH}, using a value of $1/z \simeq 0.17$. The dashed curve represents the scaling function $c \tilde \Hcal$ in \cref{eq_hscalf_asympt_MH}, with a prefactor $c\simeq 2$ determined from a fit.}
	\label{fig_avgprof_c_Dir_eq}
\end{figure}

In contrast to standard Dirichlet \bcs, which entail a fixed chemical potential  at the boundaries (see \paperI) and thus a non-conserved mass, the no-flux condition [\cref{eq_H_noflux}] ensures mass conservation for the MH equation.
In fact, due to the initial condition in \cref{eq_init_cond}, the mass $\mass([h],t)$ [\cref{eq_mass}] vanishes at all times.
\Cref{fig_hittingpdf_c_Dir} shows the probability distribution $\Pcal_1\DirNoFl(x_M)$ of the first-passage location $x_M$. We find that the essential predictions of WNT [see Fig.~I-7] are recovered by the simulations. 
Asymptotically in the transient regime ($\Mred\to 0$), $\Pcal_1$ is generally constant as a function of $x_M$ for $0<x_M<L$. At the boundaries, $\Pcal_1\DirNoFl$ vanishes as a consequence of Dirichlet \bcs. Upon increasing $\Mred$ towards values of $\Ocal(1)$, a peak develops in the central region of $\Pcal_1\DirNoFl$. 
Upon increasing $\Mred$ further, this peak diminishes, while two symmetric peaks develop near the location $x_M\DirNoFl$ [\cref{eq_xM_DirNoFl}] predicted by WNT.
One expects $\Pcal_1\DirNoFl(x_M)\to \delta\left(x_M\pm x_M\DirNoFl\right)$ as $\Mred\to\infty$, which represents a particular realization of the weak-noise limit.

In \cref{fig_avgprof_c_Dir_neq,fig_avgprof_c_Dir_eq}, the profile dynamics obtained from simulations in the transient and equilibrium regimes, respectively, is illustrated.
In panels (a), the averaged time evolution of the peak, $\bra h(x_M,\dt)\ket$, is shown as a function of the time $\dt$ until the first-passage event. 
In order to account for the spread in the distribution of $x_M$, in these two panels $\bra h(x_M,\dt)\ket$ is computed according to \cref{eq_def_avgprof} by shifting the individual profiles $h^{(s)}$ to the common first-passage location $L/2$, such that $h^{(s)}(L/2,T^{(s)})=M$. 
The peak is found to evolve algebraically, $M - \bra h(x_M,\dt)\ket\propto \dt^\alpha$, with $\alpha= \alpha_0\simeq  0.5$ for times $\dt \lesssim \tau\cro$ and  $\alpha\simeq 0.16-0.17$ for $\tau\cro \lesssim \dt\lesssim \tau\DirNoFl$.
These values for $\alpha$ practically coincide with the ones for periodic \bcs [\cref{sec_MH_pbc}] and are further discussed in \cref{sec_discussion}.
In order to estimate the crossover time $\tau\cro$ [see \cref{eq_crossover_time}], we assume that the largest mode which can be accommodated by the system is given by $k\cro\simeq 0.5 L/\lattsp$ (see \cref{app_bench} for further discussion). This renders the estimates $\tau\cro\simeq 2.4\times 10^{-7} \tau_D\ueff$ and $\tau\cro\simeq 3.2\times 10^{-9}\tau\DirNoFl$ in the transient and equilibrium regimes, respectively, which are seen to agree with the simulation data within an order of magnitude.

The time-dependent averaged profile $\bra h(x,\dt)\ket$ in the transient regime is illustrated in \cref{fig_avgprof_c_Dir_neq}(b).
The average [see \cref{eq_def_avgprof}] is computed here again by translating each profile $h^{(s)}$ to the common first-passage location $L/2$.
This transformation does not significantly affect the profile shape because the profiles are strongly localized and the distribution $\Pcal_1\DirNoFl(x_M)$ is approximately flat in the transient regime [see \cref{fig_hittingpdf_c_Dir}].
In the equilibrium regime, in contrast, $\Pcal_1\DirNoFl(x_M)$ is symmetric around $L/2$ and the first-passage event is most likely to occur at either of the two locations given in \cref{eq_xM_DirNoFl}.
In this case, the averaged profile $\bra h(x,\dt)\ket$ shown in \cref{fig_avgprof_c_Dir_eq}(b) is obtained by mirroring at $x=L/2$ all profiles $h^{(s)}$ which belong to a simulation with $x_M^{(s)}> L/2$.
The spatio-temporal evolution of the profile displayed in the plots qualitatively agrees with the predictions of WNT (see \paperI). 
As a consequence of the finite width of $\Pcal_1\DirNoFl$ around each of its two peaks, the maximum of $\bra h(x,\dt)\ket$ in \cref{fig_avgprof_c_Dir_eq}(b) is smaller than $M$, despite the fact that each stochastic realization fulfills $h^{(s)}(x_M^{(s)},T^{(s)})=M$. 

Close to the first-passage event, WNT predicts a universal dynamic scaling behavior of the profile, as expressed in \cref{eq_h4_shortTdyn_asympt,eq_h_lateT_dynscal_MH}.
As shown in Figs.\ \ref{fig_avgprof_c_Dir_neq}(c) and \ref{fig_avgprof_c_Dir_eq}(c), this property is recovered in the simulations: upon accounting for the renormalized dynamic exponent $1/z\to \alpha\simeq 0.17$, the profiles superimpose onto the scaling function $c \tilde H$ [\cref{eq_hscalf_asympt_MH}] within an inner region, where $c$ is a fit parameter. 

\section{Discussion}
\label{sec_discussion}

As demonstrated in the preceding sections, a crucial difference between the results of the Langevin simulations and the predictions of WNT arises in the time-dependence of the averaged profile.
Both in simulations and within WNT, the peak of the profile $\bra h(x_M,\dt)\ket$ approaches the first-passage height $M$ algebraically,
\beq M-\bra h(x_M,\dt)\ket\propto 
\begin{cases}
\dt^{\alpha_0} ,\qquad \qquad &\dt\lesssim \tau\cro,\\
\dt^\alpha,\qquad &\tau\cro\lesssim \dt \lesssim \tau.
\end{cases}
\label{eq_peakevol}\eeq 
However, simulations yield the values
\beq 
\alpha_0 \simeq 0.5,\qquad 
\alpha\simeq \begin{cases} 
0.27-0.3,\qquad &\text{EW},\\
0.16-0.17,\qquad &\text{MH},
\end{cases}
\label{eq_peakevol_exp}
\eeq 
for the dynamic exponents, while WNT predicts (see \paperI)
\beq \alpha\st{0,WNT}=1,\qquad  \alpha\st{WNT} = 1/z = \begin{cases}
                             1/2,\qquad &\text{EW},\\
                             1/4,\qquad &\text{MH}.
                            \end{cases}
\label{eq_peakevol_exp_WNT}\eeq 
We emphasize that these results are independent of the \bcs and apply both in the transient and in the equilibrium regime.
The crossover time  $\tau\cro$ [\cref{eq_crossover_time}] and the roughening time $\tau$ [\cref{eq_relaxtime}] correspond to the relaxation time of the shortest and largest fluctuation wavelengths, respectively, that can be accommodated by the system. 
Since $\tau\propto L^z$, the intermediate asymptotic regime characterized by the exponent $\alpha$ dominates for sufficiently large systems.
As detailed in the preceding sections, we furthermore recall that the time-evolution of the peak $\bra h(x_M,\dt)\ket$ is determined based on a slightly different averaging procedure than the one used for the full profile [see also \cref{eq_def_avgprof}].

In order to gain a basic understanding of the discrepancy between \cref{eq_peakevol_exp} and \eqref{eq_peakevol_exp_WNT}, we first consider a \emph{(Markovian) Brownian walker} $h(t)$, initially at $h(t=0)=0$, in the presence of an absorbing boundary at a fixed height $h=M$ (see \cref{sec_fbm_MPL,app_walker} for details).
Within WNT, the averaged path of the walker between the points $(t=0,h=0)$ and $(T,M)$, with $T$ fixed, is the one minimizing the associated action (see \cref{sec_fbm_MPL}). 
This results in a linear time-dependence of the walker approaching the absorbing boundary [see \cref{eq_MLP_Markov}], 
\beq M-\bra h(\dt)\ket\st{WNT}\propto \dt.\qquad \text{(standard Brownian motion)}
\label{eq_BMpath_WNT}\eeq 
As before, the average is defined here such that the first-passage event occurs at $\dt=0$.
For a Markovian Brownian walker, the averaged path to an impenetrable boundary can however also be calculated exactly, i.e., including all corrections beyond WNT (see \cref{app_avgpath_BM}). For a fixed endpoint $(T,M)$, this yields
\beq M-\bra h(\dt)\ket\propto \dt^{1/2}\qquad \text{(standard Brownian motion, fixed $T$)}
\label{eq_BMpath_avg_fixedT}\eeq 
as $\dt\to 0$.
The difference between the dynamic exponents in \cref{eq_BMpath_WNT,eq_BMpath_avg_fixedT} arises from the ``entropic repulsion'' (cf., e.g., Refs.\ \cite{fisher_walks_1984,redner_guide_2001}) exerted by the absorbing boundary onto fluctuations of the walker \emph{around} the most-likely path described by WNT.
Averaging also over the first-passage time distribution results in [see \cref{sec_Brownian_firstpsg}]
\beq M-\bra h(\dt)\ket \propto \dt^{1/2}\qquad \text{(standard Brownian motion, first-passage path)}
\label{eq_BMpath_avg}\eeq 
and accordingly does not alter the trajectory asymptotically close to the boundary compared to \cref{eq_BMpath_avg_fixedT}. Far from the boundary, however, significant changes in the walker path are induced by this additional average (see \cref{fig_firstpsg_BM,fig_avgpath_Brownian}).

The preceding results can be extended to \emph{fractional Brownian motion}, i.e., to a Gaussian random process $h(t)$ characterized by the correlation function in \cref{eq_fbm_autocorrel}.
On its most-likely path, the walker approaches the endpoint $(t=T,h=M)$  algebraically [see \cref{eq_MLP_FBM}]:
\beq M-\bra h(\dt)\ket\st{WNT}\propto \dt^{2H},\qquad \text{(fractional Brownian motion)}
\label{eq_FBMpath_WNT}\eeq 
where $H$ is the Hurst exponent of the process ($H=1/2$ for standard Brownian motion).
Beyond the weak-noise approximation, numerical simulations [see \cref{sec_avgpath_nonMarkov}] show that the actual first-passage path of a fractional Brownian walker behaves as
\beq M-\bra h(\dt)\ket \propto \dt^{H}.\qquad \text{(fractional Brownian motion)}
\label{eq_FBMpath_avg}\eeq 
Note that, as in \cref{eq_BMpath_avg}, the average is performed here also over the first-passage time distribution.
\Cref{eq_FBMpath_WNT,eq_FBMpath_avg} are straightforward generalizations of the Markovian expressions in \cref{eq_BMpath_WNT,eq_BMpath_avg}.
We conclude that taking into account fluctuation-induced interactions with the absorbing boundary effectively leads to a reduction of the dynamic exponent characterizing the averaged path of a Brownian walker from the value $2H$ predicted by WNT to the value $H$ \footnote{Note that, in contrast to the Markovian case, for fBM the influence of the entropic repulsion effect and the random character of the first-passage time could not be separated here. This requires a new simulation method and is left for future work.}.

We now apply these insights to a fluctuating profile $h(x,t)$.
To this end, we recall that a tagged monomer $h(x_M,t)$ follows a Gaussian stochastic process characterized by the Hurst exponents
\beq H_0 =1/2\qquad \text{and} \qquad H = \frac{1}{2z},\qquad \text{(profile)}
\label{eq_profile_Hurst}\eeq 
which, \emph{inter alia}, determine the variance as [see \cref{eq_hr_correl}]
\beq  
\bra [\delta h(x_M,t)]^2\ket^{1/2} \sim 
\begin{cases}
t^{H_0},\qquad &t\lesssim \tau\cro\\
t^{H},\qquad &\tau\cro\lesssim t \lesssim \tau.
\end{cases}
\label{eq_profile_subdiff}\eeq 
For times $t\gtrsim \tau$, a tagged monomer experiences the ``self-generated'' effective potential of the mass-conserving profile, as reflected by the Gaussian equilibrium variance [see \cref{eq_var_Dir,eq_var_pbc}]. 

We first turn to equilibrium initial conditions, for which the stochastic process described by \cref{eq_profile_subdiff,eq_profile_Hurst} is actually a fractional Brownian motion [see \cref{eq_hr_fbm}].
In this case, \cref{eq_FBMpath_WNT} predicts, based on \cref{eq_profile_Hurst}, the values $\alpha\st{0,WNT}=2H_0=1$ and $\alpha\st{WNT}=2H=1/z$ for the dynamic exponents in \cref{eq_peakevol}, in agreement with the explicit WNT results in \cref{eq_peakevol_exp_WNT}. (Note that the weak-noise approximation here is insensitive to the presence of an impenetrable boundary.)
Beyond WNT, \cref{eq_FBMpath_avg} accordingly predicts 
\beq \alpha_0 =H_0=\onehalf,\qquad \alpha=H= \frac{1}{2z} = 
\begin{cases} 
1/4,\qquad \text{EW},\\ 
1/8,\qquad \text{MH}
\end{cases}
\qquad \text{(prediction)}
\label{eq_peakevol_exp_pred}\eeq 
for the dynamic exponents of a profile near a first-passage event.
These values are indeed close to the simulation results in \cref{eq_peakevol_exp}, especially in the case of the short-time exponent $\alpha_0$. Possible reasons for the discrepancy of the late-time exponent $\alpha$ are discussed below.

For non-equilibrium initial conditions, corresponding to the transient first-passage regime ($\Mred\ll 1$), the stochastic process underlying \cref{eq_profile_subdiff} is not a fractional Brownian motion [see \cref{eq_hr_rough}].
However, the above reasoning concerning the averaged profile essentially relies only on the Hurst characterization of the dynamics of a tagged monomer.
In particular, this process has the same subdiffusive scaling behavior in the equilibrium and the transient regime, suggesting \cref{eq_peakevol_exp_pred} to apply also in the latter.
Indeed, the values for $\alpha$ obtained from the simulations in the two regimes are practically identical.

\begin{figure}[t]\centering
	\subfigure[]{\includegraphics[width=0.45\linewidth]{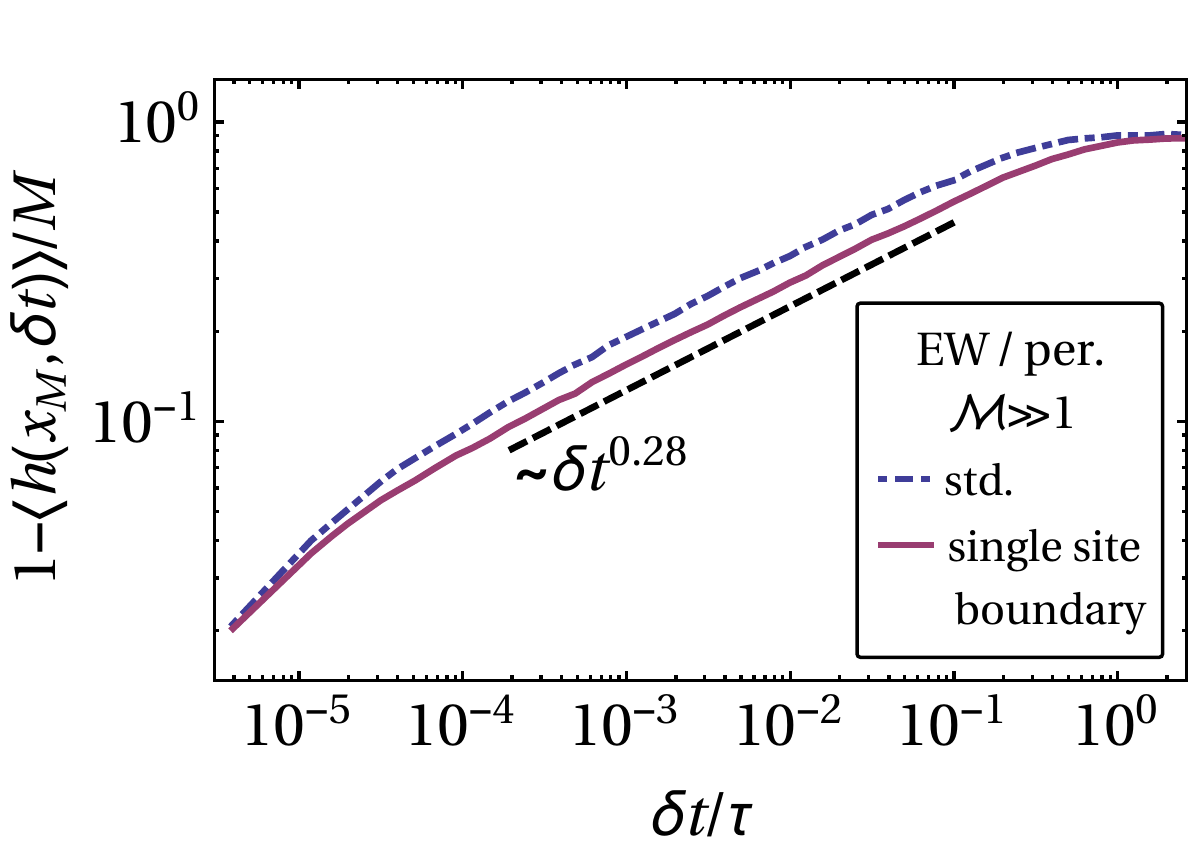}} \hfill
	\subfigure[]{\includegraphics[width=0.45\linewidth]{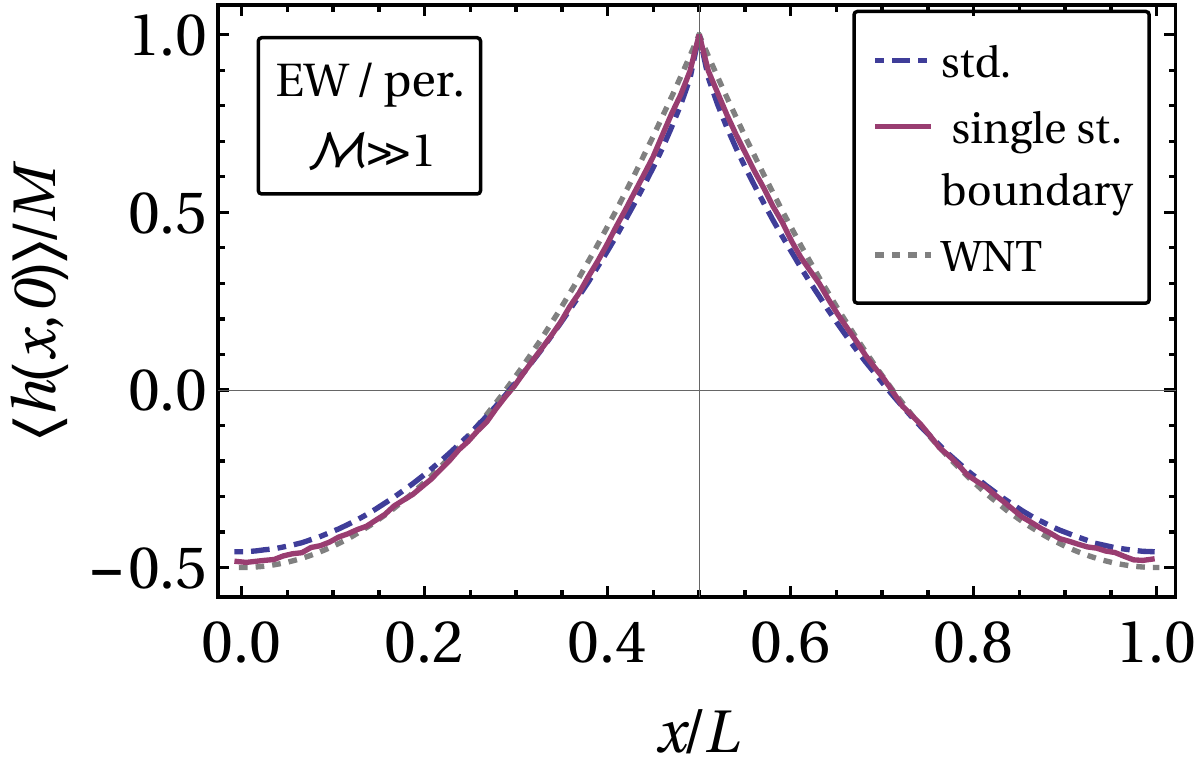}}
	\caption{Effect of the spatial extension of the absorbing boundary condition on the first-passage behavior of a profile governed by the EW equation with periodic \bcs. Panel (a) shows the averaged time evolution of the peak $\bra h(x_M,\dt)\ket$, while (b) shows the averaged profile at the first-passage event. The dash-dotted curves are obtained for an absorbing boundary acting on the whole profile [\cref{eq_firstpsg_cond}], which is the standard case in the present study. The solid curves, instead, correspond to an absorbing boundary acting only on the monomer at $x_M$ (``single site boundary''). The dotted curve in (b) represents the prediction of WNT [\cref{eq_opt2_finalprof_eq_pbc}].}
	\label{fig_boundeff}
\end{figure}

The prediction in \cref{eq_peakevol_exp_pred} is based on the equivalence of a fractional Brownian walker and a tagged monomer of an \emph{unconstrained} interface.  
However, for the first-passage dynamics considered here, the absorbing boundary condition at the height $M$ [\cref{eq_firstpsg_cond}] is essential.
This boundary condition constrains the profile as a whole and, owing to the long-range correlations of the profile, it can in principle lead to deviations in the behavior of a tagged monomer from the behavior expected for a single fractional Brownian walker.
To which extent this effect is responsible for the discrepancy between the values for $\alpha$ reported in \cref{eq_peakevol_exp} and the predictions in \cref{eq_peakevol_exp_pred} demands further studies.

Here it is possible to clarify at least the impact of the spatially extended nature of the absorbing boundary condition.
To this end, we perform simulations in which an absorbing boundary acts only on a monomer at a \emph{single} location $x_M$. 
\Cref{fig_boundeff}(a) shows $\bra h(x_M,\dt)\ket$ as a function of time obtained in this case for the EW equation with periodic \bcs (solid curve). 
One observes that $\bra h(x_M,\dt)\ket$ still follows the algebraic behavior in \cref{eq_peakevol}, with a value of $\alpha$ that is essentially identical to the one obtained for an absorbing boundary acting on all monomers [dash-dotted curve; see also \cref{fig_peakevol_nc_pbc_eq}].
As \cref{fig_boundeff}(b) shows, also the averaged profile \emph{at} the first-passage event is not significantly affected by the spatially extended character of the absorbing boundary condition. 
This insensitivity can be attributed to the rather sharply peaked shape of the first-passage profile, which is already predicted by WNT [cf.\ \cref{fig_avgprof_nc_pbc_eq}(b)]. 
Overall, the results in \cref{fig_boundeff} suggest that the spatial extension of the absorbing boundary has a negligible influence on the behavior of the averaged profile. 

We finally remark that, in principle, also insufficiently large values of the system size $L$ or of the reduced height $\Mred$ can contribute to the deviations between the observed dynamic exponent and the prediction of the fBM model. 
In fact, the crossover to the short-time diffusive regime in \cref{eq_profile_subdiff} happens earlier for smaller systems, which can result in an artificially large effective value of $\alpha$ (see, e.g., \cref{fig_peakevol_nc_pbc_eq}).
A similar effect can also be observed in the case of roughening [see, in particular, \cref{fig_bench_roughen}(c)].
However, for the largest values of $L$ used here, we have not observed a significant $L$-dependence of the effective dynamic exponent. This indicates that the residual finite-size corrections to the values in \cref{eq_peakevol_exp} are rather small (see, e.g., \cref{fig_peakevol_nc_pbc_eq}). 
Note furthermore that, within the applicability of its underlying approximations, WNT is expected to become exact in the two limits $\Mred\ll 1$ and $\Mred\gg 1$ \cite{gross_first-passage_2017}.
Indeed, the spatial profile shapes are accurately captured by WNT in these limits.
However, since WNT disregards by construction some fundamental aspects of the first-passage process (see the above discussion), we expect no convergence of the values of $\alpha$ to the predictions of WNT.

\section{Summary}
In the present study, the first-passage dynamics of an interfacial profile governed by the EW or MH equation [\cref{eq_EW,eq_MH}] has been analyzed based on numerical solutions.
We have considered here periodic as well as Dirichlet \bcs. In the case of the MH equation, the latter are imposed in conjunction with a no-flux condition in order to ensure conservation of the mass [\cref{eq_mass}].
For the EW equation with periodic \bcs, mass conservation is explicitly imposed during the time evolution via the rule in \cref{eq_height_redef}.
The first-passage event is defined as the instant at which the profile reaches a given height $M>0$ for the first time.
Accordingly, an absorbing boundary condition acts at the height $M$ [\cref{eq_firstpsg_cond}].

The obtained results are compared here to weak-noise theory (WNT) as well as to effective Brownian walker models describing the anomalous diffusion of a tagged ``monomer'' of the profile.
WNT can be considered as a saddle-point approximation to the first-passage problem and thus neglects the entropic repulsion effect of the impenetrable boundary and the random character of the first-passage time.
The present study elucidates the accuracy of WNT for the description of the noise-activated dynamics of a spatially extended, finite and highly correlated stochastic system.

We find that the \emph{shape} of the averaged profile $\bra h(x,\dt)\ket$ is in general well described by WNT. In particular, the dynamic scaling behavior predicted by WNT is qualitatively recovered in the simulations. 
In the transient regime (corresponding to small reduced heights, $\Mred\ll 1$ [see \cref{eq_hm_red}]), the averaged profile is sharply peaked and independent from the \bcs.
In the equilibrium regime (corresponding to $\Mred\gg 1$), the profile is insensitive to the \bcs only in an inner region, where a dynamic scaling behavior applies.
The associated scaling function and scaling exponents are universal.
Consistent with WNT, the roughening time $\tau$ [see \cref{eq_relaxtime}] sets the characteristic time scale for the creation of the first-passage fluctuation. 

A significant difference between WNT and the fully stochastic model [\cref{eq_EW,eq_MH}] concerns the dynamic exponent $\alpha$, which characterizes the approach of the profile towards first-passage event at the height $M$ via $M-\bra h(x,\dt)\ket\propto \dt^\alpha$.
Here, instead of the value $\alpha=1/z$ predicted by WNT [see \cref{eq_peakevol_exp_WNT}], a value close to $1/(2z)$ is found in the simulations [see \cref{eq_peakevol_exp,eq_peakevol_exp_pred}], with $z=2$ for the EW and $z=4$ for the MH equation. 
This ``renormalization'' of the dynamic exponent can be understood based on the equivalence between a tagged monomer in equilibrium and a fractional Brownian walker with Hurst index $H=1/(2z)$.
For the walker, it is shown here analytically and via dedicated numerical simulations, that the dynamic exponent $n$ describing the averaged trajectory near an absorbing boundary at height $M$, $M-\bra h(\dt)\ket\propto \dt^n$, changes from $n=2H$ within WNT to $n=H$ when fluctuation-induced (entropic) interactions between the walker and the boundary are taken into account.
Accordingly, the renormalization of the profile exponent $\alpha$ can be attributed to the  fluctuations of the profile around its most-likely path as it approaches the first-passage event (see discussion in \cref{sec_discussion}). 
We remark that our numerical solutions yield a value for $\alpha$ slightly larger [see \cref{eq_peakevol_exp}] than the prediction $\alpha=1/(2z)$ [\cref{eq_peakevol_exp_pred}], which might be related to the fact the mapping between a tagged monomer and a Brownian walker is formally obtained in the absence of an absorbing boundary. 
This aspect deserves further studies.

The inadequacy of WNT to capture the exact time-dependence of the first-passage dynamics becomes particularly clear for standard Brownian motion, in which case the problem can be solved exactly (see \cref{sec_Brownian_firstpsg}). 
A Brownian path with fixed endpoints is sensitive to the presence of the absorbing boundary only close to it [see \cref{eq_avgpathBM_shortt}].
In the weak-noise limit, the effect of the absorbing boundary diminishes, such that the averaged path reduces to the classical one [see \cref{eq_avgpathBM_longt}]. 
Upon averaging over the first-passage distribution, the influence of the boundary effectively ``spreads'' over the whole path [see \cref{eq_avgpath_Brownian_asympt,eq_avgpath_Brownian_asympt_far}].
However, in the absence of noise, the first-passage distribution trivially vanishes, as does the first-passage path [see \cref{eq_avgpath_F1_res}]. 
For future studies it would be interesting to improve WNT by taking into account the distribution of first-passage times and to include the fluctuations around the most-likely path in the presence of an impenetrable boundary. 
This would allow one to rigorously assess the various approximations involved in WNT.

As a by-product of our simulations, we have obtained the mean first-passage time $\bra T\ket$. 
In the equilibrium regime, $\bra T\ket$ is found to grow exponentially with the square of the reduced height $\Mred^2$ [\cref{eq_hm_red}]. This  reflects the self-generated harmonic potential in which a tagged monomer of an equilibrated profile moves.
In the transient regime, instead, we find an algebraic dependence of $\bra T\ket$ on the actual height $M$, which reflects the sub-diffusive motion of a tagged monomer.
It turns out that mass conservation [\cref{eq_zero_vol}] as well as the extended nature of the absorbing boundary [\cref{eq_firstpsg_cond}] can significantly affect the first-passage distribution [see \cref{app_fp_distr}].

\acknowledgements{The author thanks G.\ Oshanin for useful discussions.}

\appendix

\section{First-passage time distribution}
\label{app_fp_distr}

\begin{figure}[t]\centering
      \subfigure[]{\includegraphics[width=0.45\linewidth]{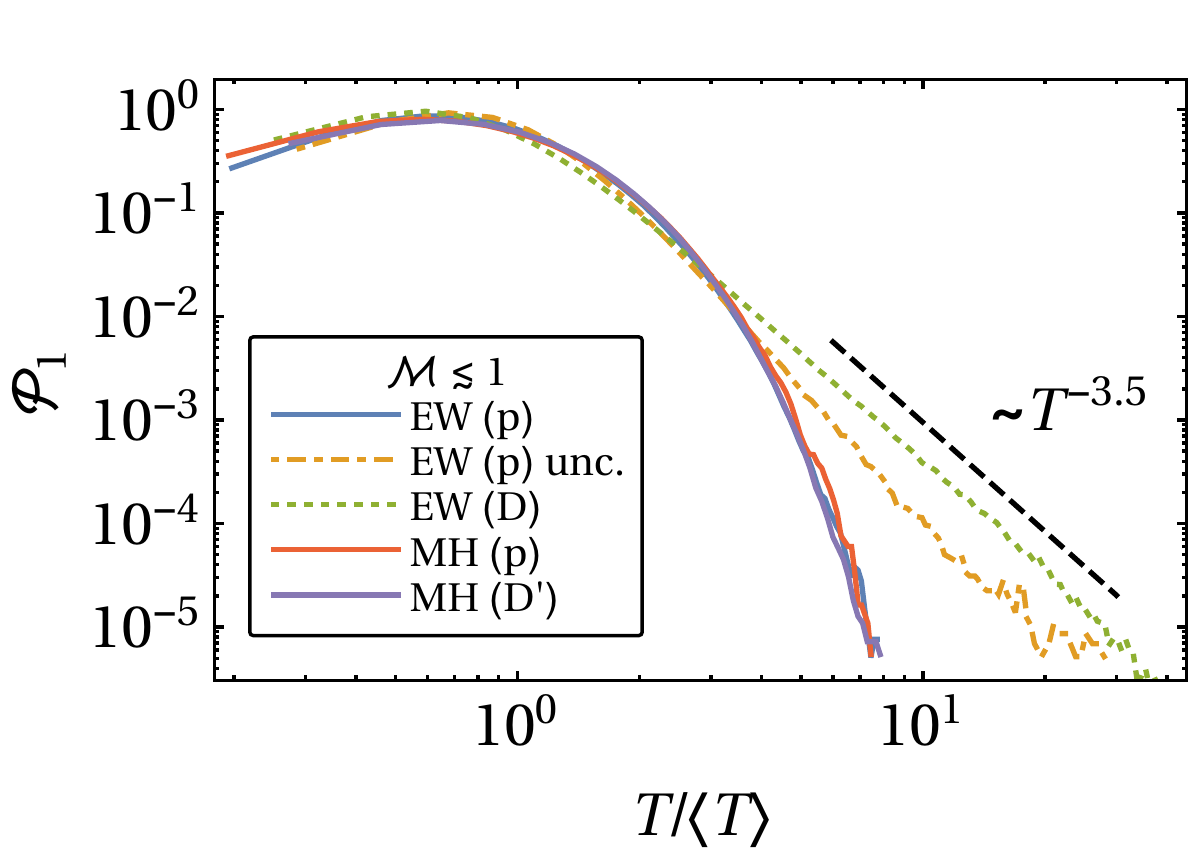} \label{fig_fpdf_tr}} \qquad 
      \subfigure[]{\includegraphics[width=0.45\linewidth]{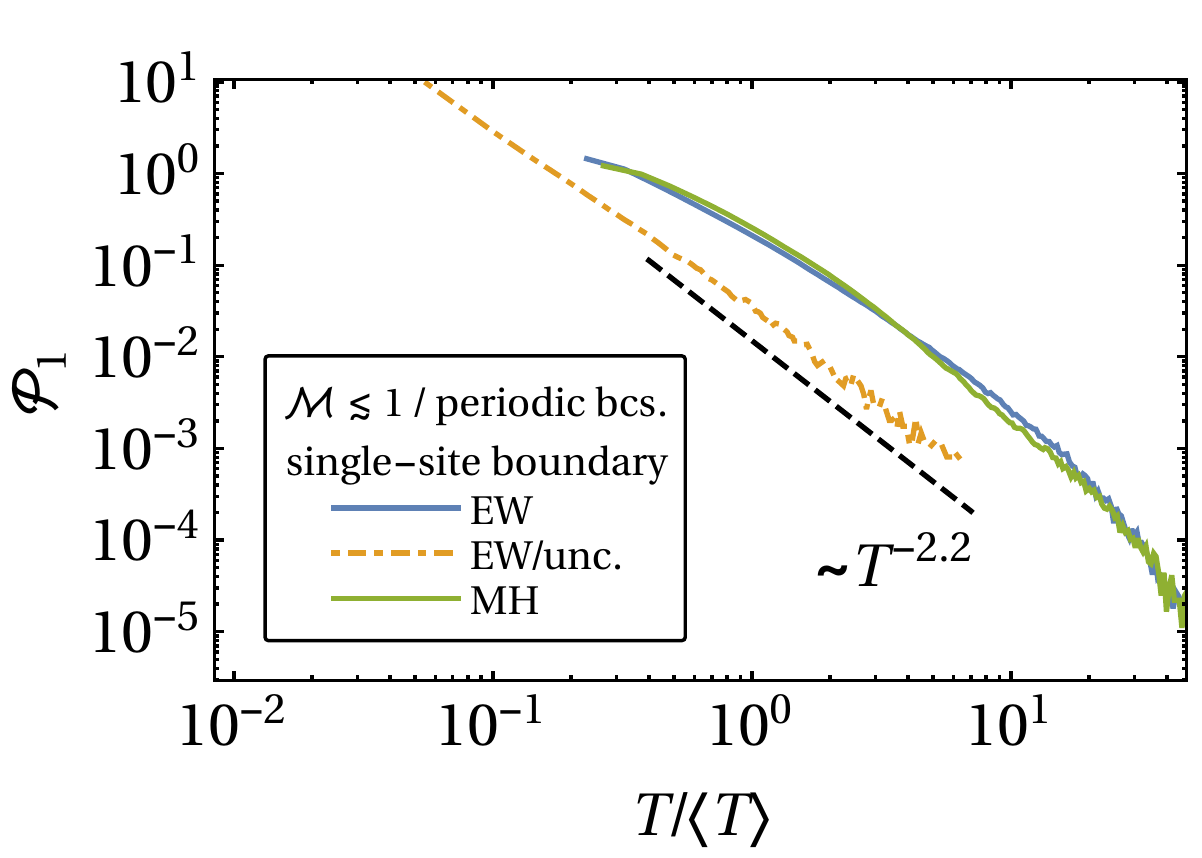} \label{fig_fpdf_singlebnd}}
      \subfigure[]{\includegraphics[width=0.45\linewidth]{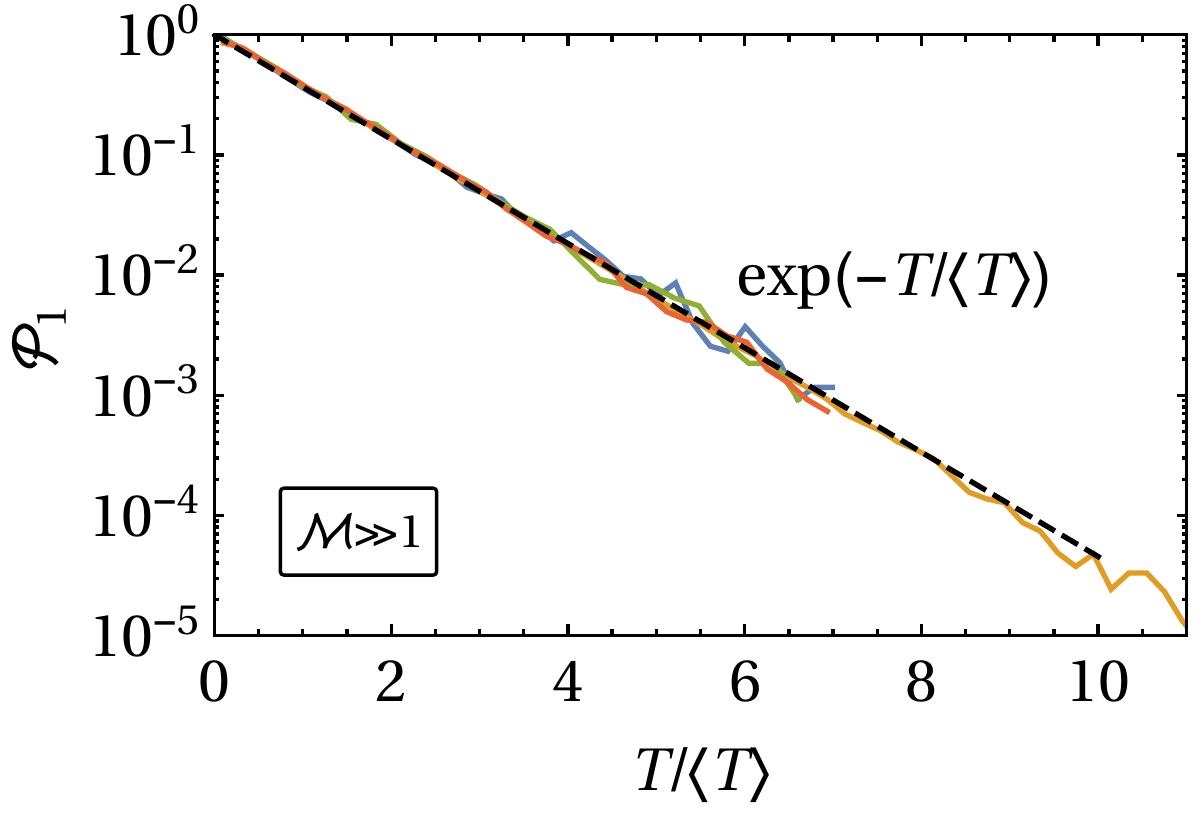} \label{fig_fpdf_eq}}
      \caption{Probability distribution of the first-passage time $T$ (normalized by its mean value $\bra T\ket$, see \cref{eq_meanfp_time}) in (a,b) the transient regime and (c) the equilibrium regime. In (a), the curve labeled by `unc.' correspond to EW dynamics in the absence of condition \eqref{eq_height_redef}, i.e., without the mass constraint [\cref{eq_zero_vol}]. In (b), the absorbing boundary condition acts only on a single monomer (at location $x_M$) instead of the whole profile. The dash-dotted curve corresponds to EW dynamics without the mass constraint [i.e., \cref{eq_height_redef} is not imposed]. In (c) the different curves, which all superimpose onto the function $\Pcal_1 \simeq \bra T\ket\exp(-T/\bra T\ket)$, correspond to EW and MH dynamics with periodic and Dirichlet (no-flux) \bcs.}
    \label{fig_fpdf}
\end{figure}

The distribution $\Pcal_1(T)$ of the first-passage time to the height $M$ obtained in the \emph{transient} regime is illustrated in \cref{fig_fpdf_tr}. Note that $T$ is normalized here by the mean first-passage time $\bra T\ket$, which is discussed separately in \cref{sec_firstpsg}.
We find that $\Pcal_1$ generally exhibits a well-defined maximum for $T\simeq \bra T\ket$.
In the case of MH dynamics, which conserves mass [see \cref{eq_zero_vol}], $\Pcal_1$ decays exponentially.
This is also found in the case of EW dynamics with periodic \bcs, in which case mass conservation is explicitly enforced via \cref{eq_height_redef}.
In contrast, if \cref{eq_height_redef} is not imposed [curve in \cref{fig_fpdf_tr} labeled by `unc.'], the first-passage distribution decays algebraically for large $T$, $\Pcal_1\sim T^{-n}$, with $n\simeq 3.5$ \footnote{Note that the considered profiles do not yet exhibit center-of-mass diffusion [see \cref{eqR_cm_diffusion}], in which case a behavior $\Pcal_1\propto T^{-3/2}$ is expected \cite{redner_guide_2001, kantor_anomalous_2007, amitai_first-passage_2010}.}.
A similar algebraic decay is also observed in the case of EW dynamics with Dirichlet \bcs, where mass is conserved only as a time average.

The behavior of $\Pcal_1$ is also sensitive to the spatially extended character of the absorbing boundary condition [see \cref{eq_firstpsg_cond}].
This is illustrated in \cref{fig_fpdf_singlebnd}, which shows $\Pcal_1$ obtained in the transient regime for an absorbing boundary acting only on the monomer at $x_M$.
Compared to \cref{fig_fpdf_tr}, $\Pcal_1$ decays here slower for large $T$, although still approximately exponentially.
Lifting, in the case of EW dynamics, additionally the mass constraint results in an algebraic decay, $P_1\sim T^{-n}$ with $n\simeq 2.2$.
This value of $n$ is smaller than the one obtained in the case of a spatially extended absorbing boundary [see \cref{fig_fpdf}(a)]. It is, however, close to the prediction $n\simeq 2.5$ given in Ref.\ \cite{krug_persistence_1997}, where the transient persistence probability of an interface has been investigated. 

In the \emph{equilibrium} regime [see \cref{fig_fpdf}(b)], both for the EW and MH equation as well as for all considered \bcs, we empirically find that the first-passage distribution is a simple exponential function of $T/\bra T\ket$:
\beq \Pcal_1(T) \simeq \bra T \ket \exp(-T/\bra T\ket).
\label{eq_P1_eqregime}\eeq 
The exponential behavior is in fact characteristic for a fractional Brownian walker in a parabolic potential \cite{sliusarenko_Kramers-like_2010} and found to persist also if the absorbing boundary condition acts only on a single monomer (data not shown). 
Removing the mass constraint in the equilibrium regime results in a simple diffusive motion of the center-of-mass of the profile, which then dominates the first-passage distribution.

\section{Equilibrium distribution of height fluctuations}
\label{app_equil_pdf}

\subsection{Periodic \bcs}

The friction and noise parameters $\frict$ and $D$ in \cref{eq_EW,eq_MH} can be determined by requiring that the ensuing steady-state probability distribution of the profile $h(x)$ is characterized by a certain temperature $\kbT$.
For periodic \bcs, \cref{eq_EW,eq_MH} yield in the steady-state a Gaussian joint-probability distribution of the form \cite{majumdar_airy_2005, majumdar_spatial_2006}
\beq P\eq\pbc[h] \sim \exp\left[ -\frac{1}{4 \kbT}\int_0^L \d x \left(\frac{\d h}{\d x}\right)^2\right]\, \delta[h(0)-h(L)]\,\delta\left[\int_0^L \d x h(x)\right],
\label{eq_Pss}
\eeq
with the temperature [see \cref{eq_FDT}] 
\beq \kbT \equiv \frac{D}{2 \frict}
\label{eq_FDT2}
\eeq
in units of $k_B$.
In \cref{eq_Pss}, the $\delta$-functions enforce the periodic \bcs and the zero-mass constraint [\cref{eq_zero_vol}].
The stationary single-site height distribution resulting from \cref{eq_Pss} is given by \cite{majumdar_airy_2005, majumdar_spatial_2006,foltin_width_1994}
\beq P\eq\pbc(h) = \sqrt{\frac{3}{\pi \kbT L}} \exp\left(-\frac{3}{\kbT L} h^2\right)\,,
\label{eq_Pss_single}\eeq  
implying the variance [see also \cref{eqR_eqvar_per}]
\beq \bra h^2\ket = \frac{\kbT L}{6}.
\label{eq_Pss_var}\eeq 
According to \cref{eq_Pss}, a profile $h(x)$ in equilibrium can be considered as a Brownian motion process for which $x$ plays the role of time.
Since the motion is required to start and end here at the same point, $h(0)=h(L)$, the process is in fact a \emph{Brownian bridge}, with the additional constraint of having zero area under it \cite{majumdar_effective_2015,mazzolo_constrained_2017}.
\Cref{eq_FDT2} is taken as a definition of the temperature throughout the present study, despite the fact that, for non-periodic \bcs, the resulting steady-state variance is different from \cref{eq_Pss_var}.

\subsection{Dirichlet \bcs}

The steady-state distribution for Dirichlet \bcs is given by the same expression as in \cref{eq_Pss}, except that $\delta[h(0)-h(L)]$ is replaced by $\delta[h(0)]\delta[h(L)]$ and that the mass constraint is present only for Dirichlet no-flux \bcs [see \cref{eq_H_Dbc,eq_H_noflux}].
Correlation functions can be readily determined with the aid of the closely-related propagator for a Brownian particle with fixed endpoints \cite{chaichian_path_2001,majumdar_brownian_2005}:
\beq G(h,x|h_0,x_0) = \int_{h(0)=h_0}^{h(x)=h} \mathcal{D}h(\xi)\, \exp\left[ -\frac{1}{4\kbT} \int_{x_0}^x \d\xi \left(\frac{\d h(\xi)}{\d\xi}\right)^2 \right] = \frac{1}{\sqrt{4\pi \kbT(x-x_0)}} \exp\left[-\frac{(h-h_0)^2}{4\kbT(x-x_0)}\right].
\label{eqBM_prop}\eeq
If, in addition to the endpoints also the area under the profile is constrained, corresponding to Dirichlet no-flux \bcs, the propagator is instead given by \cite{majumdar_airy_2005, burkhardt_semiflexible_1993}
\begin{multline} G(h,x,A| h_0, x_0, A_0) =  \int_{h(0)=h_0}^{h(x)=h} \mathcal{D}h(\xi)\, \delta\left( \int_{x_0}^x \d\xi h(\xi) -A\right)\, \exp\left[ -\frac{1}{4\kbT} \int_{x_0}^x \d\xi \left(\frac{\d h(\xi)}{\d\xi}\right)^2 \right]\\
	= \frac{\sqrt{3}}{2\pi \kbT (x-x_0)^2} \exp\left[-\frac{1}{\kbT}\left(\frac{3}{(x-x_0)^3}\left\{A-A_0-(x-x_0)h\right\}\left\{A-A_0-(x-x_0)h_0\right\} + \frac{1}{x-x_0}(h-h_0)^2\right)\right].
\label{eBM_prop_areaconstr} \end{multline}
$G(h,A,x|h_0,x_0,A_0)$ represents the joint probability to observe a Brownian particle at location $(h,x)$, having covered the area $A=A_0+\int_{x_0}^x \d x\, h(x)$, given that the particle previously was at the location $(h_0,x_0)$ and had covered the area $A_0$.
In the case of standard Dirichlet \bcs, the equilibrium variance of a fluctuating profile is given by
\beq\begin{split} 
\bra h^2(x) \ket = \frac{\int_{-\infty}^\infty \d h'\, G(h', x|0, 0)\, h'^2 \, G(0, L| h', x )}{G(0, L | 0, 0)} 
= 2\kbT L \frac{x}{L}\left(1 - \frac{x}{L}\right),
\end{split}\label{eqBM_var_bridge}\eeq
while for the averaged path, $\bra h(x)\ket=0$.
\Cref{eqBM_var_bridge} also represents the variance of a Brownian bridge (see, e.g., Refs.\ \cite{mazzolo_constrained_2017,szavits-nossan_inequivalence_2015}).
For a Dirichlet profile whose area is constrained to vanish, the averaged path results instead as
\beq 
\bra h(x) \ket_{A} = \frac{\int_{-\infty}^\infty \d A' \int_{-\infty}^\infty \d h'\, G(h',x, A'|0, 0, 0)\, h' \, G(0, L, 0| h', x, A')}{G(0, L, 0 | 0, 0, 0)} 
= 6 \frac{A}{L} \frac{x}{L}\left(1-\frac{x}{L}\right),
\label{eqBM_avgpath_areaconstr}\eeq
while its variance is given by (we consider here only $A=0$, such that $\bra h(x)\ket_{A=0}=0$)
\beq\begin{split} 
\bra h^2(x) \ket_{A=0} = \frac{\int_{-\infty}^\infty \d A' \int_{-\infty}^\infty \d h'\, G(h',x, A'|0, 0, 0)\, h'^2 \, G(0, L, 0| h', x, A')}{G(0, L, 0 | 0, 0, 0)} 
= 2\kbT L \frac{x}{L} \left(1-\frac{x}{L}\right) \left(1+ 3\frac{x}{L} \left(\frac{x}{L}-1\right)\right).
\end{split}\label{eqBM_var_areaconstr}\eeq
The above results rely on the Markovian nature of the respective stochastic process.
In particular, the normalization in \cref{eqBM_avgpath_areaconstr,eqBM_var_areaconstr} follows from the Markovian nature of the joint stochastic process $(h,A)$, i.e., $G(h,x,A| h_0, x_0, A_0) = \int_{-\infty}^\infty \d A' \int_{-\infty}^\infty \d h'\, G(h,x,A|h',x',A') G(h',x',A'|h_0,x_0,A_0) $ for any $x_0<x'<x$.

\section{Averaged path for a single Brownian walker}
\label{app_walker}

\subsection{Averaged path with constrained endpoints}
\label{app_avgpath_BM}

\begin{figure}[t]\centering
	\subfigure[]{\includegraphics[width=0.41\linewidth]{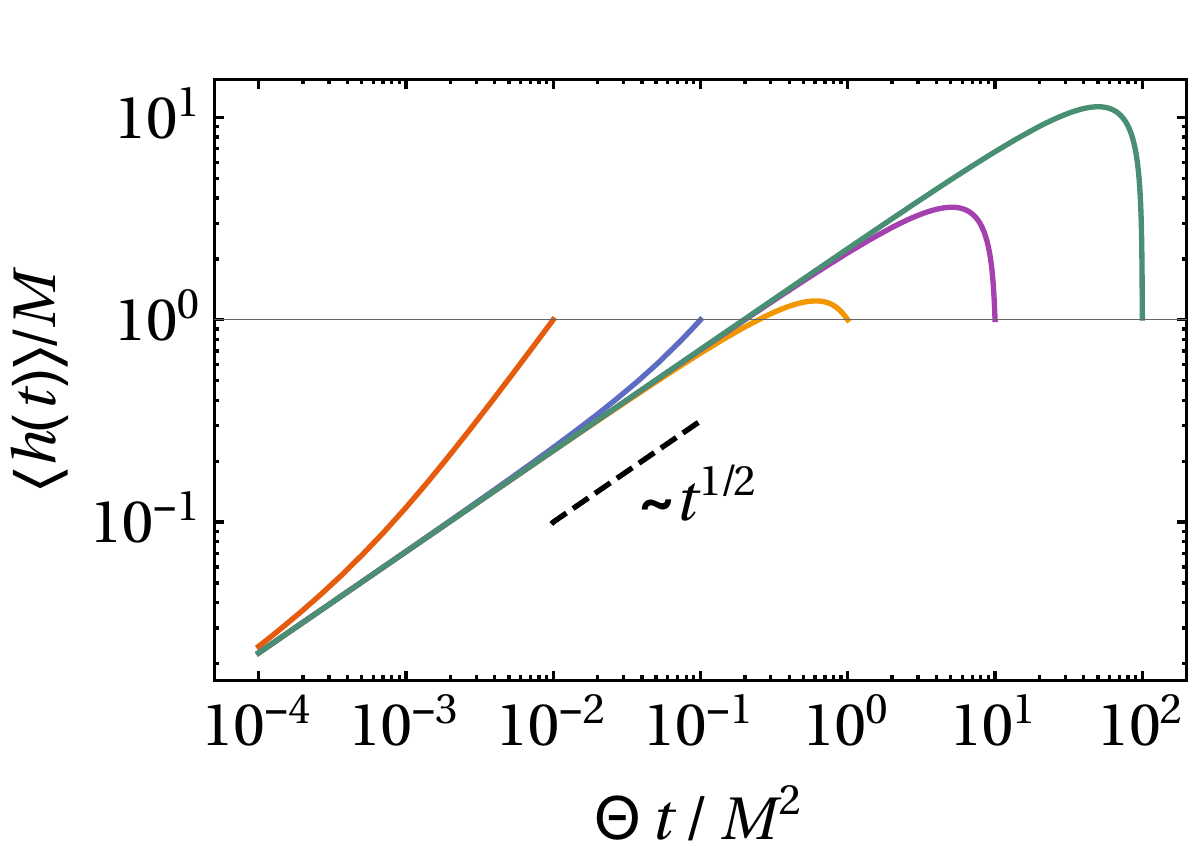}}\qquad 
	\subfigure[]{\includegraphics[width=0.42\linewidth]{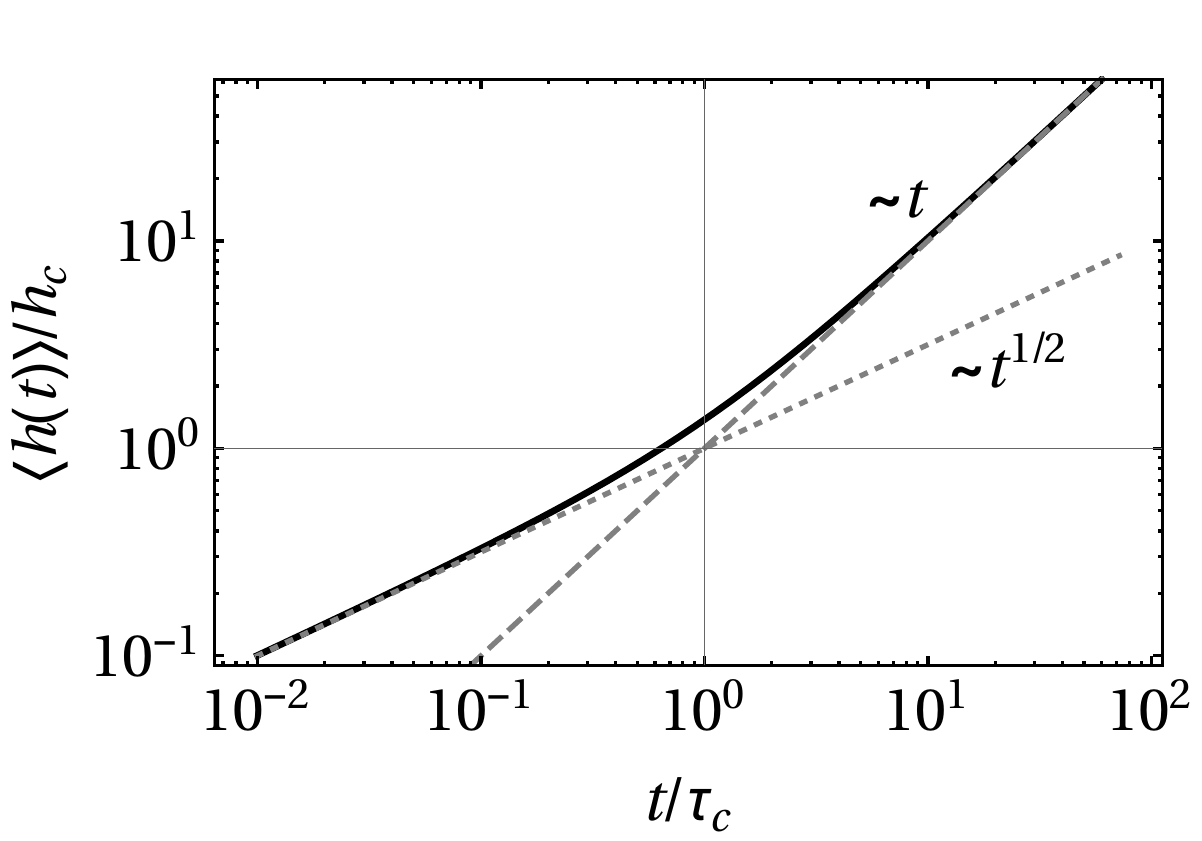}}
	\caption{Averaged path $\bra h(t)\ket_{M,T}$ [\cref{eq_BM_avgpath}] of a Brownian walker starting at $(h,t)=(\epsilon,0)$ and ending at $(M>0,T>0)$ in the presence of an absorbing boundary at $h=0$. Time is made dimensionless by rescaling with $\kbT/M^2$. (a) Dependence of $\bra h(t)\ket_{M,T}$ on the final time $T$ for a fixed final height $M$ (for the central endpoints from left to right: $\kbT T/M^2=10^{-2}, 10^{-1},1,10,100$). (b) Time dependence of $\bra h(t)\ket_{M,T}$ in the regime $M^2\gg \kbT T$ [see \cref{eq_avgpathBM_bow} and the associated discussion]. The dotted and dashed lines represent the asymptotic behaviors in \cref{eq_avgpathBM_shortt,eq_avgpathBM_longt}. The characteristic length and time scales $\tau_c$ and $h_c$ are reported in \cref{eq_avgpathBM_crossover,eq_avgpathBM_charlen}, respectively.}	
	\label{fig_avgpath_Brownian}
\end{figure}

We place an absorbing boundary at height $h=0$ and consider a (Markovian) Brownian walker that departs from $(h,t)=(\epsilon,0)$ to some distant position $(M,T)$. 
The infinitesimal quantity $\epsilon$ is required as a regularization and the limit $\epsilon\to 0$ will be performed at the end of the calculation \cite{redner_guide_2001}.
Owing to the Markovian property of the process, the averaged trajectory of the walker can be expressed as (see also Refs.\ \cite{baldassarri_average_2003, colaiori_average_2004,bhat_intermediate-level_2015})
\beq \bra h(t)\ket_{(0,T)\to (M,T)} = \lim_{\epsilon\to 0}\frac{\int_0^\infty \d h \,  G_+(M,T|h,t)\, h\, G_+(h,t|\epsilon,0)}{\int_0^\infty \d h  G_+(M,T|h,t)G_+(h,t|\epsilon,0)} =  \lim_{\epsilon\to 0}\frac{\int_0^\infty \d h G_+(M,T|h,t) \, h\,G_+(h,t|\epsilon,0)}{ G_+(M,T|\epsilon,0) }.
\label{eq_avg_path_abs}\eeq 
The propagator $G_+(h,t|h_0,t_0)$ represents the conditional probability for the walker to move from $(h_0,t_0)$ to $(h,t)$ without $h$ becoming negative and is given by the well-known expression
\beq G_+(h,t|h_0,t_0) = \frac{1}{\sqrt{4\pi \kbT (t-t_0)}}\left[ \exp\left({-\frac{(h-h_0)^2}{4\kbT (t-t_0)}}\right) - \exp\left({-\frac{(h+h_0)^2}{4\kbT (t-t_0)}}\right) \right],
\label{eq_BM_prop_abs}\eeq
which follows, e.g., by applying the image method to the propagator in \cref{eqBM_prop} (replacing $x\to t$) \cite{redner_guide_2001}.
Equation \eqref{eq_avg_path_abs} can be evaluated analytically, yielding 
\beq \begin{split} 
\bra h(t)\ket_{M,T} &= \frac{2}{M \sqrt{\pi}} \sqrt{\kbT  t \left(1-\frac{t}{T}\right)} \exp\left(\frac{M^2 t}{4 \kbT  T(t -T)}\right) + \left( \frac{M t}{T} + \frac{2d}{M} (T-t)\right)\mathrm{erf}\left[\frac{M t}{2\sqrt{d t T(T-t)}}\right] \\
&= M \left\{ \frac{2}{\sqrt{\pi}U^2} \sqrt{U^2-V^2} \exp\left(\frac{V^4}{4(V^2-U^2)}\right)  + \left(V^2 + \frac{2}{V^2} - \frac{2}{U^2}\right) \mathrm{erf}\left(\frac{V^2}{2\sqrt{U^2-V^2}}\right) \right\},
\end{split}\label{eq_BM_avgpath}\eeq
where, in the last equation, the dimensionless scaling variables $U\equiv M/\sqrt{\kbT  t}$, $V\equiv M/\sqrt{\kbT  T}$ have been introduced.
The behavior of the averaged path is illustrated in \cref{fig_avgpath_Brownian}(a) as a function of $\kbT t/M^2 = 1/U^2$.
For small times $t$, one asymptotically has 
\beq \bra h(t\to 0)\ket_{M,T} \simeq 4 \sqrt{\frac{\kbT }{\pi}t}  + \Ocal(t^{3/2}).
\label{eq_avgpathBM_shortt}\eeq 
At late times ($t\simeq T$), the behavior of the averaged path depends on the value of $T$ and $M$. The associated control parameter can be determined by noting that, for $U\sim \Ocal(V)$ (with $U>V$), the first term in the curly brackets in \cref{eq_BM_avgpath} is small, while the error function in \cref{eq_BM_avgpath} is approximately equal to one. Accordingly, values $\bra h\ket_{M,T}/M\gg 1$ are possible if $V^2\lesssim 1$, i.e., the averaged path develops a ``bow'' as seen in \cref{fig_avgpath_Brownian}(a) if
\beq \frac{M^2}{\kbT  T} \lesssim 1.
\label{eq_avgpathBM_bow}\eeq 
If, on the other hand, $M^2/\kbT T\gtrsim 1$, the averaged path behaves linearly for $t\simeq T$:
\beq \bra h(t\to T)\ket_{M,T} \simeq M\frac{t}{T} .
\label{eq_avgpathBM_longt}\eeq 
As shown in \cref{sec_fbm_MPL}, this expression, being independent of the noise $\kbT$, is simply the most-likely path of the walker [see \cref{eq_MLP_Markov}].
The cross-over time $\tau_c$ between the two regimes can be defined as the time where the two asymptotic laws in \cref{eq_avgpathBM_shortt,eq_avgpathBM_longt} are equal, yielding 
\beq \tau_c\simeq \frac{16 \kbT  T^2}{M^2\pi}.
\label{eq_avgpathBM_crossover}\eeq 
The two asymptotic laws can only be distinguished as long as $\tau_c<T$, which gives an estimate consistent with \cref{eq_avgpathBM_bow}.
Inserting \cref{eq_avgpathBM_crossover} into \cref{eq_avgpathBM_longt} yields the length scale 
\beq h_c\simeq \frac{16 \kbT  T}{\pi M},
\label{eq_avgpathBM_charlen}\eeq
which characterizes the range of influence of the absorbing boundary. 
As a reflection of the scale-free nature of the Brownian process, this length depends on coordinates ($T$ and $M$) arbitrarily far away from the boundary.
The averaged path given in \cref{eq_BM_avgpath} is illustrated in \cref{fig_avgpath_Brownian}(b) in the limit $M^2/\kbT T\gg 1$. 

In passing, we remark that the averaged trajectory $\bra h(t)\ket$ of a \emph{free} Brownian walker (i.e., in the absence of an absorbing boundary) between two points is a straight line, \beq \bra h(t)\ket \propto t.
\label{eq_avgpath_Brownian_free}\eeq 
This result follows from \cref{eq_avg_path_abs} by replacing therein $G_+$ by the standard diffusion propagator $G$ given in \cref{eqBM_prop}.
For free Brownian motion, the averaged path [\cref{eq_avgpath_Brownian_free}] coincides with the ``classical'' (most-likely) path which follows from the minimization of the corresponding action [see \cref{eq_MLP_Markov} below].

\subsection{First-passage path}
\label{sec_Brownian_firstpsg}

\begin{figure}[t]\centering
	\subfigure[]{\includegraphics[width=0.35\linewidth]{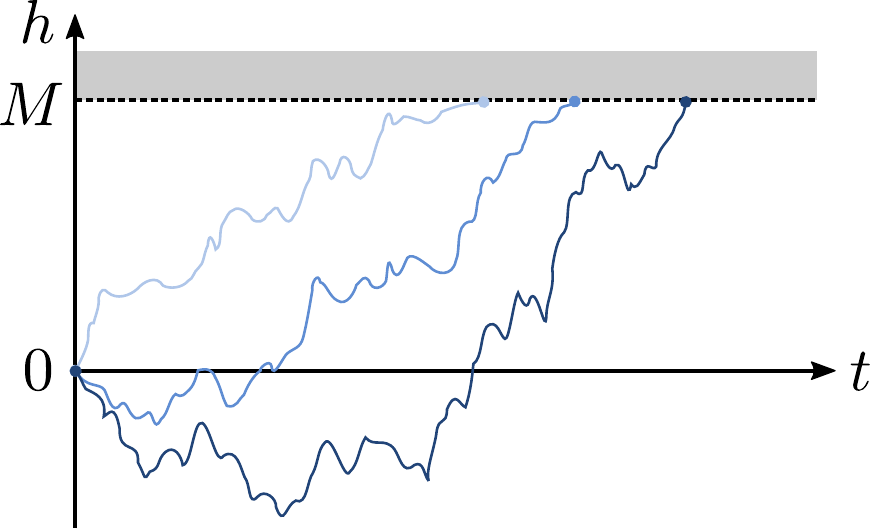}}\qquad\qquad
	\subfigure[]{\includegraphics[width=0.35\linewidth]{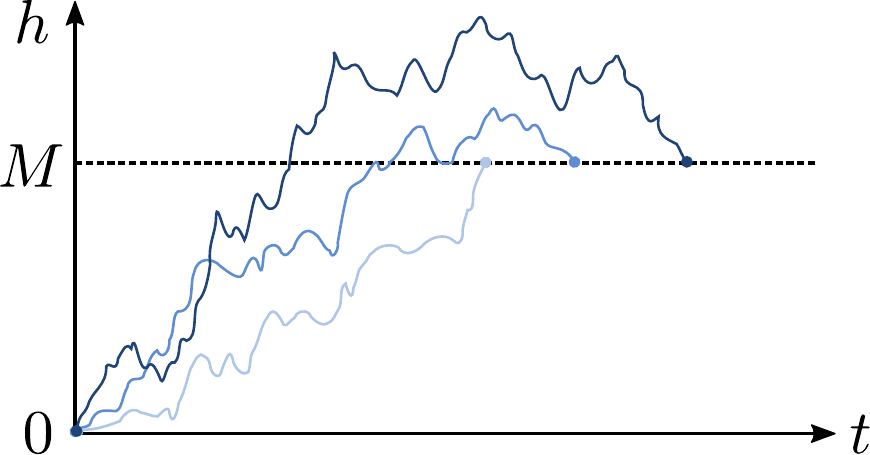}}
	\caption{Transformation of the sample paths of random walkers as considered in \cref{sec_Brownian_firstpsg}. (a) In the actual setup, a random walker $h(t)$ starts at the origin, $h(t=0)=0$, and moves until it hits a boundary at height $M>0$ (shaded bar) for the first time. (b) According to \cref{eq_avgpath_F1}, the paths are transformed such that the first-passage event occurs at the space-time origin.
	In the Markovian case, the first-passage path can be directly obtained by placing an absorbing boundary at $h=0$ and considering paths that start at $h(t=0)=0$ and are conditioned to end at $h(T)=M$, with a random time $T$ governed by $F_1(M,T)$.}
	\label{fig_walker_paths}
\end{figure}

Consider a Brownian (but not necessarily Markovian) walker starting at $(h,t)=(0,0)$ in the presence of an absorbing boundary at $h=M>0$. Let $F_1(M,T)$ be the corresponding probability distribution of the first-passage time $T$ to the height $M$ (see below) and 
$\bra h(t)\ket_{M,T}$ be the averaged path between the (fixed) points $(0,0)$ and $(M,T)$. 
We then define the averaged ``first-passage path'' of the walker, i.e., its averaged path near the first-passage event (see also Ref.\ \cite{bhat_intermediate-level_2015}), by 
\beq \bra h(t)\ket = M-\frac{\int_t^\infty \d T\, F_1(M,T) \bra h(T-t)\ket_{M,T}}{\int_t^\infty \d T\, F_1(M,T)}.
\label{eq_avgpath_F1}\eeq 
The associated transformation of the sample paths is illustrated in \cref{fig_walker_paths}.
Exact analytical expressions for $F_1$ and $\bra h(t)\ket_{M,T}$ are available only for Markovian Brownian walkers [see \cref{eq_BM_avgpath,eq_Brownian_F1pdf}]. In the non-Markovian case, we shall therefore resort to numerical calculations.

\subsubsection{Markovian case}
\label{sec_avgpath_Markov}

\begin{figure}[t]\centering
	\subfigure[]{\includegraphics[width=0.4\linewidth]{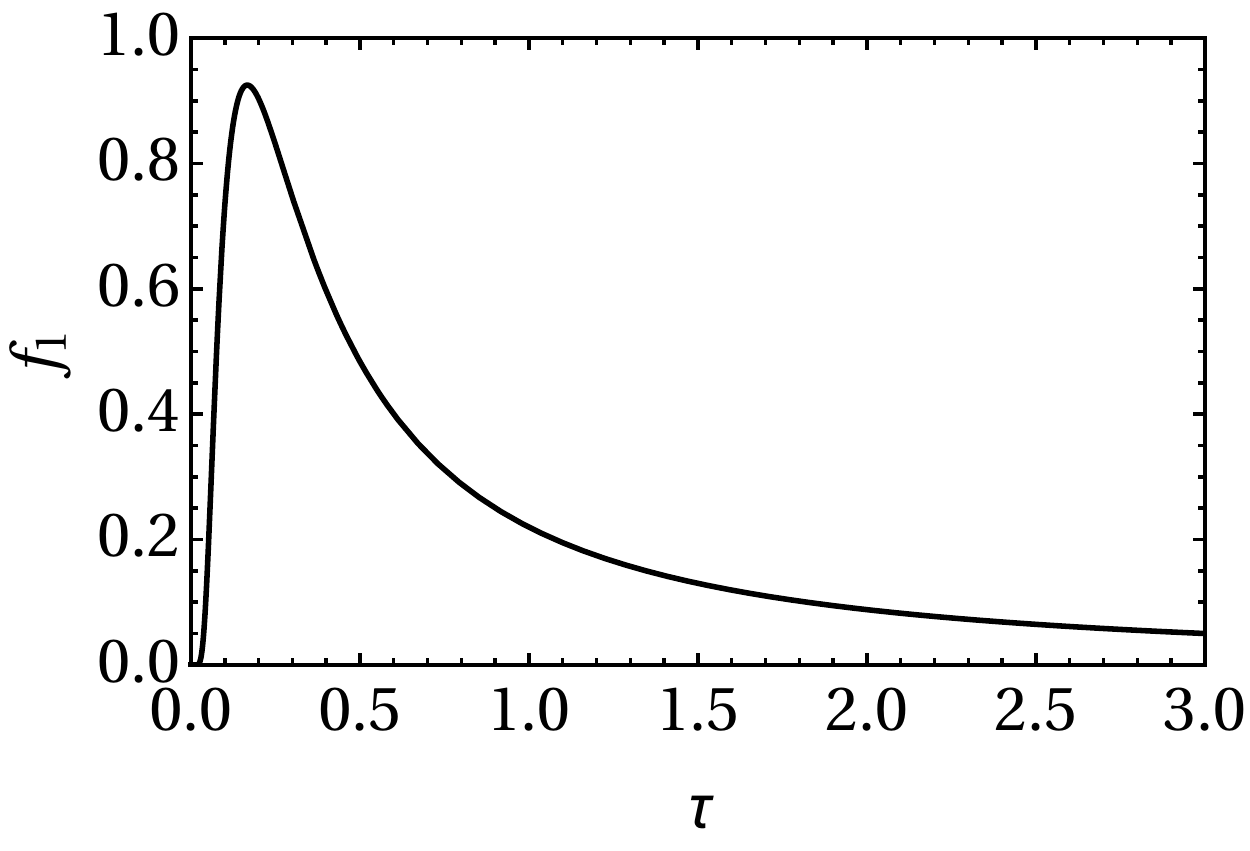}}\qquad 
	\subfigure[]{\includegraphics[width=0.4\linewidth]{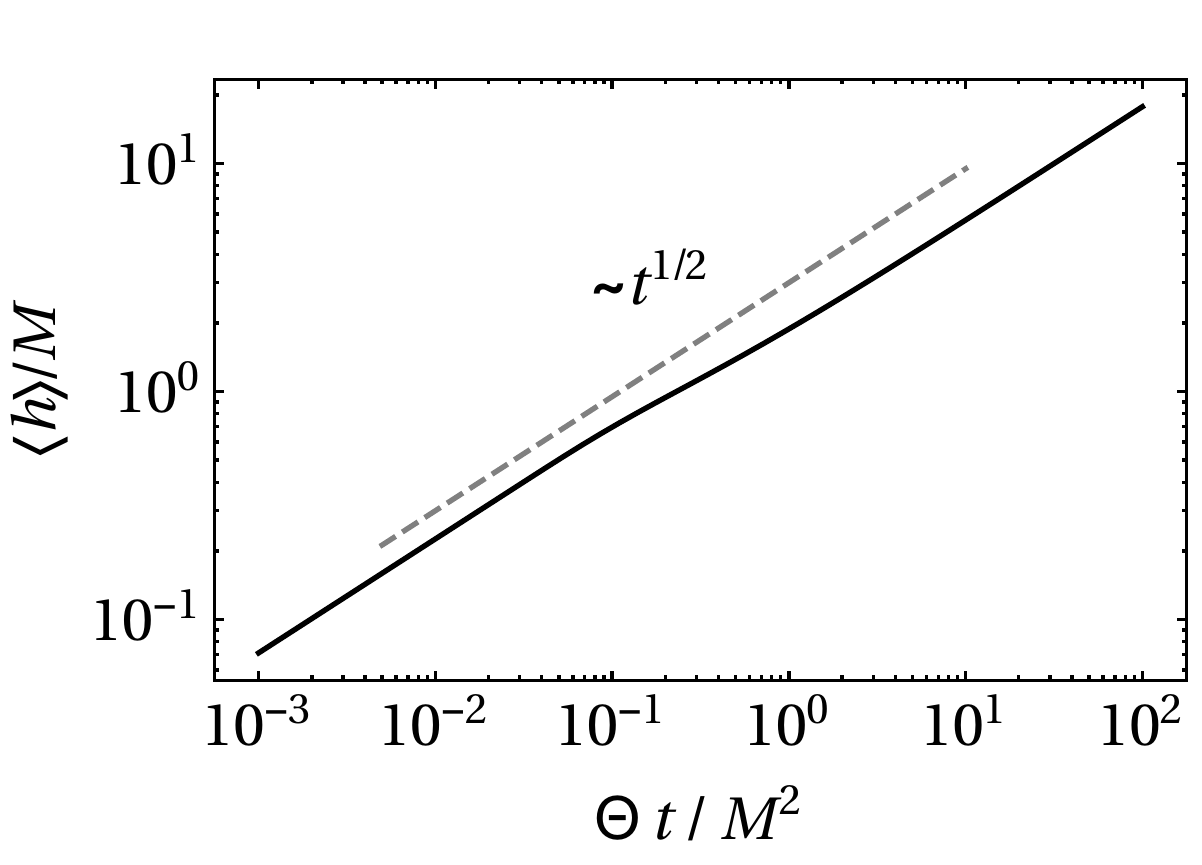}}
	\caption{Markovian Brownian motion: (a) Scaling function $\mathfrak{f}_1$ [\cref{eq_Brownian_F1pdf_scalfnc}] of the first-passage distribution as a function of $\tau\equiv \kbT T/M^2$. The maximum occurs at $\tau= 1/6$. (b) Averaged first-passage path $\bra h(t)\ket$ [\cref{eq_avgpath_F1_res}] of a Markovian Brownian walker in the presence of an absorbing boundary at $h=0$.}
	\label{fig_firstpsg_BM}
\end{figure}

In the Markovian case, the mapping implied by \cref{eq_avgpath_F1} can be implemented by placing an absorbing boundary at $h=0$ and considering a walker which starts at $(\epsilon,0)$ (with infinitesimal $\epsilon>0$) and ends at $(M,T)$ at a random time $T$ governed by $F_1(M,T)$.
Accordingly, using $\bra h(t)\ket_{M,T}$ as defined in \cref{eq_avg_path_abs}, the first-passage path in \cref{eq_avgpath_F1} reduces to
\beq \bra h(t)\ket = \frac{\int_t^\infty \d T\, F_1(M,T) \bra h(t)\ket_{M,T}}{\int_t^\infty \d T\, F_1(M,T)}.
\label{eq_avgpath_F1_Markov}\eeq 
For a Markovian Brownian walker, the first-passage time $T$ from $h=0$ to a height $h=M$ is governed by the probability distribution \cite{redner_guide_2001}
\beq F_1(M,T) = \frac{\kbT  |M|}{\sqrt{4\pi \kbT  T^3}} \exp\left(-\frac{M^2}{4\kbT T}\right) =  \frac{M^2}{\kbT } \mathfrak{f}_1(\kbT  T /M^2),
\label{eq_Brownian_F1pdf}\eeq 
where in the last equation the scaling function
\beq \mathfrak{f}_1(\tau) = \frac{\exp\left(-1/4\tau\right)}{\sqrt{4\pi\tau^3}}
\label{eq_Brownian_F1pdf_scalfnc}\eeq
has been introduced.
For $G_+$ as defined in \cref{eq_BM_prop_abs}, one has $\int_{0}^\infty \d t\, G_+(\epsilon,t|M,0) = \epsilon/\kbT $, implying 
\beq F_1(M,T) = \lim_{\epsilon\to 0} \frac{\kbT }{\epsilon}G_+(\epsilon,T|M,0).
\label{eq_firstpsg_Gp_rel}\eeq
Furthermore, noting $\int_0^\infty \d T\, F_1(M,T) = 1$, the quantity
\beq \int_t^\infty \d T\, F_1(M,T)= 1-\int_0^t \d T\, F_1(M,T)  = \mathrm{erf}\left(\frac{M}{2\sqrt{\kbT  t}}\right)
\label{eq_survival_prob}\eeq 
represents the survival probability.
Using \cref{eq_survival_prob} as well as \cref{eq_avg_path_abs,eq_firstpsg_Gp_rel}, the first-passage path defined in \cref{eq_avgpath_F1_Markov} can be calculated analytically:
\beq\begin{split} 
\bra h(t)\ket &= \frac{1}{\int_t^\infty \d T\, F_1(M,T)}  \lim_{\epsilon\to 0} \frac{\kbT }{\epsilon} \int_0^\infty \d h \int_t^\infty \d T\, G_+(h,t|\epsilon,0)\, h\, G_+(M,T|h,t)  \\
&= M \left\{ \frac{1 + \frac{4}{\xi \sqrt{\pi}} \left[1-\exp\left(-\sfrac{\xi^2}{4 }\right)\right]}{\mathrm{erf}\left(\xi/2\right)} - 1 \right\},\qquad \xi\equiv \frac{M}{\sqrt{\kbT  t}},
\end{split}\label{eq_avgpath_F1_res}\eeq 
where the integral over $T$ has been performed before the one over $h$.
Note that the term in the curly brackets is solely a function of the scaling variable $\xi$.
For small $t$, i.e., near the absorbing boundary, \cref{eq_avgpath_F1_res} reduces to 
\beq \bra h(t\to 0)\ket \simeq 4 \sqrt{\frac{\kbT  t}{\pi}} .
\label{eq_avgpath_Brownian_asympt}\eeq 
The essential reason for recovering in \cref{eq_avgpath_Brownian_asympt} the asymptotic behavior of the path with fixed endpoints $\bra h(t)\ket_{M,T}$ [see \cref{eq_avgpathBM_shortt}] is that, very close to the absorbing boundary, $\bra h(t)\ket_{M,T}$ is independent of the final time $T$ and thus can be moved out of the integral in \cref{eq_avgpath_F1_Markov} in this limit. 
For $\kbT t/M^2\gg 1$, i.e., far from the absorbing boundary, the first-passage path behaves as
\beq \bra h(t)\ket \simeq \sqrt{\pi \kbT  t}.
\label{eq_avgpath_Brownian_asympt_far}\eeq  
The non-monotonic behavior of the path $\bra h(t)\ket_{M,T}$ [\cref{eq_BM_avgpath}] for $t$ near $T$ [see \cref{fig_avgpath_Brownian}(a)] is reflected by a gentle ``bump'' of the first-passage path for $\kbT t/M^2\simeq 1$ [see \cref{fig_firstpsg_BM}(b)].
Overall, the asymptotic trajectory of a Brownian walker to its first passage point however remains at all times close to a power-law, $\bra h(t)\ket\sim t^{1/2}$.

Note that, in the weak-noise limit ($\kbT\to 0$), the first-passage path [\cref{eq_avgpath_F1_res}] vanishes.
This is in contrast to the path with fixed endpoints [\cref{eq_BM_avgpath}], for which the ``classical'' contribution, being independent of $\kbT$, prevails as $\kbT\to 0$ [see \cref{eq_avgpathBM_longt} as well as \cref{eq_MLP_Markov} below].
The time-dependence of the first-passage evolution is thus an intrinsic \emph{finite-noise} property. According to \cref{eq_avgpath_Brownian_asympt_far}, this applies even far from the absorbing boundary.

\subsubsection{Non-Markovian case}
\label{sec_avgpath_nonMarkov}

\begin{figure}[t]\centering
	\includegraphics[width=0.46\linewidth]{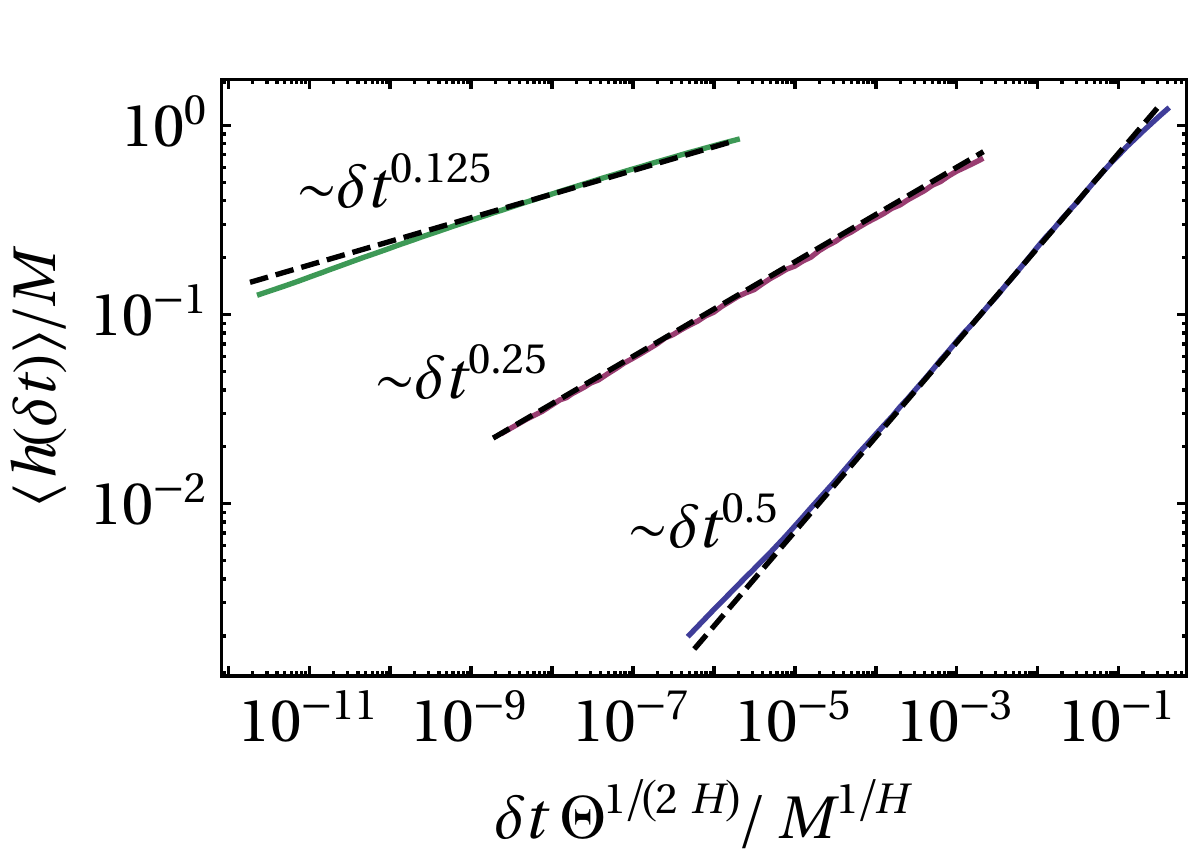}
	\caption{Numerical results (solid curves) for the averaged first-passage path $\bra h(\dt)\ket$ [\cref{eq_avgpath_F1_sim}] of a fractional Brownian walker approaching an absorbing boundary, for values of the Hurst index of $H=1/2$, $1/4$, and $1/8$. Note that the first-passage event occurs at $\dt=0$ for the averaged path, i.e., $\bra h(0)\ket=0$.
	The dashed curves represent the power law $C \delta t^{H}$, where the constant $C=4\sqrt{\kbT/\pi}$ follows from \cref{eq_avgpath_Brownian_asympt} in the case $H=1/2$, while it is obtained from a fit otherwise. }
	\label{fig_avgpath_fbm}
\end{figure}

As a specific realization of a non-Markovian random walk relevant for interfacial roughening, we consider \emph{fractional Brownian motion} (fBM).
FBM is a Gaussian process $h(t)$ with correlation function \cite{mandelbrot_fractional_1968, McCauley_hurst_2007, jeon_fractional_2010}
\beq \bra h(t) h(s)\ket = \kbT  \left( t^{2H} + s^{2H} - |t-s|^{2H} \right),
\label{eq_fbm_autocorrel}\eeq
characterized by the Hurst index $H$.
The correlation function of the relative height fluctuations $\delta h(x,t)=h(x,t)-h(x,0)$ of an equilibrated one-dimensional interface governed by \cref{eq_EW} or \eqref{eq_MH} takes the same form as in \cref{eq_fbm_autocorrel}  [see \cref{eq_hr_fbm} and, e.g., Ref.\ \cite{krug_persistence_1997}].
Standard Markovian Brownian motion results for $H=1/2$, in which case the stochastic increments $h(t+\d t)-h(t)$ are uncorrelated. For $H<1/2$ ($H>1/2$), instead, the increments are anti-correlated (positively correlated).
In the non-Markovian case, it is known that the distribution of the first-passage time $T$ to a single boundary asymptotically behaves as \cite{ding_distribution_1995,krug_persistence_1997, molchan_maximum_1999}
\beq F_1(T\to\infty) \sim T^{-2+H}.
\label{eq_fbm_firstpsg_asympt}\eeq 
Recently, an expression for the propagator of fBM with absorption has been derived perturbatively \cite{wiese_perturbation_2011,delorme_extreme-value_2016, delorme_perturbative_2016}.
However, since closed analytical results are neither available for $F_1$ nor $\bra h(t)\ket_{M,T}$, we resort in the following to numerical simulations in order to determine the first-passage path defined in \cref{eq_avgpath_F1}.

We seek the averaged path of a fractional Brownian walker starting at $h(t=0)=0$ and being absorbed at a boundary at height $M>0$  [see \cref{fig_walker_paths}(a)].
To this end, an ensemble of trajectories $\{h_{i=0,1,\ldots,N}^{(k)}\}$, each of around $N\simeq 10^7$ steps, are created and the step $T^{(k)}$, where $h^{(k)}_{T^{(k)}}\geq M$ for the first time, is determined for each trajectory $h^{(k)}$. 
Owing to the long-time tail of $F_1$ [see \cref{eq_fbm_firstpsg_asympt}], the mean first passage time $\bra T\ket$ to a single absorbing boundary is infinite. This is essentially a consequence of the fact the the walker can perform arbitrary large excursions in the negative half space ($h_i<0$) before hitting the boundary at $M$ \cite{redner_guide_2001}.
By checking different values of $N$ and $M$, we find that, in the present case, these excursions have negligible influence on the behavior of the averaged path near the boundary. 
The averaged first-passage path $\bra h_i\ket$, $i=0,\ldots,N-1$ (with $i=0$ now corresponding to the first-passage event) is then obtained as \footnote{Usually, some ``overshoot'', $h_{T}>M$, is observed, in which case the whole trajectory is shifted, i.e., $h_i \to h_i-(h_{T}-M)$, in order to ensure that $h_{T}=M$ is exactly fulfilled. The averaged path turns out to be insensitive to the overshoot correction.}
\beq \bra h_{i}\ket = N_{T\geq i}^{-1} \sideset{}{^{(T\geq i)}}\sum_k \left(M-h_{T^{(k)}-i}^{(k)}\right),
\label{eq_avgpath_F1_sim}\eeq 
where the sum is defined to run over all $N_{T\geq i}$ paths that end at times $T\geq i$.
Furthermore, the individual trajectories are shifted such that their respective first-passage times coincide (see \cref{fig_walker_paths}). 

The equivalence of \cref{eq_avgpath_F1_sim} and \cref{eq_avgpath_F1} is readily proven: the averaged path of walkers between the fixed endpoints $(h=0,t=0)$ and $(M,T)$ is given by the restricted average $\bra h_i\ket_{M,T} = \sum_{k}^{(T)} h_i^{(k)} / N_T$, where $N_T$ is the total number of such paths and the sum runs over precisely these paths. 
The discrete first-passage time distribution can be expressed as $F_1(M,T) = N_T / \sum_T N_T = N_T/N$, where $N$ is the total number of paths considered in the sample.
Using $N_{T\geq i}=\sum_{T\geq i} N_T$, the discrete analogue of \cref{eq_avgpath_F1} for the averaged path can accordingly be written as
\begin{equation} M-\bra h_i\ket = \frac{1}{\sum_{T\geq i} F_1(M,T)} \sum_{T\geq i} \bra h_{T-i}\ket_{M,T} F_1(M,T) = \frac{1}{\sum_{T\geq i} F_1(M,T)} \sum_{T\geq i} \sideset{}{^{(T)}}\sum_{k} \frac{h_{T-i}^{(k)} }{N } = \frac{1}{N_{T\geq i}} \sideset{}{^{(T\geq i)}}\sum_{k} h_{T-i}^{(k)},
\label{eq_avgpath_F1_proof}\end{equation} 
which coincides with \cref{eq_avgpath_F1_sim}.

Simulation results obtained for the averaged first-passage path defined in \cref{eq_avgpath_F1_sim} are shown in  Fig.~\ref{fig_avgpath_fbm}. For convenience, we revert to the notation of continuous time $\dt$.
FBM is simulated based on the ``circulant method'' \cite{davies_tests_1987, wood_simulation_1994} (see Ref.\ \cite{dieker_homepage_nodate} for a practical implementation).
From the plot one infers that the walker approaches the first-passage height algebraically,
\beq \bra h(\dt)\ket \sim \dt^{H}\,,
\label{eq_fbm_asymp_approach}\eeq 
with an exponent essentially coinciding with the Hurst index $H$ of the underlying fBM process.
This behavior is consistent with \cref{eq_avgpath_Brownian_asympt} in the Markovian case ($H=1/2$).
Since $M^2/\kbT \sim \Ocal(10^6)$, the slight change of the logarithmic slope of the path for $H=1/2$ observed in \cref{fig_firstpsg_BM}(b) is only partly visible in \cref{fig_avgpath_fbm}.
This applies also to the data for $H\neq 1/2$, if one assumes [as suggested by dimensional analysis of \cref{eq_fbm_autocorrel}] that the crossover time generalizes to $M^{1/H}/\kbT^{1/(2H)} $ for general fBM.
Note that $\bra h(\dt)\ket$ can be larger than $M$ for large $\dt$ because the walker can make excursions to the lower half-space [cf.\ \cref{fig_walker_paths}(a)]. 
Slight deviations from a pure algebraic behavior are noticeable in Fig.~\ref{fig_avgpath_fbm} for small times, which are found to be independent of the variance of the noise increments used in the numerical simulation as well as of the overshoot correction.

It is illuminating to consider here also the \emph{most-likely} path of a fBM between two locations $h=0$ and $M$.
The most-likely path minimizes the dynamic action of the associated probability functional and thus represents the weak-noise approximation of the averaged path. 
As shown in \cref{sec_fbm_MPL}, within WNT, one finds
\beq h\st{MLP}(\dt )\sim \dt^{2H}
\label{eq_fbm_asymp_approach_WNT}\eeq 
near the endpoint.
The different exponents in \cref{eq_fbm_asymp_approach,eq_fbm_asymp_approach_WNT} can be attributed to the repulsive effect exerted by the absorbing boundary on the fluctuations around the most-likely path (see \cref{sec_discussion} for further discussion).

\section{Most-likely path of a Gaussian random process}
\label{sec_fbm_MPL}

\begin{figure}[t]\centering
	\includegraphics[width=0.42\linewidth]{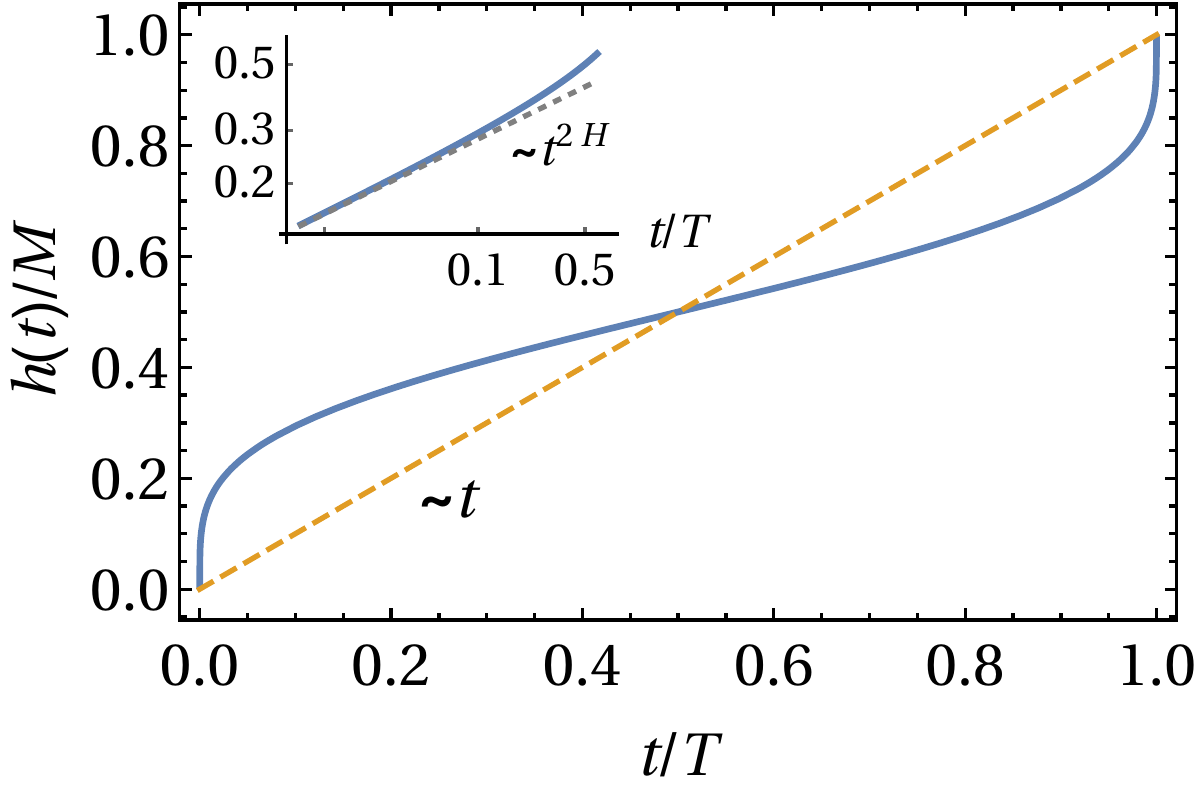}
	\caption{Most-likely path $h(t)$ of a fractional Brownian walker between the points $(h,t)=(0,0)$ and $(M,T)$ [\cref{eq_MLP_FBM}]. For illustrative purposes, we have chosen a Hurst exponent of $H=1/8$. The dashed line in the main plot represents a linear path for comparison. The inset demonstrates that the asymptotic departure from the initial point follows a power-law $h(t)\propto t^{2H}$ (dashed line). A similar behavior is obtained near the final point.}
	\label{fig_fbm_MLP}
\end{figure}

We determine here the most-likely path of a Gaussian random process $h(t)$, $0\leq t\leq T$, subject to the constraints 
\beq h(0)=M_0\qquad \text{and}\qquad h(T)=M_T,
\label{eq_gauss_constr}\eeq
where $M_0$ and $M_T$ are constants.
The following discussion is in fact a straightforward application of the constrained minimization of a quadratic functional (see, e.g., \cite{fletcher_practical_2000}).
The Gaussian process is taken to have zero mean $\bra h(t)\ket=0$ and correlation function
\beq 
G(t,t') \equiv \bra h(t) h(t')\ket.
\label{eq_gauss_cov}\eeq 
Accordingly, the joint probability distribution is given by
\beq P[h] \sim \exp(-\Scal[h]),
\label{eq_multigauss_pdf}\eeq
with the ``action''
\beq \Scal[h] \equiv \onehalf  \int_0^T \d t \int_0^T \d t'\, h(t) G^{-1}(t,t') h(t').
\label{eq_gauss_action}\eeq 
The inverse $G^{-1}$ of the correlation function is defined (in an operator sense) by
\beq \int_0^T \d t\,  G(s,t)G^{-1}(t,s') = \delta(s-s').
\label{eq_gauss_invrel}\eeq 
For a Markovian process (i.e., standard Brownian motion), one has $G^{-1}(t,t') = - \onehalf \pd_t^2 \delta(t-t')$, while the correlation function $G(t,t') = 2\text{min}(t,t')$ \cite{kleinert_path_2009}. 
For fractional Brownian motion, the explicit form of $\Scal$ is known only perturbatively \cite{wiese_perturbation_2011}.
In passing, we remark that the continuous time description used here should formally be understood as the limit of a multivariate Gaussian process of random variables $\{ h_i\}$ defined at discrete times $i=0,\Delta t,\ldots, (T/\Delta t)-1$, analogously to the definition of a path integral \cite{kleinert_path_2009}.
Imposing the constraints in \cref{eq_gauss_constr} gives rise to the augmented action
\beq \tilde \Scal([h],\lambda_1,\lambda_2) \equiv S[h] - \lambda_1[ h(0)-M_0 ] - \lambda_2[ h(T)-M_T ]
\eeq 
with the Lagrange multipliers $\lambda_{1,2}$.
Minimization of $\tilde \Scal$ with respect to $h(\tau)$ yields
\beq 0 = \frac{\delta \tilde \Scal}{\delta h(\tau)} = 2 \int_0^T \d t\, h(t) G^{-1}(t,\tau) - \lambda_1 \delta(\tau) - \lambda_2 \delta(T-\tau),
\label{eq_gauss_action_min}\eeq 
where we used the symmetry property $G(t,t')=G(t',t)$.
Multiplying \cref{eq_gauss_action_min} with the inverse correlation function $G(s,\tau)$ and integrating over $\tau$, using \cref{eq_gauss_invrel}, one obtains $h(s) = \frac{1}{2} \left[G(s,0) \lambda_1 + G(s,T)\lambda_2 \right]$.
Satisfaction of the constraints in \cref{eq_gauss_constr} provides the values of $\lambda_{1,2}$ and eventually yields the expression of the constrained minimum-action path of a general Gaussian process (see also Ref.\ \cite{norros_busy_1999}):
\beq h(s) = G(s,0) \sum_k (\bv{Q}^{-1})_{1k} M_k + G(s,T) \sum_k (\bv{Q}^{-1})_{2k} M_k,
\label{eq_gauss_MLP}\eeq 
where 
\beq \bv{Q} \equiv \begin{pmatrix} G(0,0) & G(0,T) \\ G(T,0) & G(T,T) \end{pmatrix}
\label{eq_gauss_cov_constr}\eeq 
is the ``covariance matrix of the constraints'' and $\bv{M}\equiv (M_0, M_T)$. 
These results naturally generalize to more than two constraints.
Notably, the time dependence of the minimum action path in \cref{eq_gauss_MLP} is essentially determined by the correlation function.

We now specialize the above results to fBM, i.e., a Gaussian process described by the correlation function in \cref{eq_fbm_autocorrel}.
Since this correlation function is trivially zero if one of its arguments vanishes [rendering a singular covariance matrix in \cref{eq_gauss_cov_constr}], the evaluation of \cref{eq_gauss_MLP} is performed with a value $\epsilon>0$ instead of 0 for the initial time.
After sending eventually $\epsilon\to 0$ and setting $M_0=0$, $M_T=M$ [see \cref{eq_gauss_constr}], \cref{eq_gauss_MLP} reduces to (see also Ref.\ \cite{delorme_extreme-value_2016}) 
\beq h(t) = \onehalf M T^{-2H}\left(t^{2H} - (t-T)^{2H} + T^{2H}\right),
\label{eq_MLP_FBM}\eeq 
where $0<t<T$.
For $t\to 0$ or $t\to T$, one has $h(t)\sim t^{2H}$ and $M-h(t)\sim (T-t)^{2H}$, respectively, showing that a fractional Brownian walker approaches the endpoints of a constrained path via a power-law with exponent $2H$. This is illustrated in Fig.~\ref{fig_fbm_MLP}.
In the Markovian case, \cref{eq_MLP_FBM} reduces to a straight line,
\beq h(t) = M \frac{t}{T},\qquad (H=1/2).
\label{eq_MLP_Markov}\eeq 

\section{Review on eigenfunctions}
\label{app_prev_results}

\begin{table}[t]\begin{center}
\begin{tabular}{l|c|c|c}
    & periodic [\cref{eq4_bcs_per}] & Dirichlet zero $\mu$ [\cref{eq4_bcs_DirCP}] & Dirichlet no-flux [\cref{eq4_bcs_DirNoFl}] ($b=1$)${}^\dagger$ \\
\hline
$\pd_x^{z}$ self-adjoint & yes & yes & no \\[2ex]
$\sigma_k$ & $\dps \frac{1}{\sqrt{L}} \exp\left(\frac{2\pi \im k}{L} x\right)$  & $\dps \sqrt{\frac{2}{L}} \sin\left(\frac{k\pi}{L}x\right)$ & $\sigma_k\DirNoFl$  \\[4ex]
$\varphi_k$ & $\sigma_k$ & $\sigma_k$ & $\varphi_k\DirNoFl$  \\[3ex]
$k$ & $0,\pm 1,\pm 2,\ldots^\ddagger$ & $1,2,3, \ldots$ & $1,2,3, \ldots$ \\[3ex]
$\gamma_k$ [Eqs.\ \eqref{eq6_eigen}, \eqref{eq6_adj_eigen}] & $\dps (-1)^{b+1}\left(\frac{2\pi k}{L}\right)^{z}$  & $\dps (-1)^{b+1}\left(\frac{k\pi}{L}\right)^{z}$ & $(\omega_k/L)^4$   \\[3ex]
$\kappa_k$ [\cref{eq6_ortho}] & 1 & 1 & $\dps \frac{L}{3}\left(1-\frac{(-1)^k}{\cosh(L\gamma_k^{1/4})} \right)$  \\[3ex]
$\epsilon_k$ [\cref{eq6_orthoDx2}] & \vtop{\hbox{\strut $\dps \left[-|\gamma_k|^{1/2}\right]^b\kappa_k$,}\hbox{\strut $\epsilon_0=0$}} & $\dps \left[-|\gamma_k|^{1/2}\right]^b\kappa_k$ & $-\gamma_k^{1/2} \kappa_k$  \\[3ex]
\end{tabular}
\end{center}\caption{Eigenfunctions and related properties of $\pd_x^z$ on the interval $[0,L]$ for various \bcs. The eigenfunctions $\sigma_k$ coincide with the corresponding adjoint eigenfunctions $\varphi_k$ if $\pd_x^z$ is self-adjoint. Furthermore, $b=0$ for EW dynamics, $b=1$ for MH dynamics, and the dynamic index $z=2b+2$. ${}^\dagger$Dirichlet no-flux \bcs are considered only for MH dynamics ($b=1$). Since $\sigma_k\DirNoFl$ and $\varphi_k\DirNoFl$ are not normalized here, the system size $L$ appears in the corresponding expression for $\kappa_k$. $^\ddagger$Due to the mass constraint [\cref{eq_zero_vol}], the zero mode ($k=0$) is absent from the actual solution for periodic \bcs.}
\label{tab_eigenfunc}\end{table}

Here, a number of relevant properties of the eigenfunctions of the (bi-)harmonic operator $\pd_x^z$ on the interval $[0,L]$ are collected (see \paperI for more details).
We introduce a complete set of (``proper'') eigenfunctions $\sigma_k$, $k\in \mathbb{Z}$, fulfilling
\beq \pd_x^z \sigma_k(x) = \gamma_k \sigma_k(x),\qquad z\in \{2,4\}
\label{eq6_eigen}\eeq 
with eigenvalues $\gamma_k$. The eigenfunctions are subject to one of the following \bcs:
\begin{subequations}\begin{align}
\text{periodic:}\qquad & \sigma_k\pbc(x,t)=\sigma_k\pbc(x+L,t), \label{eq4_bcs_per}\\
\text{Dirichlet zero-$\mu$:}\qquad & \sigma_k\DirCP(0,t)=0=\sigma_k\DirCP(L,t),\qquad \pd_x^2\sigma_k\DirCP(0,t)=0= \pd_x^2\sigma_k\DirCP(L,t),\label{eq4_bcs_DirCP} \\
\text{Dirichlet no-flux:}\qquad & \sigma_k\DirNoFl(0,t)=0=\sigma_k\DirNoFl(L,t),\qquad \pd_x^3 \sigma_k\DirNoFl(0,t)=0= \pd_x^3 \sigma_k\DirNoFl(L,t). \label{eq4_bcs_DirNoFl} 
\end{align}\label{eq4_bcs}\end{subequations}
The symbol $\mu$ refers to the chemical potential, which vanishes at the boundary for standard Dirichlet \bcs (see \paperI). For this reason, the latter are also called Dirichlet zero-$\mu$ \bcs here.
Associated with $\sigma_k$ are a set of \emph{adjoint} eigenfunctions $\varphi_k$, which fulfill
\beq \pd_x^z \varphi_k(x) = \gamma_k \varphi_k(x)
\label{eq6_adj_eigen}\eeq 
as well as one of the following adjoint boundary conditions:
\begin{subequations}\begin{align}
\text{periodic:}\qquad & \varphi_k\pbc(x,t)=\varphi_k\pbc(x+L,t), \label{eq4_adjbcs_per}\\
\text{Dirichlet zero-$\mu$:}\qquad & \varphi_k\DirCP(0,t)=0=\varphi_k\DirCP(L,t),\qquad \pd_x^2\varphi_k\DirCP(0,t)=0=\pd_x^2\varphi_k\DirCP(L,t), \label{eq4_adjbcs_DirCP} \\
\text{Neumann zero-$\mu$:}\qquad & \pd_x \varphi_k\DirNoFl(0,t)=0= \pd_x \varphi_k\DirNoFl(L,t),\qquad \pd_x^2 \varphi_k\DirNoFl(0,t) = 0 = \pd_x^2 \varphi_k\DirNoFl(L,t). \label{eq4_adjbcs_NeuCP} 
\end{align}\label{eq4_adjbcs}\end{subequations}
Note that proper and adjoint eigenfunctions in general have an identical set of eigenvalues $\gamma_k$.
For periodic and Dirichlet zero-$\mu$ \bcs, the operator $\pd_x^{z}$ is self-adjoint on $[0,L]$, implying that 
\beq \varphi_k\ut{(p,D)}=\sigma_k\ut{(p,D)} .
\eeq 
In contrast, for Dirichlet no-flux \bcs on $\sigma_k$ [\cref{eq4_bcs_DirNoFl}], the operator $\pd_x^{z}$ is not self-adjoint and the associated adjoint eigenfunctions $\varphi_k\DirNoFl$ are required to satisfy the distinct \bcs in \cref{eq4_adjbcs_NeuCP}.
The eigenfunctions $\sigma_m$ and $\varphi_n$ are mutually orthogonal:
\beq \int_0^L \d x\, \sigma_m^*(x)\varphi_n(x)  = \kappa_n\delta_{mn}
\label{eq6_ortho}\eeq 
with a real number $\kappa_n$.
The star denotes complex conjugation, which is necessary in order to deal with complex-valued eigenfunctions, such as those for periodic \bcs.
One furthermore has
\beq \int_0^L \d x\, \varphi_m^*(x) \varphi_n''(x) = \epsilon_n\delta_{mn}
\label{eq6_orthoDx2}\eeq 
with a real number $\epsilon_n$.
The eigenvalues of $\pd_x^4$ [see \cref{eq6_eigen}] for Dirichlet no-flux \bcs are given by
\beq \gamma_k\DirNoFl = \left(\frac{\omega_k\DirNoFl}{L}\right)^4,
\eeq 
where $\omega_k\DirNoFl$ denotes a solution to the transcendental equation
\beq \cos (\omega_k\DirNoFl) \cosh (\omega_k\DirNoFl) = 1.
\label{eq6_DirNoFl_eigenvaleq}\eeq 
Numerically one obtains
\beq \omega_k\DirNoFl = 4.7300,\, 7.8532,\, 10.9956,\ldots \qquad (k=1,2,3,\ldots)
\label{eq6_DirFl0_eigenval}\eeq 
Since $\sigma_{k=0}\DirNoFl(x)=0$, $\sigma_k\DirNoFl(x) = \sigma_{-k}\DirNoFl(x)$, and $\omega_k\DirNoFl=\omega_{-k}\DirNoFl$, we restrict the eigenspectrum to $k\geq 1$.
For $k\gtrsim 4$, an accurate approximation is provided by
\beq \omega_k\DirNoFl \simeq \pi \left( k+ \onehalf \right),
\label{eq6_DirFl0_eigenval_approx}\eeq 
which becomes asymptotically exact.
Explicit expressions and relevant properties of $\sigma_k$, $\varphi_k$ are summarized in \cref{tab_eigenfunc}. (Expressions for the eigenfunctions $\sigma\DirNoFl_k$ and $\varphi\DirNoFl_k$ are reported \paperI.)

\section{Roughening}
\label{sec_roughening}
In the absence of an impenetrable wall, the EW and the MH equation can be solved analytically. 
In the context of roughening, so far mainly bulk systems or systems with periodic \bcs have been considered \cite{majaniemi_kinetic_1996, abraham_dynamics_1989,racz_scaling_1991,antal_dynamic_1996, barabasi_fractal_1995, majaniemi_kinetic_1996, flekkoy_fluctuating_1995,flekkoy_fluctuating_1996,krug_origins_1997,darvish_kinetic_2009,taloni_correlations_2010,taloni_generalized_2012,gross_interfacial_2013, halpin-healy_kinetic_1995,de_villeneuve_statistics_2008}. 
Here, we provide a general series solution in terms of the corresponding eigenfunctions, which  can be readily specialized to various \bcs. 
We begin by casting \cref{eq_EW,eq_MH} into the common form 
\beq \pd_t h = (-1)^b \frict \pd_x^{z} h + \hat\zeta
\label{eqR_dyneq}\eeq
with $b=0,1$ for the EW and MH equation, respectively, and $z=2b+2$. The noise $\hat\zeta\equiv \pd_x^b \zeta$ is correlated as [cf.\ \cref{eq_noise}]
\beq \bra \hat\zeta(x,t) \hat\zeta(x',t') = (-1)^b  2D \pd_x^{2b} \delta(x-x')\delta(t-t').
\label{eqR_noise_correl}\eeq 

To proceed, the field $h$ and the noise $\hat\zeta$ are expanded in terms of the eigenfunctions $\sigma_k(x)$ defined in \cref{eq6_eigen}:
\begin{subequations}\begin{align}
h(x,t) &= \sum_k h_k(t) \sigma_k(x), \\
\hat\zeta(x,t) &= \sum_k \hat\zeta_k(t) \sigma_k(x).
\end{align}\label{eqR_expansions}\end{subequations}
The expansion coefficients follow from the orthogonality relation in \cref{eq6_ortho} as
\begin{subequations}\begin{align}
h_k(t) &= \int_0^L \d x\, h(x,t) \varphi_k^*(x)/\kappa_k, \\
\hat\zeta_k(t) &= \int_0^L \d x\, \hat\zeta(x,t) \varphi_k^*(x)/\kappa_k,
\end{align}\end{subequations}
where $\varphi_k(x)$ are the adjoint eigenfunctions [\cref{eq6_adj_eigen}] and $\kappa_k$ is reported in \cref{tab_eigenfunc}.
Accordingly, upon using \cref{eq6_ortho,eq6_orthoDx2}, the correlation of the noise modes follows as
\beq\begin{split}
\bra \hat\zeta_m(t) \hat\zeta_n^*(t')\ket &= \left\bra  \int_0^L \d x \int_0^{L} \d x'\, \varphi_m^*(x)\varphi_n(x') \pd_x^b \zeta(x,t) \pd_{x'}^b\zeta(x',t') \frac{1}{\kappa_m \kappa_n} \right \ket  \\
 &= (-1)^b 2D \delta(t-t') \frac{\tilde \epsilon_m}{\kappa_m^2} \delta_{mn},
\end{split}
\label{eqR_noise_correl_mode}\eeq 
where $\tilde \epsilon_k\equiv \kappa_k$ for $b=0$ and $\tilde \epsilon_k\equiv \epsilon_k$ for $b=1$. The partial integrations required in the case $b=1$ have generated the factor $(-1)^b$ in the last line of \cref{eqR_noise_correl_mode}; the same result is obtained upon using \cref{eqR_noise_correl}. All boundary terms vanish for the \bcs considered here.
The mass-conserving property of the noise for MH dynamics ($b=1$) is reflected in \cref{eqR_noise_correl_mode} by the fact that $\tilde \epsilon_0/\kappa_0^2=0$ in this case (see \cref{tab_eigenfunc}) \footnote{For Dirichlet no-flux \bcs, this is readily proven by considering the limit $\omega\to 0$.}.
For EW dynamics, instead, $\tilde \epsilon_0/\kappa_0^2=1$, such that the noise in principle contributes to the zero mode.
Upon inserting the expansions given in \cref{eqR_expansions} into \cref{eqR_dyneq} and using \cref{eq6_eigen}, one obtains  
\beq \pd_t h_m(t) = -\Lambda_m h_m(t) + \hat\zeta_m(t),\qquad \Lambda_m \equiv -(-1)^b \frict \gamma_m ,
\label{eqR_dyneq_coeff}
\eeq 
with $\Lambda_m\geq 0$ and the eigenvalues $\gamma_m$ (see \cref{tab_eigenfunc}).
For an arbitrary initial profile $h_m(0)$, the solution of \cref{eqR_dyneq_coeff} is given by
\beq h_m(t) = e^{-\Lambda_m t}\left(h_m(0) + \int_0^t \d t' e^{\Lambda_m t'} \hat\zeta_m(t') \right).
\label{eqR_solh_mode}\eeq
For the EW equation with periodic \bcs, the zero mode $h_{m=0}$ (for which $\Lambda_0=0$) is absent from the spectrum due to the mass constraint [\cref{eq_zero_vol}] enforced by \cref{eq_height_redef}. The dynamics of $h_0$ obtained in the case of an unconstrained profile is discussed separately below [see \cref{eqR_cm_diffusion}].
In the long-time, equilibrium limit, the equal-time correlation function follows as 
\beq \bra h_m(t) h_n^*(t) \ket\Big|_{t\to\infty} = \frac{D (-1)^b \tilde \epsilon_m}{\Lambda_m\kappa_m^2} \delta_{mn} \equiv V_m \delta_{mn}.
\label{eqR_eqcorrel}\eeq 
Note that $V_m= \sfrac{D |\tilde\epsilon_m|}{|\Lambda_m|\kappa_m^2} \geq 0$, as is readily shown using \cref{tab_eigenfunc}.
\Cref{eqR_eqcorrel} does not apply to a zero mode, in which case \cref{eqR_solh_mode} directly yields $\bra h_0(t) h_n^*(t) \ket=0$ for all $n$ [see also \cref{eqR_noise_correl_mode}].

Assuming uncorrelated initial conditions, $\bra h_m(0) h_n^*(0)\ket \propto \delta_{mn}$, the two-time correlation function of a \emph{relative} height fluctuation $\delta h_m(t)\equiv h_m(t)-h_m(0)$ follows from \cref{eqR_solh_mode} as
\beq\begin{split}
\bra \delta h_m(t) \delta h_n^*(s)\ket &= \Big\{ \bra |h_m(0)|^2\ket \left[ 1-e^{-\Lambda_m t} + 1 - e^{-\Lambda_m s} - \left(1-e^{-\Lambda_m (t+s)}\right)  \right] \\
& \qquad + V_m \left[1-e^{-\Lambda_m (t+s)} - \left(1-e^{-\Lambda_m|t-s|}\right) \right] \Big\}\delta_{mn}.
\end{split}\label{eqR_gencorrel}\eeq
If the profile is initially flat, $h_m(0)=0$, only the second term in \cref{eqR_gencorrel} remains:
\beq \bra\delta h_m(t) \delta h_n^*(s)\ket\st{flat} = V_m \left[1-e^{-\Lambda_m (t+s)} - \left(1-e^{-\Lambda_m|t-s|}\right) \right] \delta_{mn}.
\label{eqR_correl_delta_flat}\eeq 
For thermal initial conditions, where according to \cref{eqR_eqcorrel} $\bra |h_m(0)|^2\ket =V_m$, \cref{eqR_gencorrel} instead becomes
\beq \bra\delta h_m(t) \delta h_n^*(s)\ket\st{th} = V_m \left[ 1-e^{-\Lambda_m t} + 1-e^{-\Lambda_m s} - \left(1-e^{-\Lambda_m|t-s|}\right) \right]\delta_{mn}.
\label{eqR_correl_delta_th}\eeq
The real-space correlation function follows as 
\beq \bra \delta h(x,t) \delta h(y,s)\ket = \sum_m \bra \delta h_m(t) \delta h_m^*(s)\ket \sigma_m(x)\sigma_m^*(y),
\label{eqR_correl_realsp}\eeq
where we used the fact that $h_{-k}=h_k^*$ for periodic \bcs, which is a consequence of $h(x,t)$ being real.

For $t=s$ and $x=y$, the real-space correlation function reduces, both for flat and thermal initial conditions, in the long-time limit to
\beq \bra |\delta h(x)|^2\ket\eq \equiv \bra |\delta h(x,t)|^2\ket_{t\to\infty} =  \sum_m V_m |\sigma_m(x)|^2.
\label{eqR_correl_realsp_eq}\eeq 
For periodic \bcs (see \cref{tab_eigenfunc}), one has $|\sigma_m(x)|^2=1/L$, and \cref{eqR_correl_realsp_eq} becomes [see also \cref{eq_Pss_var}]
\beq \bra |\delta h\pbc|^2\ket\eq = \frac{D L}{(2\pi)^2 \frict }\sum_{m,m\neq 0} \frac{1}{m^2} = \frac{DL}{12\frict} = \frac{\kbT L}{6},
\label{eqR_eqvar_per}\eeq 
where we used \cite{gradshteyn_table_2014} $\sum_{m=1}^\infty m^{-2} = \pi^2/6$ and introduced the temperature $\kbT=D/(2\frict)$ according to \cref{eq_FDT}. 
For Dirichlet zero-$\mu$ \bcs, instead, \cref{eqR_correl_realsp_eq} becomes [see also \cref{eqBM_var_bridge}]
\beq \bra |\delta h\DirCP(x)|^2\ket\eq = \frac{2 D L}{\pi^2 \frict}\sum_{k=1}^\infty \frac{1}{k^2}\sin^2\left(\frac{k\pi x}{L}\right) = 2\kbT L \frac{x}{L}\left(1 - \frac{x}{L}\right),
\label{eqR_eqvar_DirCP}\eeq 
where we used $\sin^2(y)=[1-\cos(2y)]/2$ and well-known Fourier series representations of trigonometric functions \cite{gradshteyn_table_2014}. 
In the case of Dirichlet no-flux \bcs, instead of directly calculating the infinite sum in \cref{eqR_correl_realsp_eq}, we invoke a mapping to Brownian motion, which according to \cref{eqBM_var_areaconstr} yields
\beq \bra |\delta h\DirNoFl(x)|^2\ket\eq = 2\kbT L  \frac{x}{L} \left(1-\frac{x}{L}\right) \left(1+ 3\frac{x}{L} \left(\frac{x}{L}-1\right)\right).
\label{eqR_eqvar_DirNoFl}\eeq
This expression is found to numerically coincide with \cref{eqR_correl_realsp_eq}.

For $x=y$, but arbitrary times, \cref{eqR_correl_realsp} becomes 
\begin{subequations}\begin{align}
\bra\delta h(x,t) \delta h(x,s)^*\ket\st{flat} &= \Ccal(t+s,x) - \Ccal(|t-s|,x),\\
\bra\delta h(x,t) \delta h(x,s)^*\ket\st{th} &= \Ccal(t,x) + \Ccal(s,x) - \Ccal(|t-s|,x),
\end{align}\label{eqR_correl_twotime}\end{subequations}
with
\beq \Ccal(t,x) \equiv \sum_k V_k \left(1-e^{-\Lambda_k t}\right) |\sigma_k(x)|^2.
\label{eqR_correl_C}\eeq
The \emph{roughness} of an interface is defined as one of the following equal time correlation functions:
\begin{subequations}\begin{align}
\bra |\delta h(x,t)|^2\ket\st{flat} &= \Ccal(2t,x), \label{eqR_roughness_flat}\\
\bra |\delta h(x,t)|^2\ket\st{th} &= 2\Ccal(t,x). \label{eqR_roughness_th}
\end{align}\label{eqR_roughness}\end{subequations}
The finiteness and the discreteness of the system imply the existence of a smallest and a largest mode index, $k\st{min}$ and $k\st{max}$. 
In order to obtain a closed expression for the correlation function $\Ccal(t,x)$, we replace the sum in \cref{eqR_correl_C} by an integral.
The error arising from this approximation is small if the summands in \cref{eqR_correl_C} vary significantly only over a few $k$. This, in turn, applies if the system size is large and $t\ll 1/\Lambda_{k\st{min}}$, since then the variation occurs for large $k$, where $\Lambda_k\sim k^z$.
For periodic \bcs one has $k\st{min}\pbc=1$ and $k\st{max}\pbc= \lceil (L/\lattsp-1)/2 \rceil$ (see \cref{tab_eigenfunc} as well as \cref{app_discr}), such that \cref{eqR_correl_C} becomes 
\beq \Ccal\pbc(t) \simeq \frac{D}{\frict \pi} \int_{p\pbc\st{min}}^{p\pbc\st{max}} \d p \frac{1-\exp(-\frict p^z t)}{k^2} \equiv \Ccal(t; p\pbc\st{min,max}),
\label{eqR_C_pbc0}\eeq
where we introduced the wave number $p\pbc \equiv 2\pi k\pbc/L$ associated with $k$.
Note that \cref{eqR_C_pbc0} is independent of $x$ owing to translational invariance.
For standard Dirichlet \bcs, instead, one has $k\st{min}\Dbc = 1$ and $k\st{max}\Dbc = L/\lattsp-1$ (see \cref{app_discr}]).
In order to evaluate \cref{eqR_correl_C}, we focus on the point $x=L/2$ and note that $\sqrt{L/2}\,\sigma_k(L/2)=1,0,-1,0,1,\ldots$ for $k=1,2,3,\ldots$, such that one obtains 
\beq\begin{split} \Ccal\DirCP(t,L/2) &= \frac{2D}{\frict L} \sum_{n=0}^{n\st{max}} \frac{1-\exp(-\frict [(2n+1)\pi/L]^z t)}{[(2n+1)\pi/L]^2} 
\simeq \frac{D}{\frict \pi} \int_{p\st{min}\DirCP}^{p\st{max}\DirCP} \d p \frac{1-\exp(-\frict p^z t)}{p^2} \\ &= \Ccal(t; p\st{min,max}\DirCP),
\end{split}\label{eqR_C_Dbc}\eeq
where $n\equiv 2k+1$, $n\st{max}\equiv  \lceil (L/\lattsp -1)/2 \rceil$, and $p\DirCP \equiv \pi k\DirCP/L$.
For Dirichlet no-flux \bcs and a sufficiently large integer $k'$, one may approximate, for $k\geq k'$, $|\Lambda_k|/\eta \simeq ((k+1/2)\pi /L)^4$, $[\sigma\DirNoFl_k(L/2)]^2\simeq 2/3$ for even $k$, while $\sigma\DirNoFl_k(L/2)=0$ for odd $k$ (see \cref{tab_eigenfunc} and \paperI).
Leaving at this point the largest mode $k\st{max}\DirNoFl$ unspecified \footnote{The numerical analysis in \cref{app_bench} suggests $k\st{max}\DirNoFl\simeq L/(2\lattsp)$}, we accordingly obtain ($n\equiv k/2$)
\beq \Ccal\DirNoFl(t,L/2) \simeq \frac{2D}{\frict L} \sum_{n=\lfloor k'/2 \rfloor}^{\lceil k\st{max}\DirNoFl/2 \rceil} \frac{1-\exp(-\frict [2\pi n/L+\pi/2L]^4 t)}{(2\pi n/L+\pi/2L)^2} \simeq \Ccal(t; p\st{min,max}\DirNoFl)
\label{eqR_C_DirNoFl}\eeq 
with $p\st{max}\DirNoFl = (k\st{max}\DirNoFl+1/2)\pi /L$. 
The freedom in the choice for the lower bound $k'$ leads to a negligible error in $\Ccal\DirNoFl$ at large times. 
We thus re-instate for the smallest wave number the exact value $p\st{min}\DirNoFl = \omega_1\DirNoFl /L$, with $\omega_1\DirNoFl $ defined in \cref{eq6_DirFl0_eigenval}.
In conclusion, the expression for $\Ccal(t;p\st{min,max})$ in \cref{eqR_C_pbc0}, which approximates the one-point correlation function in \cref{eqR_correl_C} at $x=L/2$, depends on the \bcs only via the integration boundaries $p\st{min,max}$.
The integral in \cref{eqR_C_pbc0} can be calculated in closed form, leading to
\beq \Ccal(t) =\frac{D}{\frict\pi}(\frict t)^{\frac{1}{z}} z^{-1} \int_{\frict t p\st{min}^z}^{\frict t p\st{max}^z} \d x\, x^{-\frac{1}{z}-1} \left(1-e^{-x}\right) = \frac{D}{\frict\pi} (\frict t)^{1/z} z^{-1} \left[ \Gamma\left(-z^{-1}, x\right)-z x^{-1/z} \right]_{x=\frict t p\st{min}^z}^{x=\frict t p\st{max}^z},
\label{eqR_C_pbc}
\eeq
with $\Gamma(n,x)$ being the upper incomplete Gamma-function.
To proceed, we introduce the crossover time $\tau\cro = 1/(\frict p\st{max}^z)$, as well as the \textit{roughening time} 
\beq   \tau_R = 
\begin{cases} \displaystyle
\left(\frac{L}{2\pi}\right)^z \frac{1}{\frict},\qquad &\text{periodic},\\ \displaystyle
\left(\frac{L}{\pi}\right)^z \frac{1}{\frict},\qquad &\text{standard Dirichlet},\\ \displaystyle
\left(\frac{L}{\omega_1\DirNoFl}\right)^z \frac{1}{\frict},\qquad &\text{Dirichlet no-flux \bcs},
\end{cases}
\label{eq_trough} \eeq
for which one has $\tau_R\simeq 1/(\eta  p\st{min}^z)$ [see also \cref{eq_relaxtime,eq_crossover_time}].
From \cref{eqR_dyneq_coeff} one infers that $\tau_R$ and $\tau\cro$ are the relaxation times of the mode with largest and smallest wavelength that can be accommodated by the system.
The correlation function $\Ccal(t)$ exhibits three distinct asymptotic regimes:
\beq \Ccal(t) \simeq \frac{2\kbT}{\pi} 
\begin{cases}\displaystyle \frac{p\st{max}^{z-1}-p\st{min}^{z-1}}{z-1} t, \quad &  t\lesssim \tau\cro, \\
\Gamma\left(1-z^{-1}\right)(\frict t)^{1/z}, &  \tau\cro \lesssim t \lesssim \tau_R,\\ 
\frac{1}{p\st{min}}-\frac{1}{p\st{max}}, &  t\gtrsim \tau_R,
\end{cases}\label{eqR_C_pbc_asympt}\eeq
i.e., an initial diffusive growth followed by a subdiffusive law characterized by the dynamic index $z$. 
For times of the order of $\tau_R$, the height variance saturates to its equilibrium value \footnote{Due to the involved approximations, the equilibrium value in \cref{eqR_C_pbc_asympt} differs from the exact results in \crefrange{eqR_eqvar_per}{eqR_eqvar_DirNoFl}.}.
In the subdiffusive regime, the two-time correlation functions in \cref{eqR_correl_twotime}, evaluated for $x=L/2$, take the form \cite{krug_persistence_1997, taloni_generalized_2012} \footnote{The equilibrium variance results from \eqref{eqR_hr_correl} as $\lim_{t\to \infty}\bra \delta h(x,t)\delta h(x,t)\ket=2\kbT L/\pi^2$ and differs from the discrete result in \cref{eq_var_pbc} due to the continuum assumption.}
\begin{subequations}\begin{align} 
\bra \delta h(x,t) \delta h(x,s)^* \ket\st{flat} &\simeq (2\kbT/\pi)\frict^{1/z}\Gamma(1-z^{-1})  \left[ (t+s)^{1/z} - |t-s|^{1/z} \right], \label{eqR_hr_rough}\\
\bra \delta h(x,t) \delta h(x,s)^* \ket\st{th} &\simeq (2\kbT/\pi)\frict^{1/z}\Gamma(1-z^{-1})   \left[ t^{1/z} + s^{1/z} - |t-s|^{1/z} \right]. \label{eqR_hr_fbm}
\end{align}\label{eqR_hr_correl}\end{subequations}
Note that the prefactors in the above expression will be different if $x\neq L/2$.
Recalling that the height fluctuations are Gaussian, \cref{eqR_hr_fbm} shows that, in the equilibrium regime, a tagged monomer of a one-dimensional profile performs a \emph{fractional Brownian motion} [cf.\ \cref{eq_fbm_autocorrel}] with Hurst index \cite{krug_persistence_1997}
\beq H=\frac{1}{2z}.
\label{eqR_Hurst_z}\eeq 
In contrast, in the transient roughening regime, the correlation function in \cref{eqR_hr_rough} describes a non-Markovian Gaussian process with non-stationary increments.
In the subdiffusive regime, a tagged monomer $h(x,t)$ traverses a distance $\Delta$ within a characteristic diffusion time $\tau_D$ determined by $\Delta \simeq \bra \delta h(x,\tau_D)^2\ket^{1/2}$. \Cref{eqR_hr_correl}, which applies to $x \simeq L/2$ , yields accordingly
\beq \tau_D \simeq \frac{\Delta^{2z}}{[(2/\pi) \kbT\Gamma(1-z^{-1})]^z c \frict},
\label{eqR_tdiff}\eeq 
with $c=2$ for flat initial conditions and $c=2^z$ for thermal ones.
Within a time $\tau_R$, a tagged monomer has covered a region of a typical extent set by the equilibrium variance, $\bra h^2\ket^{1/2}$ [see \crefrange{eqR_eqvar_per}{eqR_eqvar_DirNoFl}].

For EW dynamics with periodic \bcs, the zero-mode ($k=0$) is absent from \cref{eqR_solh_mode} as a consequence of enforcing the mass constraint [see \cref{eq_zero_vol,eq_height_redef}].
Without this constraint, the two-time correlation function of $h_0$ [cf.\ \cref{eqR_eqcorrel}] results as
\beq \bra h_0(t) h_0^*(s)\ket = 2D \, \text{min}(t,s),
\label{eqR_cm_diffusion}\eeq 
representing standard Brownian diffusion of the center-of-mass of the profile.

\section{Numerical implementation of the Langevin simulations}
\label{app_implement}
\subsection{Discretization}
\label{app_discr}

\begin{figure}[t]\centering
	\includegraphics[width=0.28\linewidth]{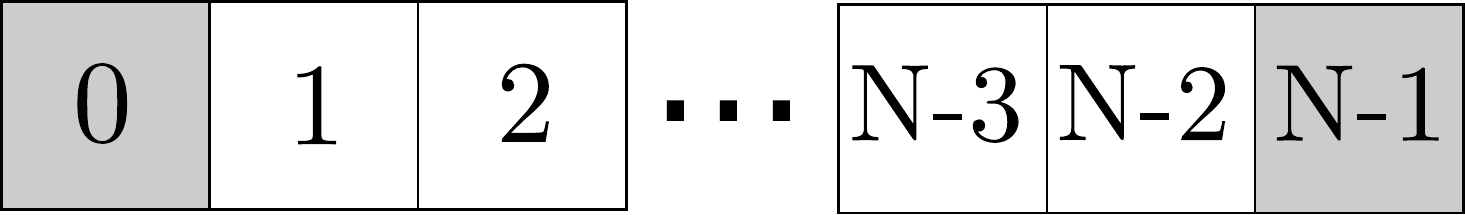}
	\caption{On a one-dimensional lattice of $N$ nodes, Dirichlet \bcs are imposed at the nodes 0 and $N-1$ (highlighted gray), i.e., $h(x_0)=0=h(x_{N-1})$.}
	\label{fig_Dirichlet_sketch}
\end{figure}

As reported in \cref{eq_discretized}, we use in our simulations a standard Euler discretization in time for \cref{eq_EW,eq_MH}:
\beq h(x_i,t+\Delta t) = h(x_i,t) -  \eta\,\Delta t\,(-\nabla^2)^{z/2} h(x_i,t) + \sqrt{2D \Delta t} \nabla^{z/2-1}\zeta(x_i,t),
\label{eq1_dyneq}\eeq 
where $i=0,1,2,\ldots,N-1$, $N=L/\lattsp$ denotes the number of lattice nodes and $\lattsp$ is the lattice spacing.
For notational simplicity, here we have dropped the tilde on $\zeta$.
The required spatial derivatives are discretized based on a standard central difference scheme \cite{zwillinger_handbook_1997}: 
\begin{subequations}\begin{align}
\nabla \zeta_i(x_i) &\equiv \onehalf \left[\zeta(x_{i+1}) - \zeta(x_{i-1})\right], \label{eq1_cengrad} \\
\nabla^2 h(x_i) &\equiv h(x_{i-1}) - 2h(x_i) + h(x_{i+1}), \label{eq1_lapl}\\
\nabla^4 h(x_i) &\equiv h(x_{i-2}) - 4h(x_{i-1}) + 6h(x_i) - 4h(x_{i+1}) + h(x_{i+2}). \label{eq1_bilapl}
\end{align}\end{subequations}
Here and  in the following, the time argument is suppressed and length is expressed in units of $\lattsp$.
We remark that the discretizations for the bi-Laplacian in \cref{eq1_bilapl} and the Laplacian in \cref{eq1_lapl} are related via $\nabla^4 = \nabla^2(\nabla^2)$.
We consider periodic \bcs,
\beq h(x_{N+i})=h(x_{i}),
\label{eq1_pbc}\eeq 
and Dirichlet \bcs 
\beq h(x_{0})=0 = h(x_{N-1}).
\label{eq1_Dir}\eeq 
For Dirichlet \bcs, \cref{eq1_dyneq} is evaluated only at the nodes $1\leq i\leq N-2$ (see \cref{fig_Dirichlet_sketch}).
Since the discretized Laplacian in \cref{eq1_lapl} requires only the values of $h$ at the nearest neighbors, the \bcs defined in \cref{eq1_pbc,eq1_Dir} fully determine the discretized EW dynamics.
In contrast, in the case of the MH equation, the discretized bi-Laplacian in \cref{eq1_bilapl} involves also next-nearest neighbors and is thus \emph{a priori} undefined at the boundary nodes $i\in \{1,N-2\}$.
For Dirichlet no-flux \bcs, an expression of the discretized bi-Laplacian at the boundary can be determined by requiring conservation of the total mass $\mass=\sum_{i=1}^{N-2} h(x_i)$ at each time step.
Within the domain $2\leq i\leq N-3$, the deterministic contribution to the change of the mass, i.e., the contribution stemming from the second term on the r.h.s.\ of \cref{eq1_dyneq}, is obtained as
\beq \sum_{i=2}^{N-3} \nabla^4 h(x_i) =  -3h(x_1) +3h(x_2) -h(x_3) -h(x_{N-4}) + 3h(x_{N-3}) -3h(x_{N-2}),
\label{eq1_bilapl_sum}\eeq 
where \cref{eq1_bilapl,eq1_Dir} have been used and any prefactors are omitted.
Accordingly, the simplest choice for $\nabla^4$ at the nodes $i\in \{1,N-2\}$ ensuring vanishing of the total deterministic mass change is given by: 
\begin{subequations}\begin{align}
\nabla^4 h(x_1)  &= 3 h(x_1) - 3h(x_2)+ h(x_3), \\
\nabla^4 h(x_{N-2})  &= h(x_{N-4}) - 3h(x_{N-3}) + 3h(x_{N-2}).
\end{align}\label{eq1_bilapl_bound}\end{subequations}
Concerning the stochastic contribution to the mass change, the last term on the r.h.s.\ of \cref{eq1_dyneq} yields
\beq 
\sum_{i=1}^{N-2} \nabla \zeta(x_i) = \onehalf \left[ -\zeta(x_0) - \zeta(x_1) + \zeta(x_{N-2}) + \zeta(x_{N-1})\right].
\label{eq1_noise_sum}\eeq
In order for this expression to vanish, a choice for $\zeta(x_i)$ at the boundary nodes $i=0,N-1$ must be made. 
Given \cref{eq1_Dir}, it appears natural to set 
\begin{subequations}\begin{align}
\zeta(x_0) = &0 = \zeta(x_{N-1}),  \\
\intertext{which, by \cref{eq1_noise_sum}, implies }
\zeta(x_1) = &0 = \zeta(x_{N-2}).
\end{align}\label{eq1_noise_bound1}\end{subequations}
The above choice is not unique, but minimizes artificial correlations.
Alternatively, one may set $\zeta(x_0) = -\zeta(x_1)$ and $\zeta(x_{N-1}) =- \zeta(x_{N-2})$. This choice been checked in a number of cases to yield similar results to the prescription in \cref{eq1_noise_bound1}.
For periodic \bcs, finally, it is straightforward to prove that mass is exactly conserved by \cref{eq1_dyneq}.

It can be readily shown that \cref{eq1_bilapl_bound} in fact implies a vanishing (discretized) flux at the boundaries, i.e., $\nabla \mu(x_i)=0$ for $i=1,N-2$, where 
\beq \mu(x_i) \equiv -\nabla^2 h(x_i) = - [ h(x_{i-1}) -2h(x_i) + h(x_{i+1}) ]
\label{eq1_chempot}\eeq 
is the chemical potential.
To this end, we introduce the forward difference 
\beq  \nabla^F h(x_i) \equiv h(x_{i+1}) - h(x_i),
\label{eq1_fwdgrad}\eeq 
in terms of which the Laplacian of $\mu$ can be written as
\beq \begin{split}
\nabla^2 \mu(x_i) = -\nabla^4 h(x_i) &= [\mu(x_{i-1}) - \mu(x_i)] + [\mu(x_{i+1}) - \mu(x_i)] \\
&= - \nabla^F \mu(x_{i-1}) + \nabla^F \mu(x_i) ,\qquad i=1,\ldots,N-1.
\end{split}\label{eq1_lapl_chempot}\eeq 
Upon imposing no-flux \bcs in the discretized form \footnote{The symmetry of \cref{eq1_noflx_chempot} is more clearly revealed by writing the second relation as $\nabla^B \mu(x_{N-1})=0$, where $\nabla^B \mu(x_i) = \mu(x_{i}) -  \mu(x_{i-1}) = \nabla^F \mu(x_{i-1})$ is the backward difference.}
\beq \nabla^F \mu(x_{0})=0=\nabla^F \mu(x_{N-2}),
\label{eq1_noflx_chempot}\eeq 
one recovers the expressions in \cref{eq1_bilapl_bound}:
\begin{subequations}\begin{align}
\nabla^4 h(x_1) &= \mu(x_1) - \mu(x_2),\\ 
\nabla^4 h(x_{N-2}) &= \mu(x_{N-2}) - \mu(x_{N-3}). 
\end{align}\end{subequations}

We finally recall some useful properties related to the eigenmode decomposition of the profile for various \bcs.
In the case of periodic \bcs, $h(x)$ can be expressed in terms of its Fourier modes as
\beq h\pbc(x) = \sum_{q=0}^{N-1} \exp\left(\frac{2\pi \im q}{L} x \right) h_q\pbc.
\label{eq1_h_Four}\eeq
Correspondingly, taking into account the discrete nature of $h(x)$, the Fourier coefficients $h_q$ are given by 
\beq h_q\pbc = \frac{1}{N} \sum_{k=0}^{N-1} \exp\left(-\frac{2\pi \im}{L}q k\lattsp\right) h\pbc(k\lattsp),
\label{eq1_h_invFour}\eeq 
where we reinstated the lattice spacing $\lattsp$.
Note that 
\beq \sum_{k=0}^{N-1} \exp\left(\frac{2\pi\im}{N} q k\right) = N\delta_{q,N\mathbb{Z}},
\eeq 
where $\delta_{q,N\mathbb{Z}}\equiv 1$ if $q$ is an integer multiple of $N$, and zero otherwise.
Since \cref{eq1_h_invFour} implies $h_{N-q}=h_q^*$, a real-valued $h(x)$ is completely determined by its (complex) Fourier coefficients $h_q$ within the first Brillouin zone, $q=0,1,\ldots,\lceil (N-1)/2\rceil$, where $\lceil x\rceil$ denotes the largest integer smaller than or equal to $x$.

In the case of standard Dirichlet \bcs, $h(x)$ can be analogously decomposed as
\beq h\Dbc(x) = \sum_{q=0}^{N-1} \sin\left(\frac{\pi q}{L} x \right) h_q\Dbc,
\label{eq1_h_Direxp}\eeq 
with the inverse relation
\beq h_q\Dbc = \frac{2}{N} \sum_{k=0}^{N-1} \sin\left(\frac{\pi q}{L} k\lattsp\right) h\Dbc(k\lattsp).
\eeq 
The orthogonality property is given by
\beq \sum_{k=0}^{N-1} \sin\left(\frac{\pi p}{N}k\right) \sin\left(\frac{\pi q}{N}k\right) = 
\begin{cases}
\frac{N}{2}\delta_{p,q},\qquad & q,p\neq 0,\\
0,\qquad & p=q=0,
\end{cases}
\label{eq1_DirCP_ortho}\eeq 
assuming $0\leq q,p\leq N-1$.
Indeed, since $\sin(\pi (2N-q) k / N) = -\sin(\pi q k /N)$, the specification of the expansion coefficients $h_q$ for $0\leq q\leq N-1$ completely determines $h\Dbc(x)$ on a lattice of $N=L/\lattsp$ points and, accordingly, one has $h\Dbc_{2N-q} = -h\Dbc_q$ \footnote{The mode $q=0$ is usually not considered as part of the spectrum for Dirichlet \bcs.}

\subsection{Benchmarks}
\label{app_bench}

\begin{figure}[t]\centering
	\includegraphics[width=0.5\linewidth]{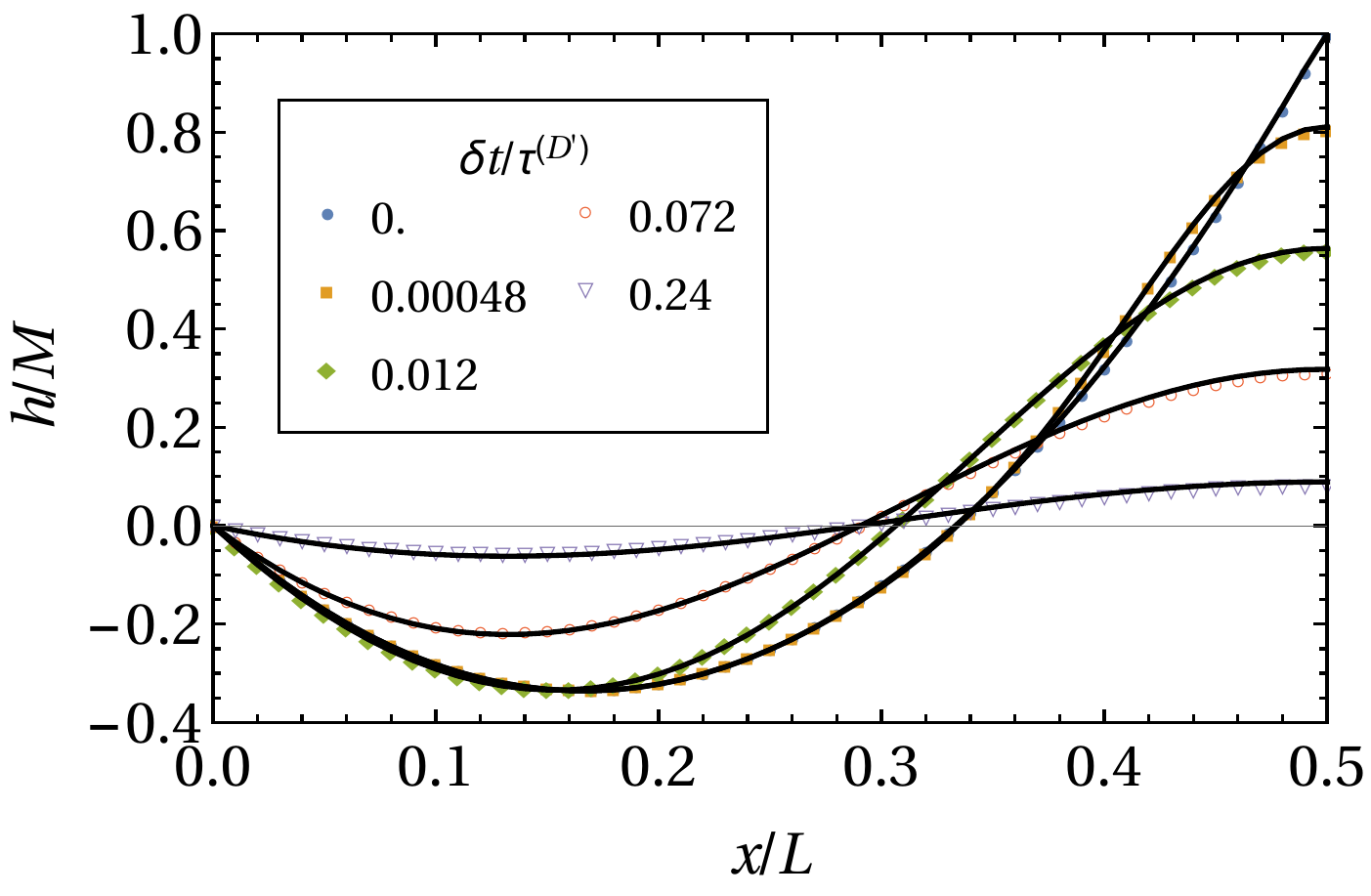}
	\caption{Deterministic relaxation of a profile with Dirichlet no-flux \bcs for the noiseless MH equation [see \cref{eq1_dyneq}]. The profile is initialized with the  static first-passage profile obtained within WNT in the equilibrium regime, given by Eq.~I-(C63) for $\dt=0$ and $x_M=L/2$. Symbols represent simulation results, while solid curves represent the time-dependent first-passage profile $h(x,\dt)$ predicted by WNT [Eq.~I-(C63)] for the same value of $x_M$. Time is expressed in terms of the relaxation time $\tau\DirNoFl$ [\cref{eq_relaxtime}].}
	\label{fig_bench_det_relax}
\end{figure}

\begin{figure}[t]\centering
    \subfigure[]{\includegraphics[width=0.44\linewidth]{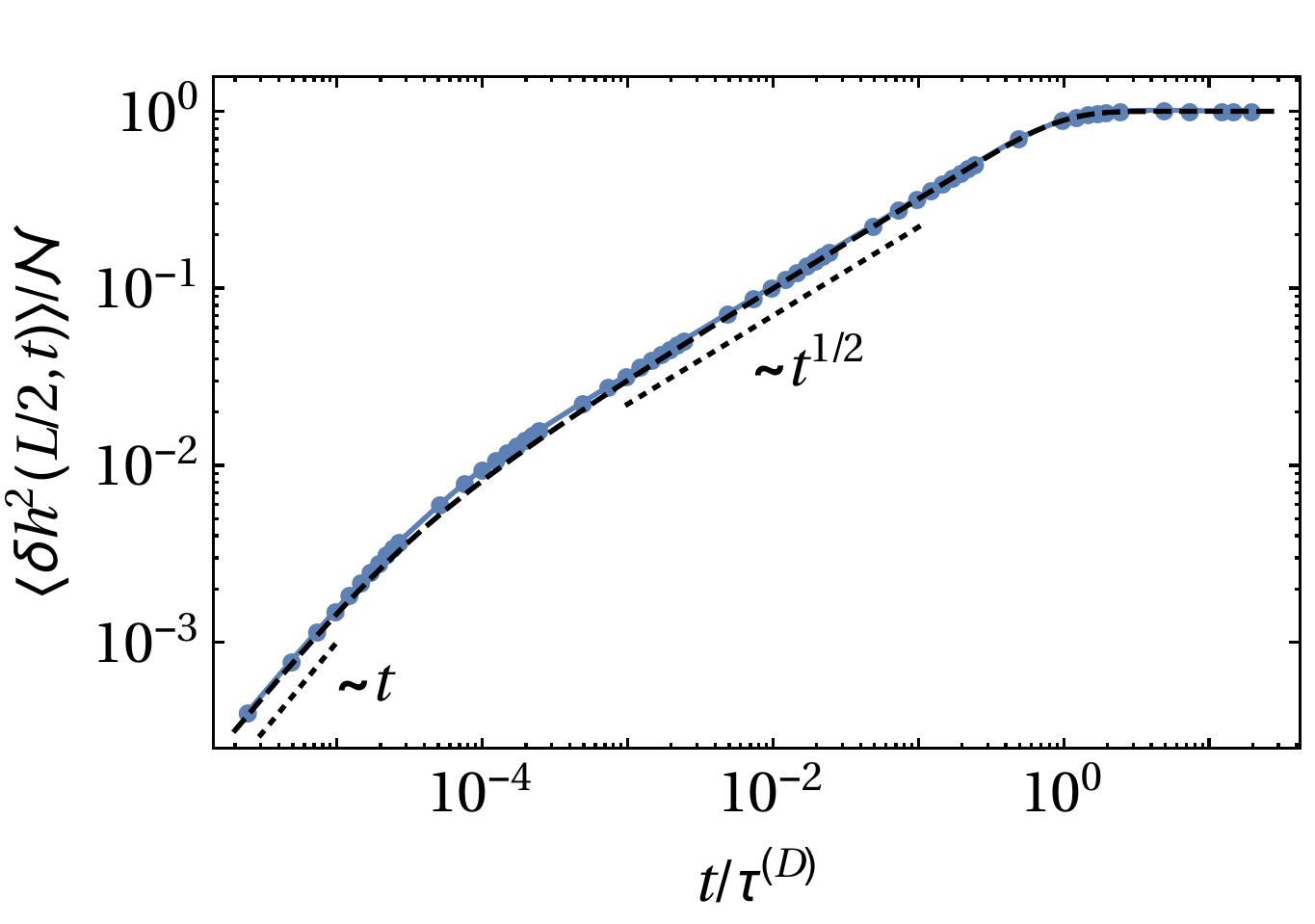}}\qquad
    \subfigure[]{\includegraphics[width=0.44\linewidth]{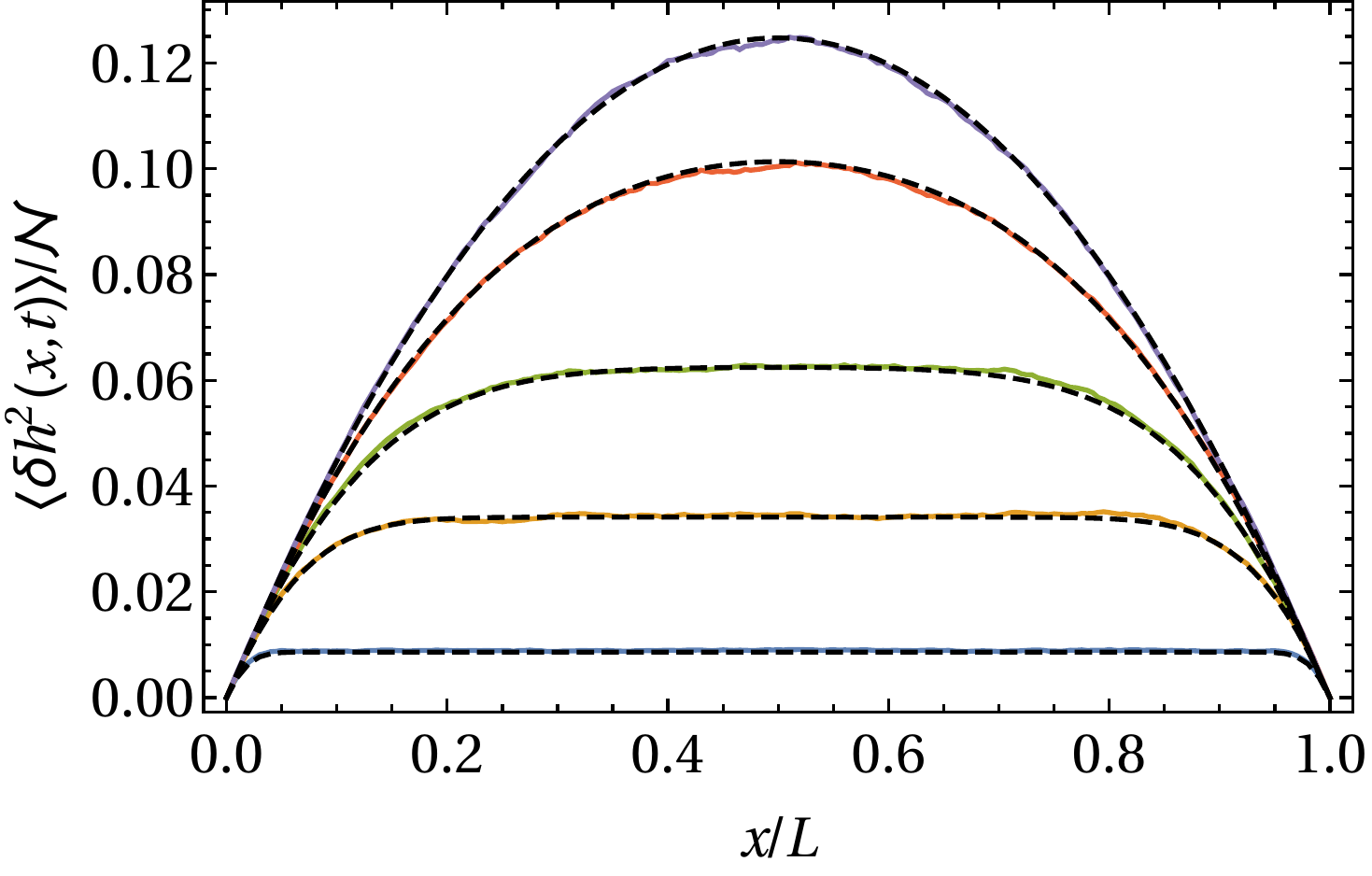}}
    \subfigure[]{\includegraphics[width=0.44\linewidth]{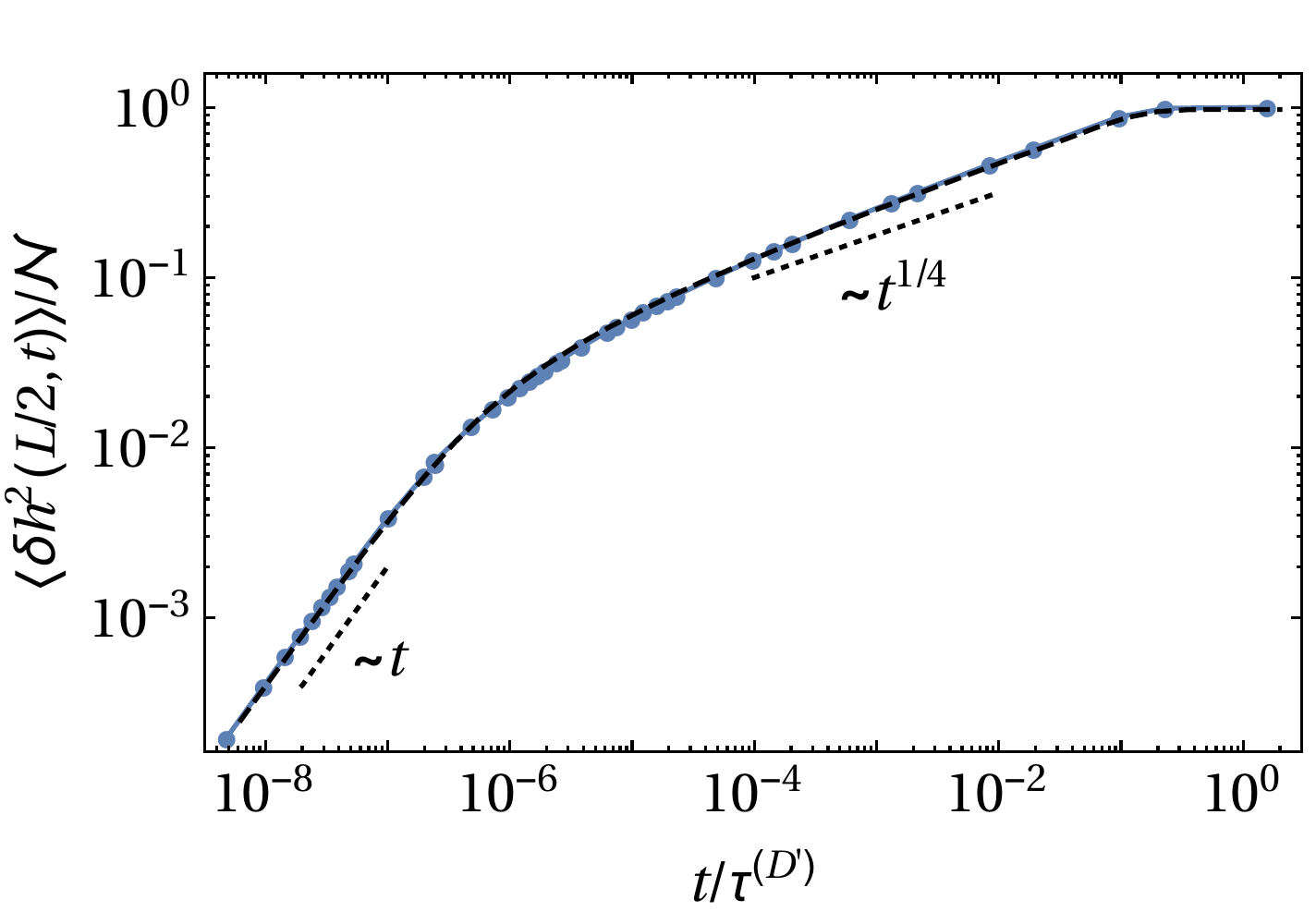}}\qquad
    \subfigure[]{\includegraphics[width=0.44\linewidth]{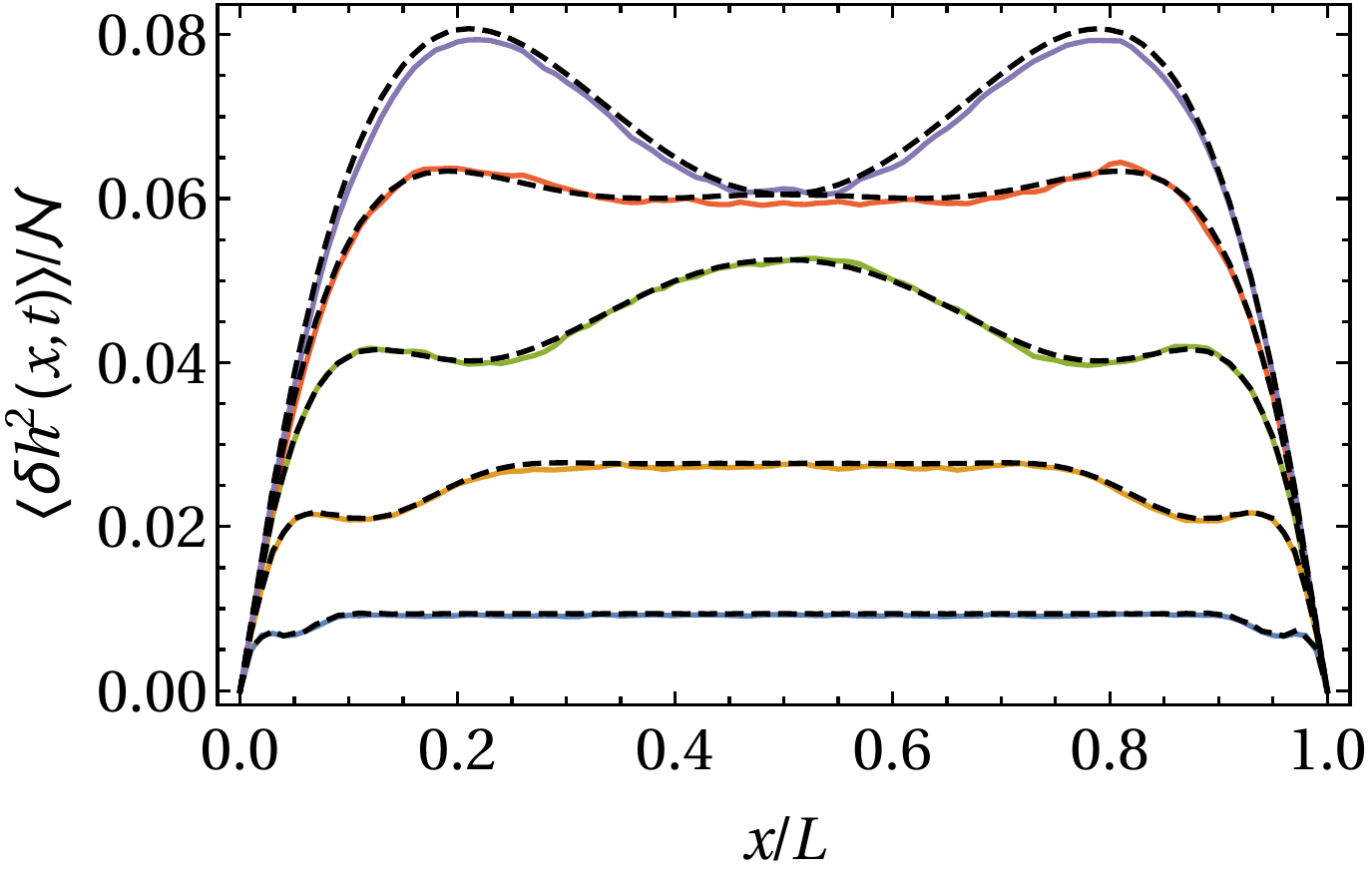}}
    \caption{Non-equilibrium roughening of a profile for (a,b) EW dynamics with (standard) Dirichlet \bcs [\cref{eq_H_Dbc}], and (c,d) MH dynamics with Dirichlet no-flux \bcs [\cref{eq_H_Dbc,eq_H_noflux}]. The profile is initialized at time $t=0$ in a flat configuration [\cref{eq_init_cond}]. Panels (a,c) show the time evolution of the roughness $\bra |\delta h(L/2,t)|^2\ket$ [\cref{eqR_roughness}] at the mid-point, while panels (b,d) show the roughness as a function of $x$ for various times. Simulation data are represented in (a,c) by the connected symbols and in (b,d) by the solid lines. The dotted lines in (a,c) represent power-laws expected asymptotically at early and intermediate times [see \cref{eqR_C_pbc_asympt}]. For sufficiently large system size $L$, the exponent characterizing the intermediate time regime is predicted to be $1/z$ independently from the \bcs. The dashed curves in (b,d) represent the analytical predictions in \cref{eqR_roughness_flat}, normalized (via the factor $\mathcal{N}$) to the corresponding long-time limits [see \cref{eqR_eqvar_DirCP,eqR_eqvar_DirNoFl}].}
    \label{fig_bench_roughen}
\end{figure}

We now assess the accuracy of the discretizations in \cref{eq1_dyneq} with a few benchmarks.
\Cref{fig_bench_det_relax} illustrates the deterministic relaxation of a profile governed by the \emph{noiseless} MH equation with Dirichlet no-flux \bcs.
As initial configuration (at time $\dt=0$) we take here the first-passage profile obtained from the solution of WNT in Eq.~I-(3.19) for $T\to\infty$ and $x_M=L/2$.
Since this profile pertains to the equilibrium regime, the relaxation solution shown in \cref{fig_bench_det_relax} is expected to be the identical to the time-inversed activation solution obtained within WNT for the same value of $x_M$. A convenient form of the activation solution, expressed in terms of the time variable $\dt=T-t$, is provided in Eq.~I-(C63).
We find close agreement between the simulation results (symbols) and WNT (solid curves). 

\Cref{fig_bench_roughen} illustrates interfacial roughening for the EW and MH equations with Dirichlet (no-flux) \bcs and a flat initial configuration [\cref{eq_init_cond}].
Simulation results (symbols and solid lines) obtained for the time-dependent variance $\bra \delta h^2(x,t)\ket$ [normalized to its long-time limit $\bra \delta h^2(x,t\to \infty)\ket$] are shown in panels (a,c) as a function of time for $x=L/2$ and in panels (b,d) as a function of $x$ for various times.
The dashed curves in \cref{fig_bench_roughen} represent the analytical predictions reported in \cref{eqR_roughness_flat} [normalized to the equilibrium variance $\mathcal{N}$ given in \cref{eqR_eqvar_DirCP,eqR_eqvar_DirNoFl}]. 
(We remark that, for periodic \bcs, the variance evolves in essentially the same fashion as in \cref{fig_bench_roughen}(a,c).)
In agreement with \cref{eqR_C_pbc_asympt}, the variance grows linearly in time for $t\lesssim \tau\cro$, followed by an algebraic growth with exponent $1/z$ for $\tau\cro\lesssim t\lesssim \tau$, where $\tau\cro$ and $\tau$ are the crossover and the relaxation time, respectively [see \cref{eq_crossover_time,eq_relaxtime}].
In the case of MH dynamics, the slight deviation of the simulation results from the expected value $1/4$ of the power-law exponent at intermediate times is found to gradually diminish upon increasing the system size $L$.

In order to determine the crossover time $\tau\cro$, recall that for standard Dirichlet \bcs, one has $L/\Delta x-1\simeq L/\lattsp$ distinct wavemodes ($k=1,\ldots,L/\Delta x-1$) on a lattice of size $L$ [see \cref{eq1_DirCP_ortho} and the related discussion]. 
As observed in \cref{fig_bench_roughen}(a), the resulting crossover time $\tau\cro$ is correctly captured by the solution in \cref{eqR_C_pbc_asympt} (dashed curve).
For Dirichlet no-flux \bcs, there is no symmetry between the eigenmodes $\sigma_k\DirNoFl(x)$ [see \cref{tab_eigenfunc}] for small and large $k$, hence the largest possible eigenmode is not easily obtained. 
In order to gain further insight into this issue, we determine the stability of an eigenmode  via numerical simulation. 
To this end, the noiseless relaxation of a profile $h(x,t)$, initialized as $h(x,0)=\sigma_k\DirNoFl(x)$, is simulated under MH dynamics.
For a system size of, e.g., $L=100\lattsp$, we find that eigenmodes with $k\lesssim L/(2\lattsp)$ typically keep their shape during the time evolution, i.e., $h(x,t)/h(x\st{ref},t)\simeq \sigma_k\DirNoFl(x)$, where $x\st{ref}$ is a suitable reference position. 
For $k\gtrsim L/(2\lattsp)$, instead, the profile is strongly disturbed during the evolution, indicating that the corresponding eigenmode is unstable under the discretization used here. 
The instability is amplified upon increasing $k$.
We find that, using for the evaluation of \cref{eqR_roughness_flat} a value of $k\simeq 0.6 L/\Delta x$ for the largest mode number, accurately captures the crossover-time observed in \cref{fig_bench_roughen}(c).


%

\end{document}